\documentclass[draftcls,onecolumn,10pt]{IEEEtran}

\usepackage[pdftex]{graphicx}
\usepackage{url}

\usepackage{geometry}
\usepackage{amssymb} 
\usepackage{cite}
\usepackage[cmex10]{amsmath}
\usepackage{amsthm}
\usepackage{dsfont}
\usepackage{color}

\usepackage{epstopdf}
\usepackage[margin=2ex,font=scriptsize]{subcaption}

\DefineNamedColor{named}{Purple}{cmyk}{0.45,0.86,0,0}

\usepackage{benstyle}
\newcommand*{\rb}{\mathrm{b}}
\newcommand*{\ri}{\mathrm{i}}
\newcommand*{\rf}{\mathrm{f}}
\newcommand*{\rp}{\mathrm{p}}
\newcommand*{\rw}{\mathrm{w}}

\begin{document}

\title{On Asymptotic Incoherence and its Implications for Compressed Sensing of Inverse Problems}
\author{A. D. Jones, B. Adcock and A. C. Hansen%
\thanks{This work was presented in part as a poster at Matheon CSA2013 in Berlin during December 2013.}%
\thanks{A. D. Jones is with the University of Cambridge. B. Adcock is with Simon Fraser University. A. C. Hansen is with the University of Cambridge and the University of Oslo.}%
}

\markboth{IEEE Transactions on Information Theory}{}

\date{}

\maketitle

\begin{abstract}
Recently, it has been shown that incoherence is an unrealistic assumption for compressed sensing when applied to many inverse problems.  Instead, the key property that permits efficient recovery in such problems is so-called local incoherence.  Similarly, the standard notion of sparsity is also inadequate for many real world problems.  In particular, in many applications, the optimal sampling strategy depends on asymptotic incoherence and the signal sparsity structure. The purpose of this paper is to study asymptotic incoherence and its implications towards the design of optimal sampling strategies and efficient sparsity bases. It is determined how fast asymptotic incoherence can decay in general for isometries. Furthermore it is shown that Fourier sampling and wavelet sparsity, whilst globally coherent, yield optimal asymptotic incoherence as a power law up to a constant factor.  Sharp bounds on the asymptotic incoherence for Fourier sampling with polynomial bases are also provided. A numerical experiment is also presented to demonstrate the role of asymptotic incoherence in finding good subsampling strategies.
\end{abstract}

\begin{IEEEkeywords}
Compressed sensing, Fourier transforms,  nonuniform sampling, polynomials, wavelet transforms.
\end{IEEEkeywords}

\section{Introduction}
\IEEEPARstart{C}{ompressed} sensing, introduced by Cand\`es, Romberg \& Tao \cite{CandesRombergTao} and Donoho \cite{donohoCS},  has been one of the major achievements in applied mathematics in the last decade \cite{candesCSMag,EldarDuarteCSReview,EldarKutyniokCSBook,FornasierRauhutCS,FoucartRauhutBook}. By exploiting additional structure such as sparsity and incoherence, one can solve inverse problems by uniform random subsampling and convex optimisation methods, and thereby recover signals and images from far fewer measurements than conventional wisdom suggests. 

\begin{figure}
\begin{center}
\includegraphics[width=7cm]{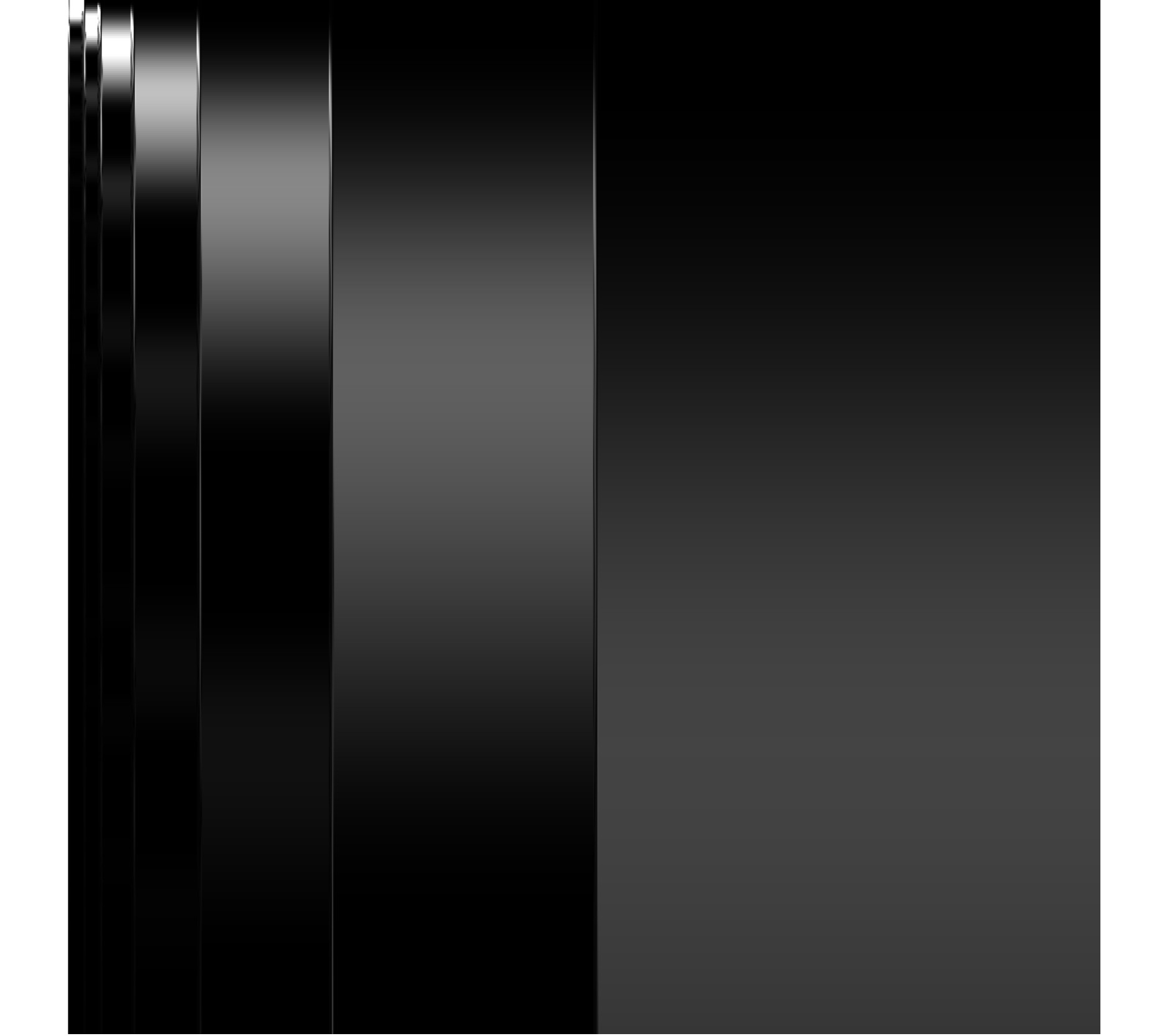}
\includegraphics[width=7cm]{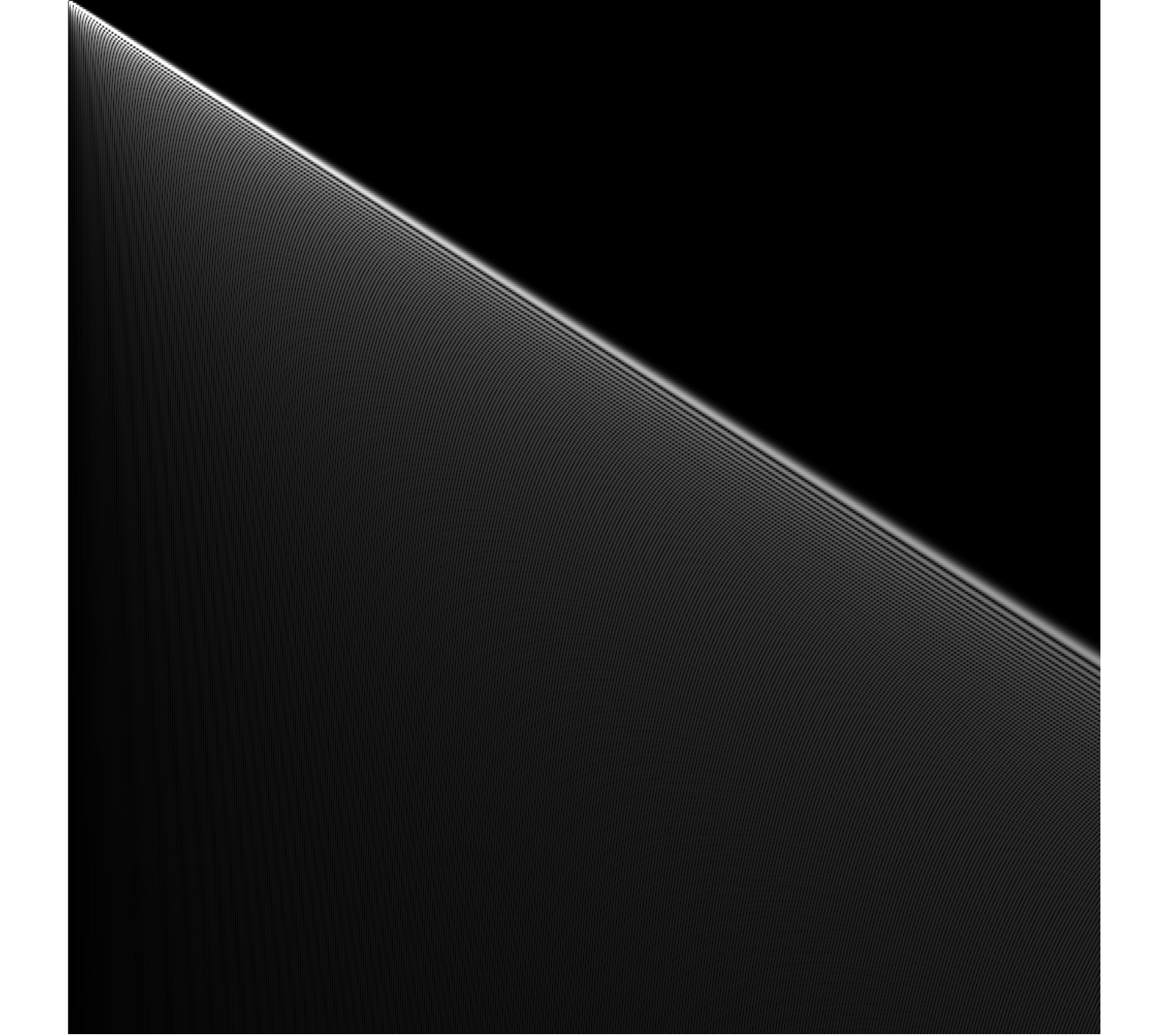}
\caption{Plots of the absolute values of the entries of the compression matrix $U$ 
corresponding to Fourier sampling with Daubechies6 boundary wavelets (left) and Legendre 
polynomials (right).  Light regions correspond to large values and dark 
regions to small values. $U_{1,1}$ corresponds to the top-left entry. Notice that if these matrices were incoherent, then these figures would have been monotone. This means an alternative to coherence is needed to study infinite dimensional compression matrices.}
\label{fig:Aplot}
\end{center}
\end{figure}

However, in many applications -- including Magnetic Resonance Imaging (MRI) \cite{Unser,Lustig3}, X-ray Computed Tomography \cite{Stanford_CT, quinto2006xrayradon}, Electron Microscopy  \cite{lawrence2012et,leary2013etcs}, etc -- incoherence is always lacking if one tries to keep the model in its original continuous form, as can be seen from Figure \ref{fig:Aplot}. 
The reason for this can be traced to the observation that many classical inverse problems are based on continuous  integral transforms such as the Fourier transform: 

\begin{equation}\label{inverse_problem}
g = \mathcal{F}f, \qquad f \in \rL^2(\mathbb{R}^d), \quad \mathcal{F}f(\omega) = \int_{\mathbb{R}^d} f(x) e^{-2\pi i \omega \cdot x }  \, dx.
\end{equation}
In this case the resulting recovery problem is that of reconstructing an unknown function $f$ from pointwise samples of $g$.

In compressed sensing, such a transform is combined with an appropriate sparsifying transformation associated to a basis or frame, giving rise to an infinite \textit{measurement} matrix. The \emph{coherence} of an infinite matrix\footnote{The notation $\mathcal{B}( \ell^2(\bbN))$ refers to the space of bounded linear maps from $\ell^2(\bbN)$ to itself.} $U \in \mathcal{B}( \ell^2(\bbN))$ or a finite matrix $W \in \bbC^{N \times N}$ is defined as
\be{ \label{Eq:CoherenceDefinitions}
\mu(U)=\max_{i,j \in \bbN} | U_{ij} |^2, \qquad \mu(W)=\max_{i,j =1,\cdots,N} | W_{ij} |^2.
} 
In the finite-dimensional case, the matrix $W$ is typically a change of basis matrix from some \emph{Sampling Basis} $(\rho(m))_{m=1}^N$ to some \emph{Reconstruction Basis} $(\tau(m))_{m=1}^N$ of $\bbC^N$:
\be{ \label{Eq:ChangeOBasisFinite}
	W_{m,n}= \langle \tau(n), \rho(m) \rangle.
}
Similarly for the infinite-dimensional case, the matrix $U$ is instead formed from a sampling basis $(\rho(m))_{m \in \bbN}$ and reconstruction basis $(\tau(m))_{m \in \bbN}$ of $L^2(\bbR^d)$:
\be{ \label{Eq:ChangeOBasisInfinite}
	U_{m,n}= \langle \tau(n), \rho(m) \rangle.
}
The term \emph{incoherence} refers to $\mu(U), \mu(W)$ being small. In the finite case, if we assume $W$ is an isometry with respect to the Euclidean norm, then the statement that $W$ is incoherent can be interpreted as $W$ being evenly flat or spread out. In the limiting case where $\mu(W)=N^{-1}$, every entry must have the same absolute size and the matrix is perfectly flat. In the infinite case, such a notion of uniform flatness is impossible for an isometry $U \in \mathcal{B}( \ell^2(\bbN))$, since its columns are also infinite and normalised.

In the search for an alternative, examples often guide the way. Wavelets, or their various generalizations, are frequently used as the sparsifying transformation, and for smooth functions, one often considers orthogonal polynomials.  As Figure \ref{fig:Aplot} reveals, although such a measurement matrix is clearly not incoherent, it is \textit{asymptotically incoherent}: that is to say, the high coherence terms (large matrix entries) are concentrated around a submatrix of $U$. 

Therefore one should not only consider the notion of the coherence of a sensing matrix $U$, but also a notion of \emph{local coherence}, which we \emph{define} as the coherence over a submatrix of $U$. One way to generate local coherences is by using the following projection operators $\pi_N, R_N: \ell^2(\bbN) \to \ell^2(\bbN)$
\be{ \label{projectiondefinitions}
\begin{aligned}
\pi_N(x)_i:= \begin{cases}
 0 & i \neq N
\\  x_i & i = N
\end{cases} ,
\qquad
R_N(x)_i:= \begin{cases}
 0 & i < N
\\  x_i & i \ge N
\end{cases} ,
\end{aligned}
}
and look at the decay of the corresponding \emph{line and block coherences} 
\be{ \label{Eq:LineAndBlock}
\mu(\pi_N U), \ \mu(U\pi_N), \ \mu(R_N U), \ \mu(U R_N).
}

The term `Line coherence' refers to $\mu(\pi_NU) \ \big(\mu(U\pi_N)\big)$ being equal to the squared absolute maximum of the $N$th row (column) of $U$. Likewise $\mu(R_NU) \ \big(\mu(UR_N)\big)$ is equal to the squared absolute maximum over $U$ without the first $N-1$ rows (columns). `Asymptotic incoherence' refers to the decay of the line/block coherences as $N\to \infty$.

Notions of local coherence have been studied before \cite{BookAHRT, AHPRBreaking, discrete}. For example, in \cite{discrete} the ``local coherence'' between two bases $\Phi=(\varphi_j)_{j=1, \cdots , N}, \Psi=(\psi_k)_{j=1, \cdots, N}$ of $\bbC^N$ was defined as

\begin{equation} \label{Eq:DiscreteCoherence}
\mu_j^{loc}(\Phi, \Psi)= \sup_{1\le k \le N} | \langle \varphi_j, \psi_k \rangle|, \quad j=1,\cdots, N.
\end{equation}
This is analogous to a discrete version of the line coherence $\mu(\pi_NU)$.

It is also worth mentioning that lack of coherence is not the only problem that people are faced with when trying to model continuous problems of Fourier type. For example in NMR based problems such as MRI samples are not taken pointwise but in paths or lines \cite{MatchingBlockDist}, restricting the ability to subsample freely.

\subsection{Order Notations and Conventions} \label{Sub:Order}

Throughout this paper we will be using the following notations and conventions to succinctly describe various types of decay; for $f,g:\bbN \to \bbR_{>0}, \ 0 \notin \bbN$:
\be{ \label{Eq:DecayDefinitions}
\begin{aligned}
f(N)=\mathcal{O}(g(N)) &\Leftrightarrow  \exists C>0 \ s.t. \ f(N) \le C \cdot g(N) \ \forall N \in \bbN,
\\
f(N)=o(g(N)) &\Leftrightarrow  \forall C>0 \ f(N) \le C \cdot g(N) \ \forall N \in \bbN,
\\
f(N)=\Theta(g(N)) &\Leftrightarrow  \exists  C_1,C_2>0 \ s.t. \ C_1 \cdot g(N) \le f(N) \le C_2 \cdot g(N) \ \forall N \in \bbN.
\end{aligned}
}
Moreover for $f,g: S \to \bbR_{>0}$ where $S$ is a set we write 
\be{ \label{Eq:LesssimDefine}
f \lesssim g \Leftrightarrow \exists C>0 \ s.t. \ f(s) \le C \cdot g(s) \ \forall s \in S.
}

\subsection{Compressed Sensing and the Coherence Barrier}
Let us now provide some background regarding compressed sensing and incoherence.

Working in the finite dimensional case, standard compressed sensing theory \cite{Candes_Plan,BAACHGSCS} says that if $x \in \mathbb{C}^N$ is $s$-sparse, i.e.\ $x$ has at most $s$ nonzero components, then, with probability exceeding $1-\epsilon$, $x$ is the unique minimiser to the problem  
\be{
\label{fin_dim_l1}
\min_{\eta \in \bbC^N} \| \eta \|_{l^1} \quad \mbox{subject to} \quad P_{\Omega} W \eta 
= P_{\Omega} Wx,
}
where $P_{\Omega}$ is the projection onto $\mathrm{span}\{e_j:j\in \Omega\}$, $\{e_j\}$ is the canonical basis, $\Omega$ is chosen uniformly at random with $|\Omega| = m$ and
\be{
\label{m_est_Candes_Plan}
m \gtrsim  \mu(W) \cdot N \cdot s \cdot \log (\epsilon^{-1}) \cdot \log (N),
}

The estimate \R{m_est_Candes_Plan} demonstrates how the three pillars of compressed sensing -- sparsity, incoherence and uniform random subsampling -- combine to allow for recovery with substantial subsampling.  Coherence-based sampling has proved effective in a number of different CS scenarios, e.g. polynomial interpolation \cite{PolyExpB}. However, now suppose that $\mu(W)$ is large; for example, $\mu(W) \cdot N = \mathcal{O}(N)$ as $N \rightarrow \infty$.  In this case, \R{m_est_Candes_Plan} suggests that no dramatic subsampling is possible: that is, we must take roughly $N$ samples to recover $x$, even though $x$ is often extremely sparse.  We refer to this phenomenon as the {\it coherence barrier}. 

\subsection{The flip test: How to get the model right}
The discussion above and Figure \ref{fig:Aplot} suggests that incoherence may not be the appropriate tool to describe compressed sensing when used in infinite dimensional problems involving basis transforms, as is the case with MRI using Fourier samples. Even though there has already been some progress on developing sampling methods dependent on local coherence and conventional sparsity \cite{discrete}, it turns out sparsity must be revised as well. The flip test is a numerical tool designed to verify what kind of sparsity structure we actually recover in practical applications such as MRI, CT, Electron microscopy etc.

The initial test was first introduced in \cite{AHPRBreaking}, and we will present variants of it here. The first flip test was designed to answer the following question: Is the success of subsampling techniques used in compressed sensing independent of the location of the coefficients to be recovered? In other words is sparsity the right model for compresses sensing?

\begin{figure}[!h]
\begin{center}
\includegraphics[width=0.3\textwidth]{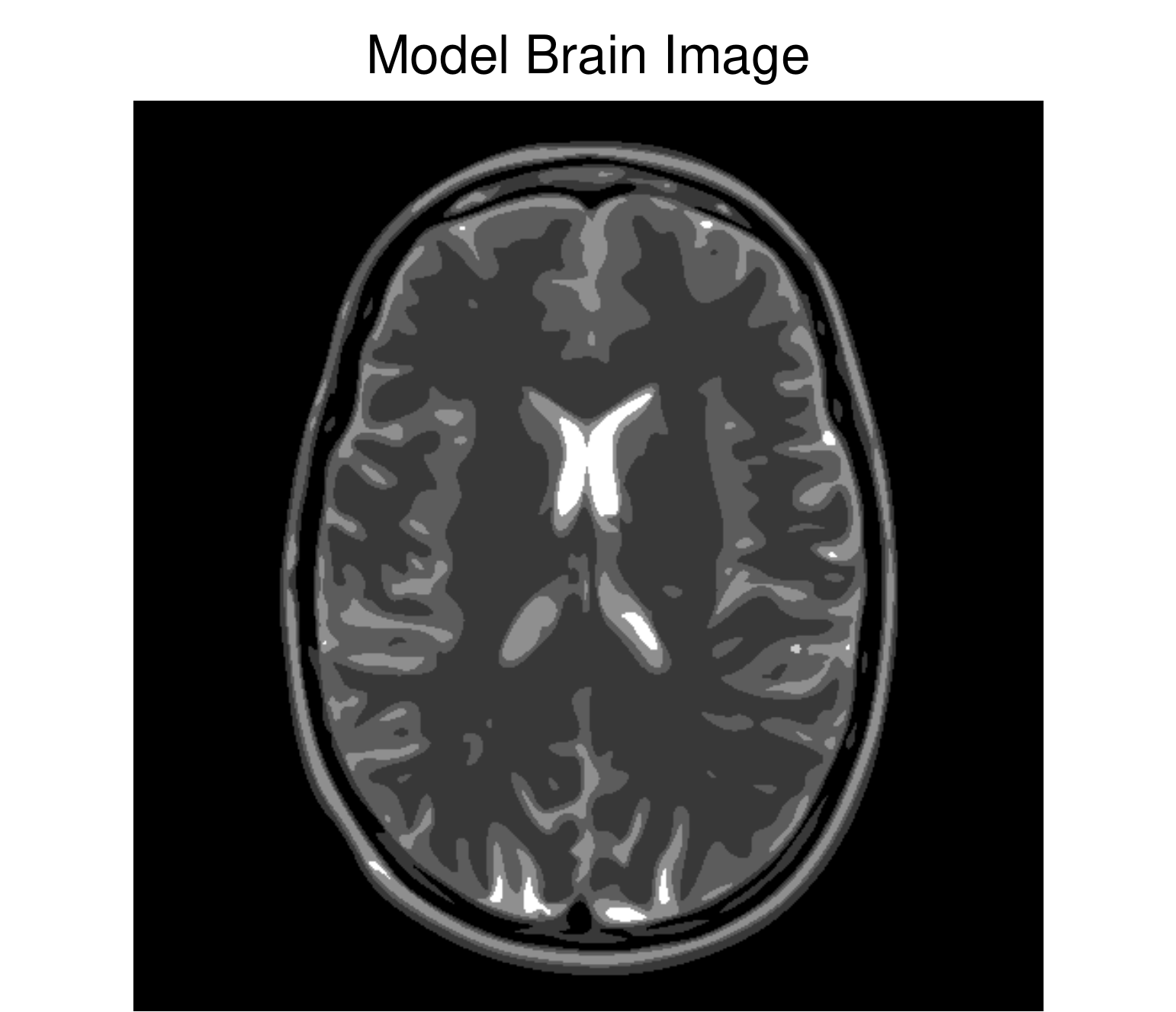}
\includegraphics[width=0.3\textwidth]{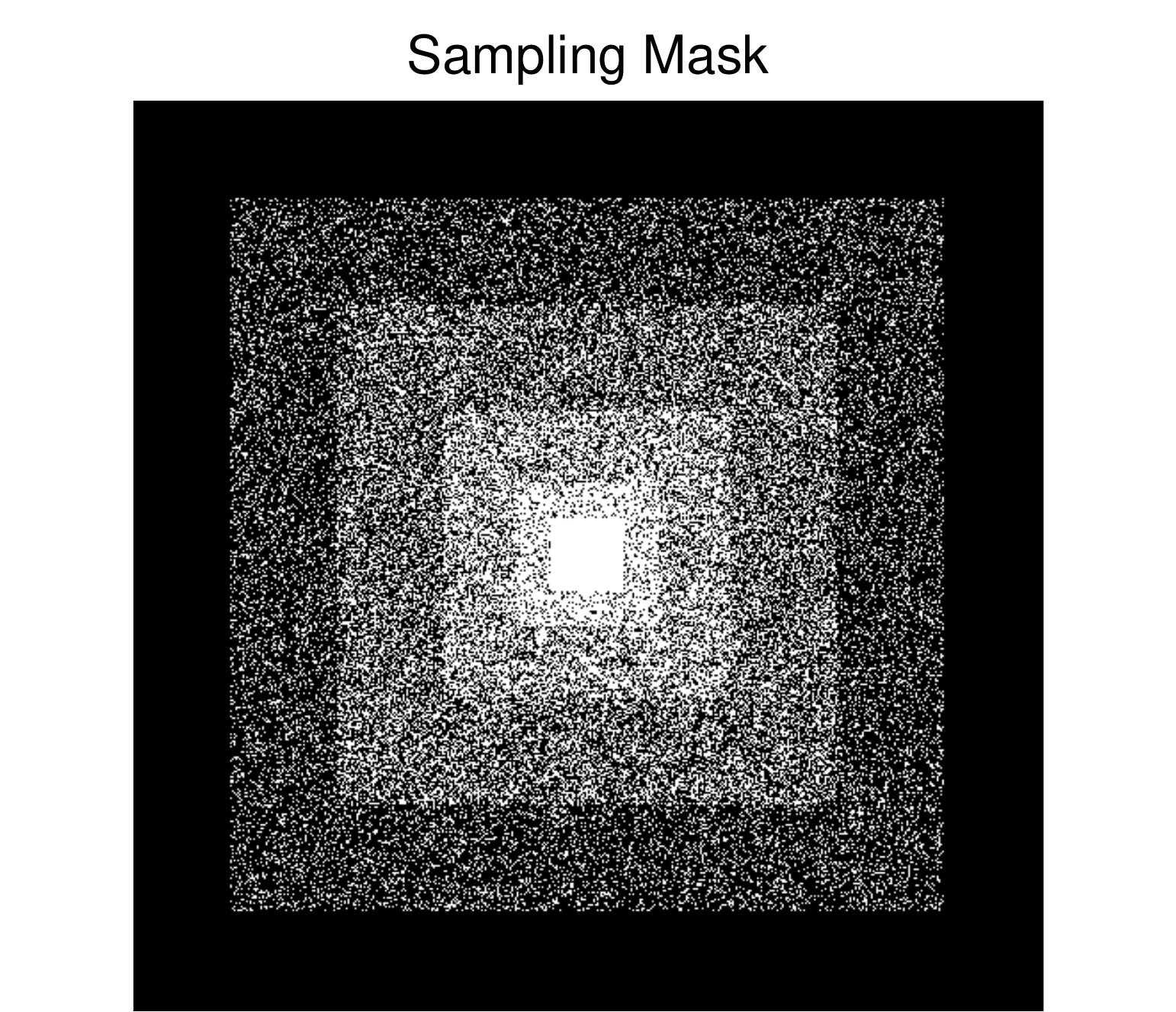}

\includegraphics[width=0.3\textwidth]{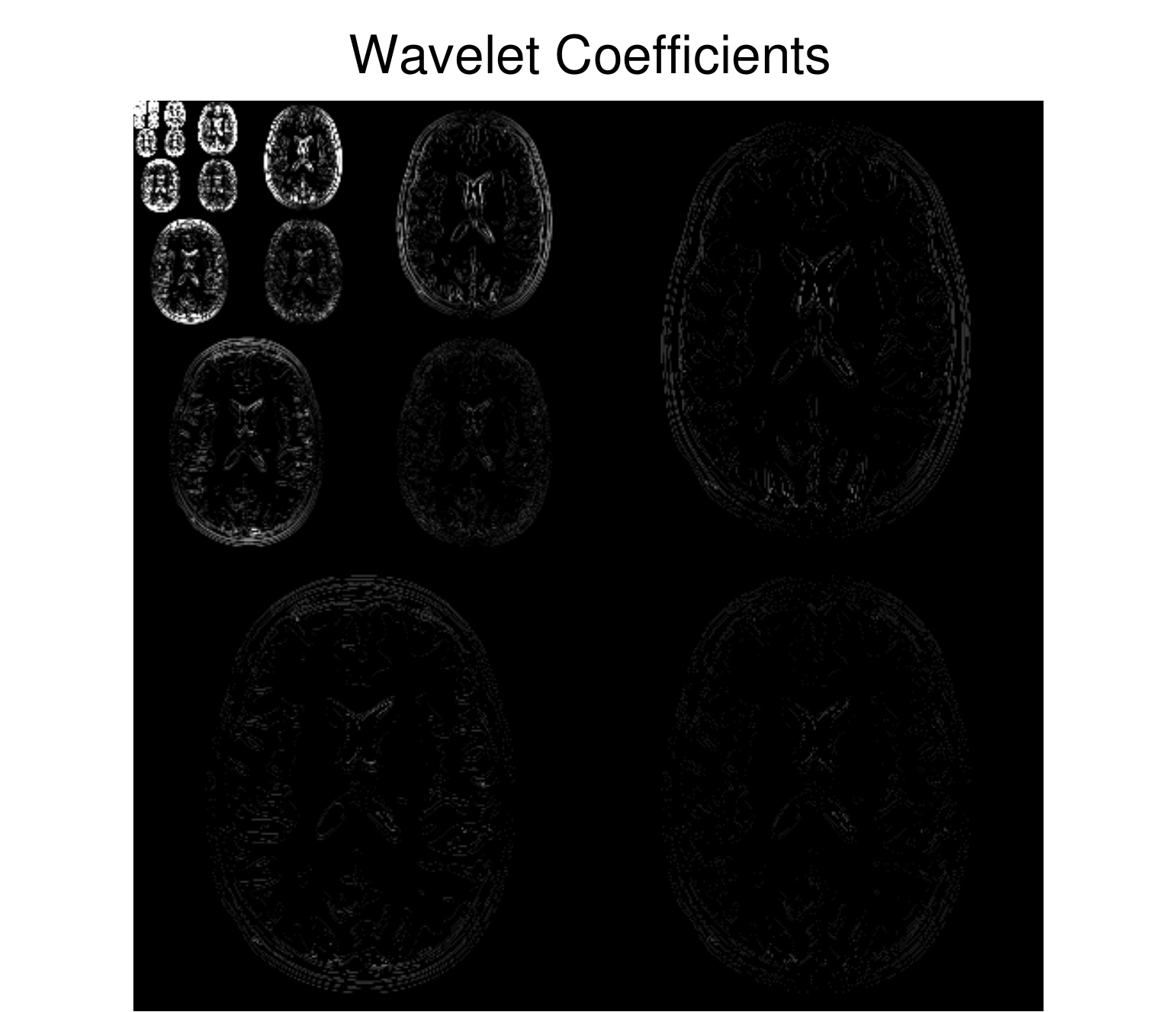} 
\includegraphics[width=0.3\textwidth]{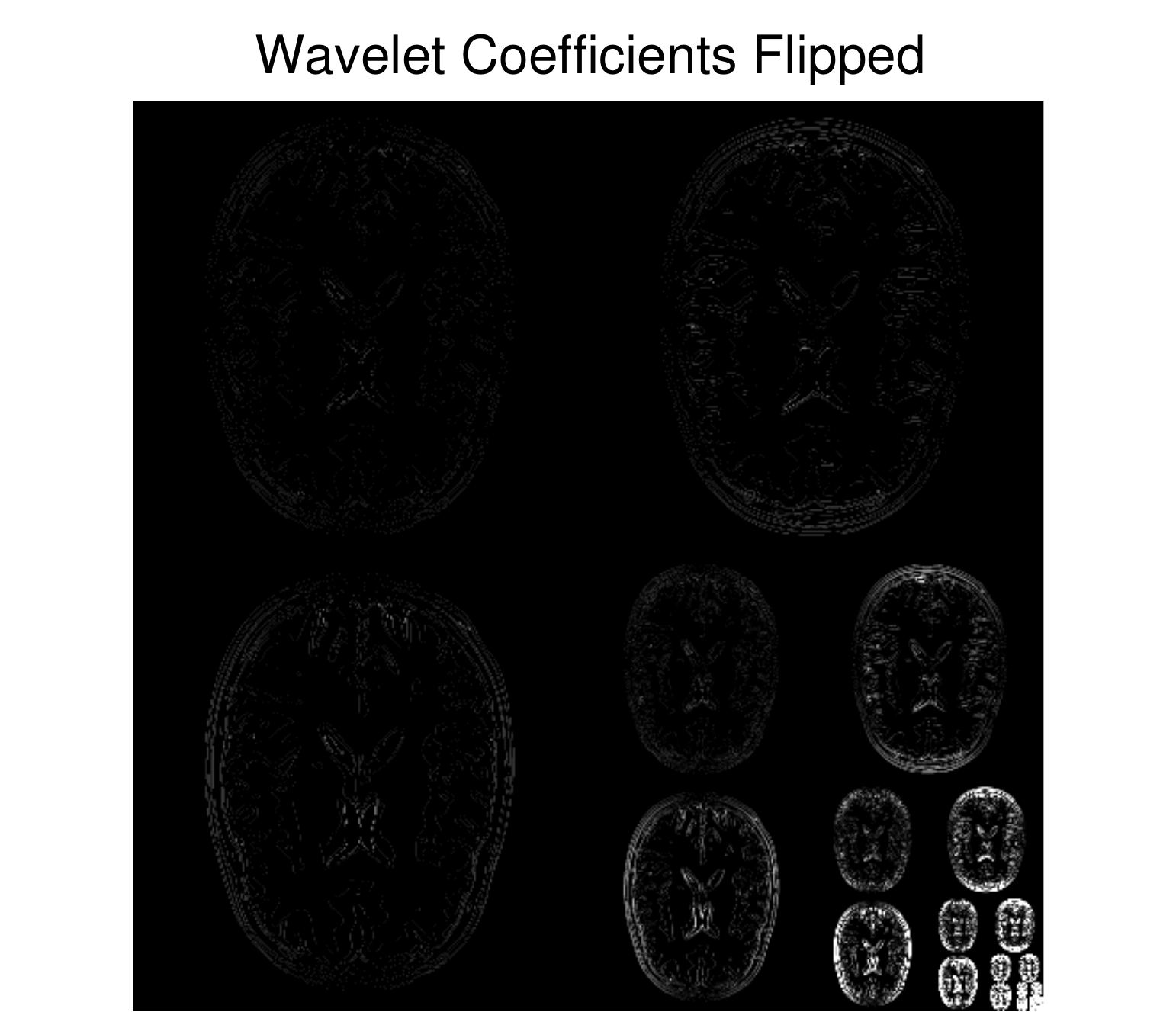} 
\includegraphics[width=0.3\textwidth]{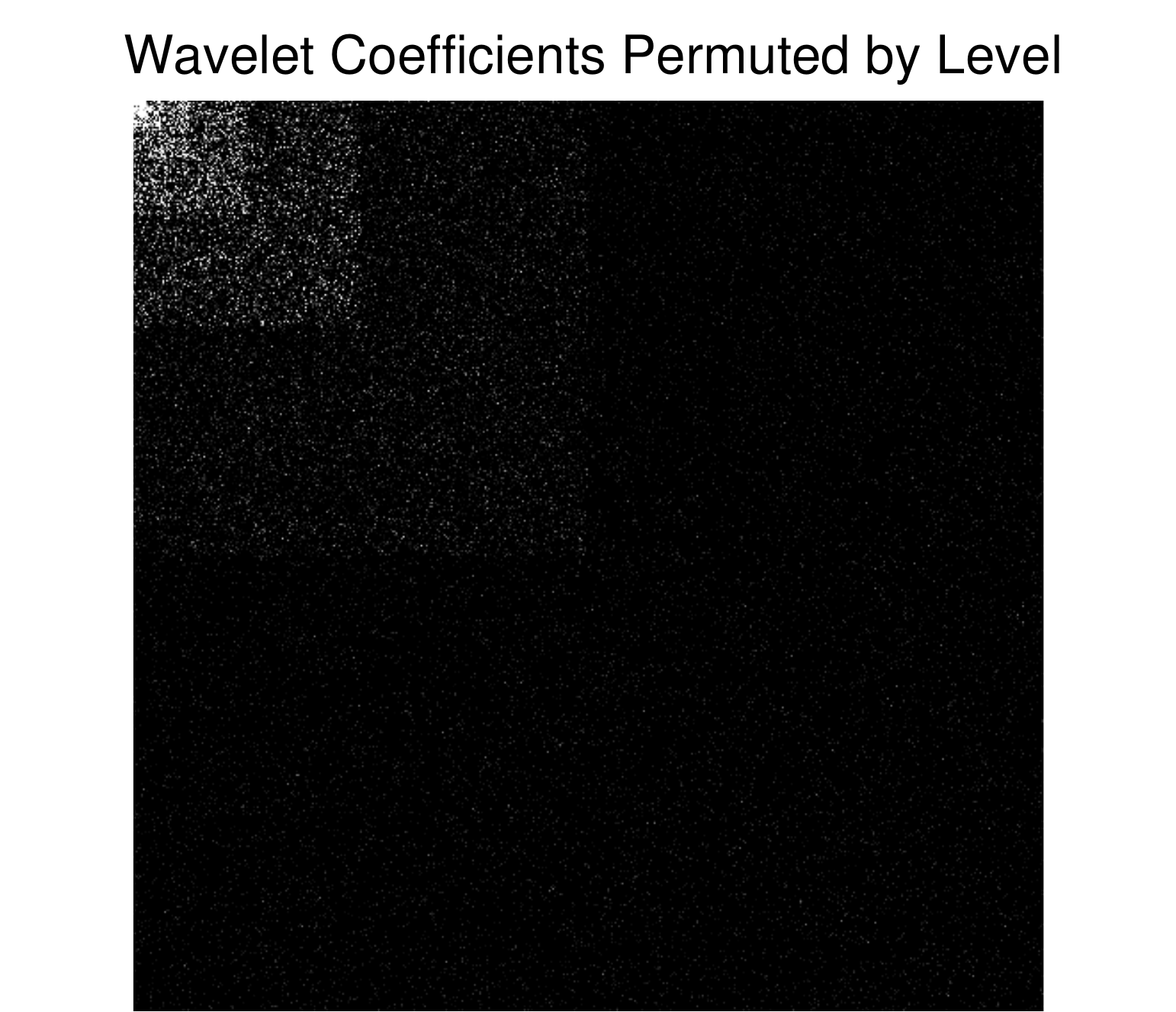}

\includegraphics[width=0.3\textwidth]{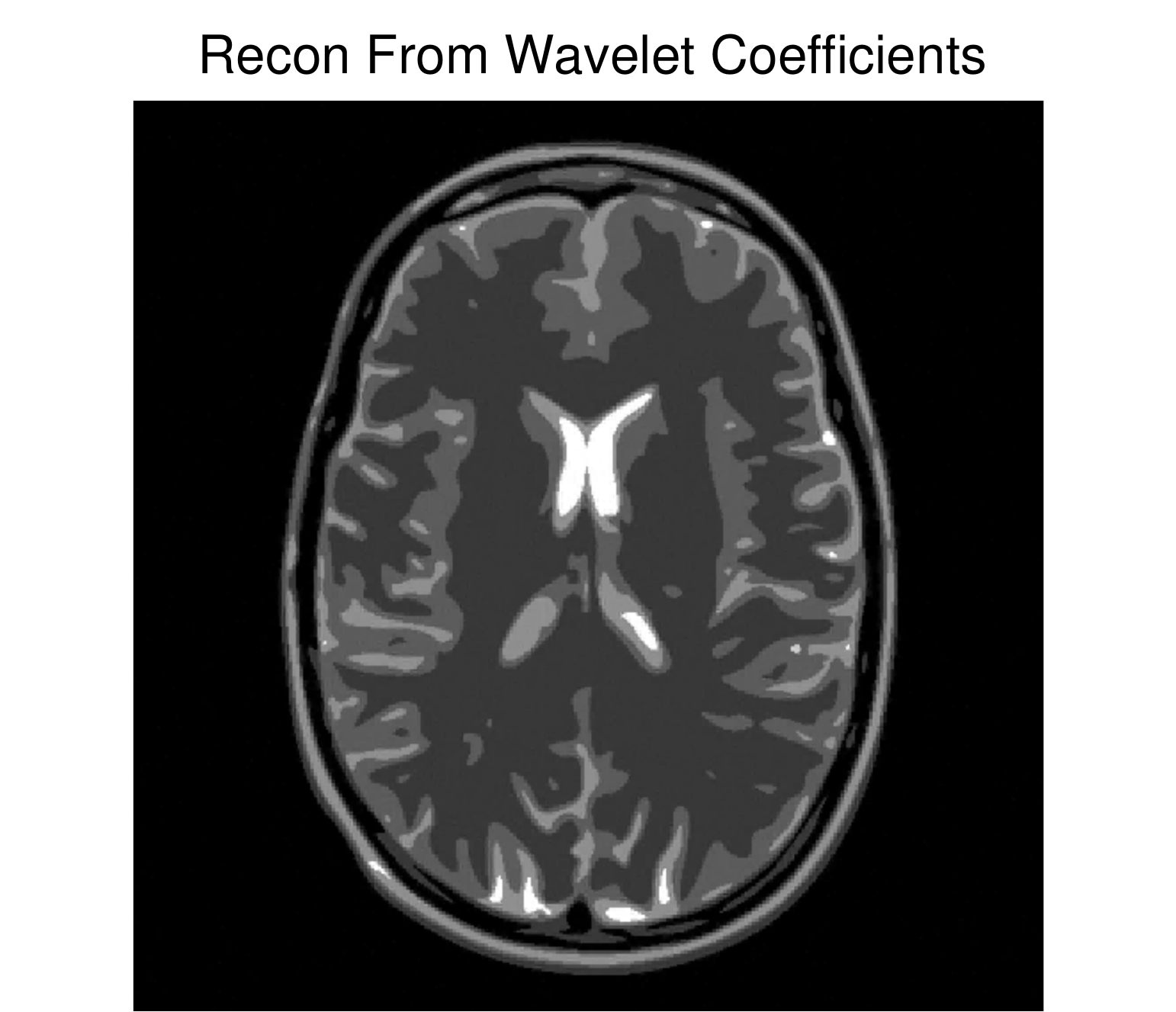}
\includegraphics[width=0.3\textwidth]{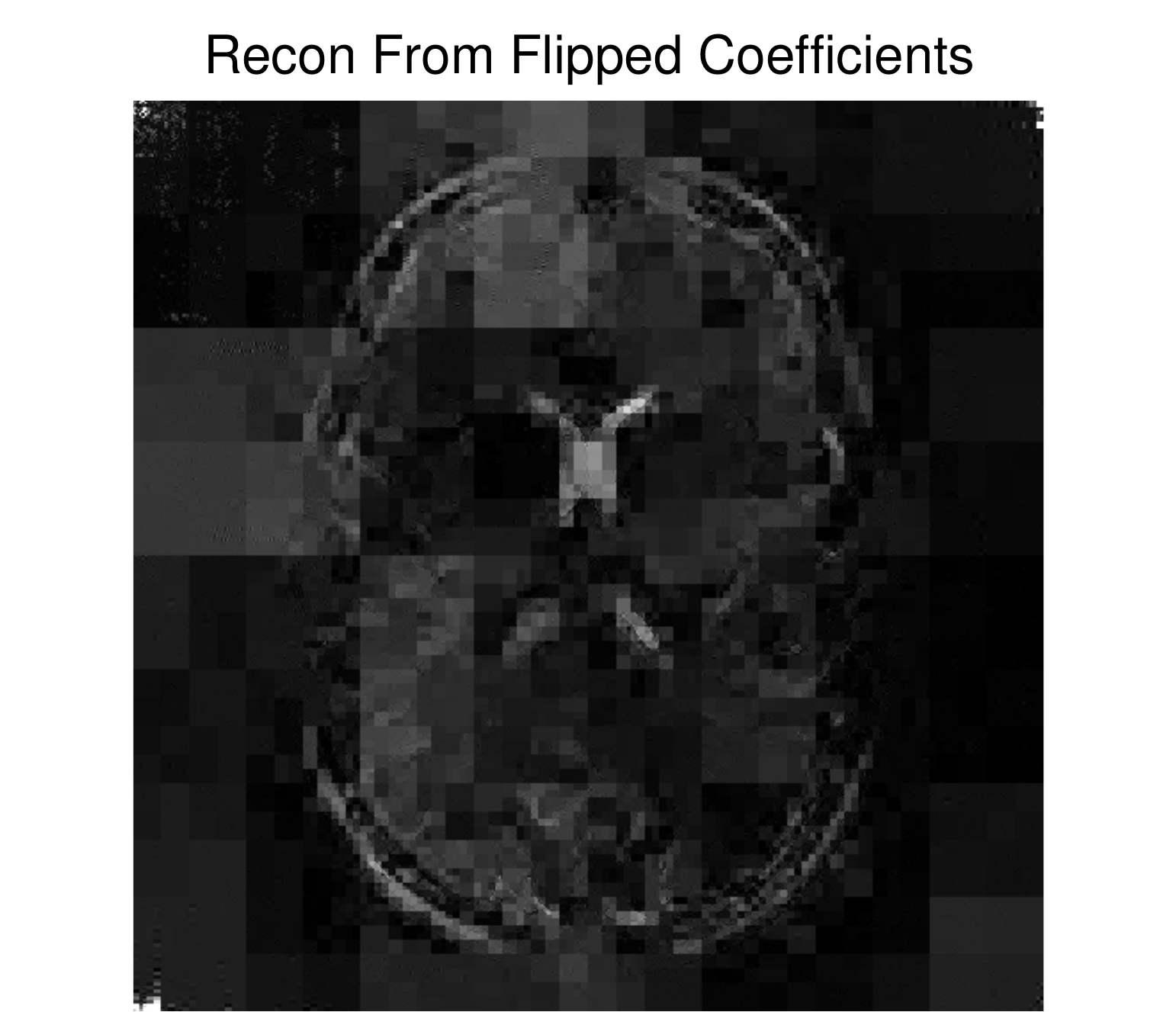}
\includegraphics[width=0.3\textwidth]{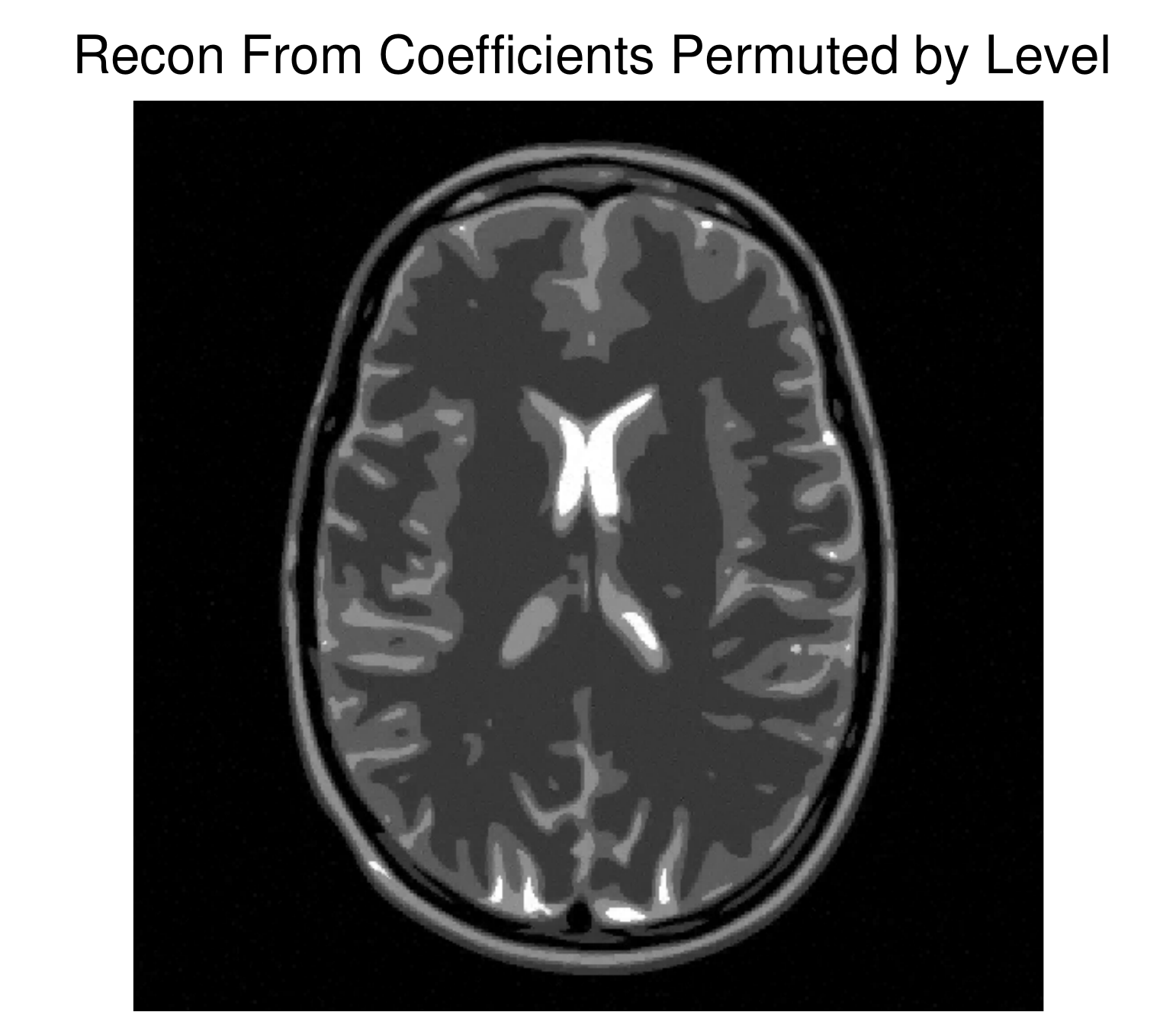} 
\end{center}
\caption{ A standard flip test demonstrating the importance of the location of wavelet coefficients on the success of CS subsampling techniques. Here we decompose the Brain image into a two-dimensional separable Haar basis and solve for these coefficients through subsampling Fourier data of the corresponding image (Resolution$=512 \times 512$). We then compare the results to the same wavelet data but the order flipped versus randomly permuting the coefficients inside each wavelet level. After using CS to solve for the flipped/permuted coefficients, the data is then flipped/permuted back and then imaged.}
\label{Fig:FlipTests}

\end{figure}

The flip test can be described as follows
Let 
$x\in\bbC^N$ be a {vector}, and $U\in\bbC^{N \times N}$ a measurement matrix.
We then sample according to some pattern $\Omega\subseteq\{1,\hdots,N\}$
with $|\Omega| = m$ and solve (\ref{fin_dim_l1}) for $x$, i.e.
$\min \|z\|_1$ s.t $P_{\Omega}Uz = P_{\Omega}Ux$ to obtain a 
reconstruction $z=\alpha$. Now we flip $x$ to obtain a vector $x'$ with 
reverse entries, $x'_i = x_{N-i}, i=1,\hdots,N$ and solve \R{fin_dim_l1} for $x'$ 
using the same $U$ and $\Omega$, i.e.\ $\min\|z\|_1$ s.t. 
$P_{\Omega}Uz = P_{\Omega}Ux'$. Assuming $z$ to be a solution, then by 
flipping $z$ we obtain a second reconstruction $\alpha'$ of the original 
vector 
$x$, where $\alpha'_i=z_{N-i}$. If the success of the sampling technique is independent of the structure of the coefficients, then we should have that $\alpha$ and $\alpha'$ are close. As Figure \ref{Fig:FlipTests} suggests, this is not the case.

Another version of the flip test is to permute the coefficients in the wavelet levels that corresponds to the different scales. In other words, instead of flipping the coefficients completely, the coefficients are permuted only within a certain level, but never across the levels. The result of this is visualised in Figure \ref{Fig:FlipTests}. Note that the results of the flip test presented here carry over to different subsampling schemes as well (see \cite{fliptest}).

{\bf Conclusion of the flip tests:}
\begin{itemize}
\item[(i)] The optimal sampling strategy depends on the signal structure.
\item[(ii)] There is no uniform recovery, hence the matrix involved does not satisfy the Restricted Isometry Property (RIP).
\item[(iii)] Theories based on sparsity will not explain the success of the recovery. Instead a notion of sparsity structure where there is a sparsity $s_k$ of important coefficients in the $k$th wavelet level is needed.
\item[(iv)] Because of this dependence on sparsity structure, both local coherences in the sampling basis $\mu(\pi_NU)$ \emph{and} the sparsity basis $\mu(U\pi_N)$ must be considered.
\end{itemize}

\begin{remark}
The flip test can be extended to show \cite{fliptest} that standard successful sampling schemes used in MRI, Electron Tomography, Neutron/$^3$He Scattering, Fluorescence Microscopy etc. do not recover all weighted sparse \cite{fliptest} vectors when using wavelets (or other X-lets). In particular, weighted sparsity suffers from similar issues as conventional sparsity, namely that the class is too big and allows for vectors that cannot be recovered  with standard sampling schemes. Hence weighted sparsity is insufficient for modelling compressed sensing with wavelets. However, theory and tests demonstrate that sparsity in levels (as in (iii)) may be a much more realistic model for compressed sensing. Motivated by this we use that as the structured sparsity model in this paper.
\end{remark}

\subsection{Overcoming the Coherence Barrier}
When faced with the coherence barrier, the standard compressed sensing approach of subsampling uniformly
at random does not work. This begs the question: do we have an alternative? Empirically, it is known that the answer to this question is yes: one can break the coherence barrier by sampling at different rates over different frequency ranges.  This was recently confirmed by mathematical analysis in \cite{BookAHRT, AHPRBreaking}.  The key to their work was to replace the three principles of compressed sensing with the three concepts of {\it sparsity in levels, multi-level sampling} and {\it local coherence} -- and prove recovery estimates akin to \R{m_est_Candes_Plan} under these more general settings.
\defn{[Sparsity in levels]
\label{d:Asy_Sparse}
Let $x$ be an element of either $\bbC^N$ or $l^2(\bbN)$.  For $r \in \bbN$ let $\mathbf{M} = (M_1,\ldots,M_r) \in \bbN^r$ with $1 \leq M_1 < \ldots < M_r$ and $\mathbf{s} = (s_1,\ldots,s_r) \in \bbN^r$, with $s_k \leq M_k - M_{k-1}$, $k=1,\ldots,r$, where $M_0 = 0$.  We say that $x$ is $(\mathbf{s},\mathbf{M})$-sparse if, for each $k=1,\ldots,r$,
\bes{
\Delta_k : = \mathrm{supp}(x) \cap \{ M_{k-1}+1,\ldots,M_{k} \},
}
satisfies $| \Delta_k | \leq s_k$.  We denote the set of $(\mathbf{s},\mathbf{M})$-sparse vectors by $\Sigma_{\mathbf{s},\mathbf{M}}$.
}

\defn{[Multi-level sampling scheme]
\label{multi_level_dfn}
Let $r \in \bbN$, $\mathbf{N} = (N_1,\ldots,N_r) \in \bbN^r$ with $1 \leq N_1 < \ldots < N_r$, $\mathbf{m} = (m_1,\ldots,m_r) \in \bbN^r$, with $m_k \leq N_k-N_{k-1}$, $k=1,\ldots,r$, and suppose that
\bes{
\Omega_k \subseteq \{ N_{k-1}+1,\ldots,N_{k} \},\quad | \Omega_k | = m_k,\quad k=1,\ldots,r,
}
are chosen uniformly at random, where $N_0 = 0$.  We refer to the set
\bes{
\Omega = \Omega_{\mathbf{N},\mathbf{m}} := \Omega_1 \cup \ldots \cup \Omega_r
}
as an $(\mathbf{N},\mathbf{m})$-multilevel sampling scheme. Observe that such sampling schemes are different to that of taking independent identically distributed (IID) samples from a common law.
}

\defn{\label{loc_coherence}
Let $U$ be an isometry of either $\bbC^{N}$ or $l^2(\bbN)$.
If $\mathbf{N} = (N_1,\ldots,N_r) \in \bbN^r$ and $\mathbf{M} = (M_1,\ldots,M_r) \in \bbN^r$ with $1 \leq N_1 < \ldots N_r $ and $1 \leq M_1 < \ldots < M_r $. In \cite{AHPRBreaking} the authors introduced their own definition of local coherence as follows; the $(k,l)^{\rth}$ local coherence of $U$ with respect to $\mathbf{N}$ and $\mathbf{M}$ is
\ea{ \label{localincoherence}
\mu_{\mathbf{N},\mathbf{M}}(k,l) &= \sqrt{\mu(P^{N_{k}}_{N_{k-1}}UP^{M_{l}}_{M_{l-1}}) \cdot  \mu(P^{N_{k}}_{N_{k-1}}U)},\qquad k,l=1,\ldots,r,
}
where $N_0 = M_0 = 1$ and $P^{a}_{b}$ denotes the projection matrix corresponding to the indices $\{a +1,\hdots, b\}$.  
}

In \cite{AHPRBreaking} a new theory of compressed sensing for changes of bases between infinite-dimensional Hilbert-spaces was introduced based on these assumptions.  In this case we solve the following problem: if $x \in \ell^2(\bbN)$ is $(s,\textbf{M})$-sparse and $U$ is an isometry of $\ell^2(\bbN)$ then we hunt for the (hopefully unique) $\eta$ that solves
\be{
\label{infinite_dim_l1}
\min_{\eta \in \ell^2(\bbN)} \| \eta \|_{l^1} \quad \mbox{subject to} \quad P_{\Omega} U \eta 
= P_{\Omega} Ux,
}
In this case, instead of a standard compressed sensing estimate \R{m_est_Candes_Plan} determining the total number of measurements, one has the following estimate regarding the local number of measurements $m_k$ in the $k^{\rth}$ level:
\be{
\label{conditions31_levels}
m_k \gtrsim (N_k-N_{k-1}) \cdot \log(\epsilon^{-1}) \cdot \left(
\sum_{l=1}^r \mu_{\mathbf{N},\mathbf{M}}(k,l) \cdot s_l\right) \cdot \log\left(N\right),\quad k=1,\ldots,r.
}
When the above condition is satisfied, \cite{AHPRBreaking} found that with probability exceeding $1-\epsilon$, $x$ can be shown to be the unique minimiser to (\ref{infinite_dim_l1}).
In particular, the number of samples $m_k$ needed to be taken in each region $\{N_{k-1}+1,\cdots N_k\}$ can be inferred through the local sparsities and coherences using the asymptotic relation (\ref{conditions31_levels}).

\subsection{Using Asymptotic Incoherence to Infer How to Sample}
This estimate begs the following question: how do the local sparsity and incoherences behave in practice?  As described in \cite{BookAHRT, AHPRBreaking}, natural images possess not just sparsity, but so-called asymptotic sparsity.  That is, the ratios $s_k / (N_k- N_{k-1}) \rightarrow 0$ as $k \rightarrow \infty$ in any appropriate basis (e.g.\ wavelets and their generalizations).

We also observe that the coherence term $\mu_{\mathbf{N},\mathbf{M}}(k,l)$ can be estimated as follows:
\begin{equation}\label{inco_est}
\mu_{\mathbf{N},\mathbf{M}}(k,l) \leq \sqrt{ \min \big( \mu(R_{N_{k-1}+1}U) , \mu(UR_{M_{l-1}+1}) \big) \cdot  \mu(R_{N_{k-1}+1}U)}.
\end{equation}

Therefore, combining this with (\ref{conditions31_levels}), it is clear from \R{inco_est} that in order to determine the appropriate sampling density one needs good estimates for the block coherences $\mu(R_NU)$ and $\mu(UR_N)$ for $N \in \mathbb{N}$.  This is the key contribution of this paper.  

\subsection{Related Results} \label{SubSec:Related}

Before stating our main results we will discuss the related work in \cite{discrete, WardFourierAndPolys, WardRowCoherIntro}. A key point regarding these results is that they are based on the sparsity model and the Restricted Isometry Property (RIP). It is clear from the flip test that sparsity is not the right model for the Fourier to wavelet case which is central to MRI and numerous other applications.  In particular, sampling strategies that seek to recover sparse vectors will have to take an unrealistically large number of samples. This is also reflected in the theoretical guarantees of \cite{discrete} which exhibit several additional logarithmic factors over traditional compressed sensing estimates. This is in contrast to the framework based on sparsity in levels \cite{AHPRBreaking} (which is the basis of this paper) that is shown both empirically \cite{RomanAsym} and theoretically \cite{fliptest} to even outperform incoherent sampling such as random sub-Gaussian, permuted Fourier, expanders etc.

In \cite{discrete}, they work with discretised Fourier and Haar wavelet elements $\varphi_k$ and $h^e_{n,l}$ defined on $\bbN^N=\bbN^{2^p}$ as follows:
\begin{equation} \label{Eq:DiscreteBasisDefs}
\begin{aligned}
& \varphi_k(t)=\frac{1}{\sqrt{N}}e^{2 \pi \ri tk/N}, \qquad -N/2+1 \le k \le N/2 \in \bbZ,
\\
& h^0(t)=2^{-p/2}, \quad h^1(t) = 
\begin{cases}
2^{-p/2}, & 1\le t \le 2^{p-1}, \\
-2^{-p/2}, & 2^{p-1} < t \le 2^p,
\end{cases} \\
& h^e_{n,l}(t)=2^{n/2}h^e(2^nt-2^pl), \quad 0<n<p, \quad 0 \le l<2^n.
\end{aligned}
\end{equation}

Using this notation the following upper bound on the discretised 1D Fourier/Haar wavelet case were derived (Lemma 6.1):
\begin{equation} \label{RachelBound1}
|\langle \varphi_k, h_{n,l} \rangle| \le \min\Big( \frac{6 \cdot 2^{n/2}}{|k|} , 3 \pi 2^{-n/2}   \Big), \qquad k=-N/2+1,\cdots,N/2 \in \bbZ, \quad n=1,\cdots p-1 \in \bbN,
\end{equation}
Result (\ref{RachelBound1}) was used to derive the following bound in Corollary 6.4:
\begin{equation} \label{RachelBound2}
|\langle \varphi_k, h_{n,l} \rangle| \le \frac{3 \sqrt{2 \pi}}{\sqrt{k}}, \qquad k=-N/2+1,\cdots,N/2 \in \bbZ.
\end{equation}

With the terminology of our paper, the result (\ref{RachelBound2}) shows that the corresponding discrete change of basis matrix $U_N \in \bbC^N \times \bbC^N$ satisfies (ordering the $\varphi_k$ by frequency c.f Def. \ref{Def:frequency_ordering})
\begin{equation} \label{Eq:DiscreteReframed}
\mu(\pi_NU_N) =\mathcal{O}(N^{-1}).
\end{equation}
The closest results to (\ref{Eq:DiscreteReframed}) in this paper are for a continuous change of basis matrix between Fourier elements and wavelets in $L^2[-1,1]$ shown in (\ref{Eq:FourierWavFastest})
\begin{equation} \label{FourierWaveOneDim}
\mu(\pi_NU) =\Theta(N^{-1}), \qquad \mu(U \pi_N) =\Theta(N^{-1}).
\end{equation}
This result covers \emph{all Daubechies wavelet bases} and not just the Haar case. Moreover the result provides asymptotic lower bounds.

The methods of proof are also worth comparing.  In \cite{discrete} explicit forms of the discrete Haar basis in Section 2.2 are used to derive (\ref{RachelBound2}). For higher order Daubechies wavelets such an explicit description is not available and this approach cannot be extended. In this paper we instead rely on the fact that the entries of $U$ can be viewed as Fourier transforms of the wavelet basis e.g. (\ref{Udef}). Since Daubechies wavelets were originally constructed through their Fourier transform, this analytical approach is very natural.

This work is also mentioned in \cite{WardRowCoherIntro} along with further results for pointwise sampling of polynomials with a weighted sparsity model. As discussed in \cite{fliptest}, weighted sparsity suffers from the same issues as sparsity and becomes an unrealistic model for Fourier samples and wavelet recovery. Moreover, the results in this paper on polynomials are with Fourier sampling which has very little connections with direct point samples considered in \cite{WardRowCoherIntro}.

\section{Main Results} \label{Sec:MainResults}

\subsection{Coherence Bounds} \label{SubSec:CoherBounds}

Our main results provide estimates for the precise convergence rates of $\mu(P_{N}U)$ and $\mu(UP_{N})$ in the case of one-dimensional Fourier-wavelet and Fourier-polynomial bases spanning $L^2[-1,1]$. For $\epsilon \in (0,1/2]$ fixed, the Fourier basis $B_\rf(\epsilon)$ is defined as
\be{ \label{Eq:FourierDefineMain}
\chi_k( x ) \ = \ \sqrt{\epsilon} \exp{(2 \pi \ri \epsilon k x)} \ \cdot \ \mathds{1}_{[(- 2 \epsilon)^{-1},(2 \epsilon)^{-1}]} (x),  \qquad x \in \bbR, \quad k \in \mathbb{Z}.
}
Notice that because $\epsilon \in (0,1/2]$, $B_\rf(\epsilon)$ is a basis of $L^2[(2\epsilon)^{-1},(2\epsilon)^{-1}]$ The standard wavelet basis $B_\rw$ for a given Daubechies scaling function $\phi$ and wavelet $\psi$ consists of functions of the form  
\be{ \label{Eq:WaveletDefineMain}
 \phi_{j,k}(x) =2^{j/2} \phi(2^j x-k) , \qquad \psi_{j,k} (x)  = 2^{j/2} \psi(2^j x - k).
}
A more precise description of these bases is provided in Section \ref{Sec:1DFW}. Suppose $U \in \mathcal{B}(\ell^2(\bbN))$ is the change of basis matrix formed by the pair of bases $(B_\rf(\epsilon),B_\rw)$ (see Definition \ref{Eq:U} for a formal definition of $U$). Since $U : \ell^2(\bbN) \to \ell^2(\bbN)$ the bases must be indexed by $\bbN$ which means we must enumerate the bases in some way using orderings (see Section \ref{Sec:CoherOrder}). For $B_\rf(\epsilon)$ we enumerate with increasing frequency (a \emph{frequency ordering}) and for $B_\rw$ we order with $j$ increasing (a \emph{leveled ordering}). 

\begin{theorem} [Fourier-Wavelet Case] \label{Thm:FourierWavFastest}
Under the above conditions (with $\epsilon \in I_{J,p}$, see Remark \ref{Rem:CorrectEpsilon}), we have 
\be{ \label{Eq:FourierWavFastest}
\mu(\pi_N U), \mu(U \pi_N) = \Theta(N^{-1}).
}
From Lemma \ref{Lem:lineToBlock} we also deduce that $\mu(R_NU), \mu(UR_N)=\Theta(N^{-1})$.
\end{theorem}

Next let $B_\rp=(\tilde{p}_n)_{n \in \bbN}$ denote the basis of $L^2$-normalised Legendre polynomials on $[-1,1]$. Suppose $U \in \mathcal{B}(\ell^2(\bbN))$ now denotes the change of basis matrix formed by the pair of bases $(B_\rf(\epsilon),B_\rp)$ with a frequency ordering of $B_\rf(\epsilon)$ ($B_\rp$ is already ordered by polynomial degree). 
\begin{theorem} [Fourier-Polynomial Case] \label{Thm:FourierPolyFastest}
In this case (given $\epsilon \in (0,0.45]$), we have
\be{ \label{Eq:FourierPolyFastest}
\mu(\pi_N U), \mu(U \pi_N) = \Theta(N^{-2/3}),
}
and we deduce $\mu(R_NU), \mu(UR_N)=\Theta(N^{-2/3})$. 
\end{theorem}

Theorems (\ref{Thm:FourierWavFastest}) and (\ref{Thm:FourierPolyFastest}) are covered by Corollaries \ref{Cor:FourierWavelet} and \ref{Cor:FourierPolynomial} respectively.

These results suggest that subsampling using compressed sensing is in general more effective for the Fourier-wavelet case than the Fourier-polynomial case, assuming similar sparsity structures.

\subsection{Optimality Results} \label{SubSec:OptimalityMain}

We also show that (\ref{Eq:FourierWavFastest}) and (\ref{Eq:FourierPolyFastest}) cannot be improved by changing the orderings of the two bases. 

\begin{theorem} [Optimality] \label{Thm:OptimalityOfBounds}
For the Fourier-wavelet case $(B_\rf(\epsilon), B_\rw)$, none of the following decay rates can achieved
\be{ \label{Eq:FourierWavFaster}
\mu(R_N U)=o(N^{-1}), \quad  \mu(U R_N) = o(N^{-1}).
}
For the Fourier-polynomial case $(B_\rf(\epsilon),B_\rp)$, the following decay rates are impossible no matter what orderings of the bases are used:
\be{ \label{Eq:FourierPolyFaster}
\mu(R_N U)=o(N^{-2/3}), \quad \mu(U R_N) = o(N^{-2/3}).
}
\end{theorem}
Theorem \ref{Thm:OptimalitySumUp} covers this result.

Finally we look at the general case we only impose that $U \in \mathcal{B}(\ell^2(\bbN))$ is an isometry and ask how fast $\mu(R_N U)$ can possibly decay. Theorem \ref{Thm:IsometryDecayLowerBound} states that 
\be{ \label{Eq:Summability}
\sum_{N\in \bbN}\mu(R_N U)<\infty,
}
must hold in general. Therefore $\mu(R_NU)=o(N^{-\alpha})$ is impossible for $\alpha>1$ showing that the Fourier-wavelet decay (\ref{Eq:FourierWavFastest}) attains the fastest theoretically possible decay rate as a power law. Furthermore Theorem \ref{Thm:IncoherenceCounter} shows that, as a statement for all isometries $U \in \mathcal{B}(\ell^2(\bbN))$, (\ref{Eq:Summability}) cannot be improved upon.

\section{Outline for the Remainder of the Paper} \label{Sec:Outline}

This outline is for those wishing to quickly extract the results and proofs in this paper. Section \ref{Sec:CoherOrder} is mandatory reading for all the other sections. Section \ref{Sec:Bases} is also important for understanding the bases and orderings needed for the one-dimensional coherence bounds. Sections \ref{Sec:1DFW} and \ref{Sec:1DFP} covering the Fourier-wavelet Fourier-polynomial cases are independent of each other. Section \ref{Sec:NumericalSubsampling} demonstrates how these two cases work in practice with some numerical data. Sections \ref{Sec:Optimality} and \ref{Sec:General} cover optimality results and theoretical limits and are independent of Sections \ref{Sec:Bases}-\ref{Sec:NumericalSubsampling}, barring the applications of Theorem \ref{Thm:OptimalitySumUp} to (\ref{Eq:FourierWavFastest}) and (\ref{Eq:FourierPolyFastest}) which is used as motivation. Finally in Section \ref{Sec:AltOpt} we discuss an alternative to the notion of optimality presented in Section \ref{Sec:Optimality}.

\section{Coherences and Orderings} \label{Sec:CoherOrder}

We work in an infinite dimensional separable Hilbert space $ \mathcal{H}$ with two closed infinite dimensional subspaces $V_1, V_2$ spanned by orthonormal bases $B_1,B_2$ respectively,
\[ V_1 = \overline{ \text{Span} \{ f \in B_1 \}}, \qquad   V_2 = \overline{ \text{Span} \{ f \in B_2 \}    }.\]
We call $(B_1,B_2)$ a `basis pair'.
\begin{definition}[Orderings]
Let $S$ be a set. Say that a function $\rho: \mathbb{N} \to S$ is an `ordering' of $S$ if it is bijective.
\end{definition}
\begin{definition}[Change of Basis Matrix]
For a basis pair $(B_1,B_2)$, with corresponding orderings $\rho:\mathbb{N} \to B_1$ and $\tau:\mathbb{N} \to B_2$, form a matrix $U \in \mathcal{B}(\ell^2(\bbN))$ by the equation
\begin{equation}\label{Eq:U}
 U_{m,n} :=  \langle \tau(n) , \rho(m) \rangle.
\end{equation}
Whenever a matrix $U$ is formed in this way we write `$U:=[(B_1,\rho),(B_2,\tau)]$'.
\end{definition}
At this point it is wise to look at how the orderings of the two bases effect the various notions of local coherence. 

\begin{lemma} \label{Lem:OneSidedDependence}
Let $U=[(B_1,\rho),(B_2,\tau)]$. For the coherence terms with the projection on the left hand side, i.e. $\mu(R_NU),\mu(\pi_NU)$, there is no dependence on the choice of ordering of the second basis $\tau$.  Likewise the coherences $\mu(UR_N),\mu(U \pi_N)$ do not depend upon the ordering of the first basis $\rho$.
\end{lemma}
This result follows immediately from the definitions of the line/block coherences (\ref{projectiondefinitions}). Even though $\mu(R_NU)$ does not depend on $\tau$, dependence on $\rho$ is so strong that arbitrarily slow decay of $\mu(R_NU)$ is possible for any isometry $U\in \ell^2(\bbN)$ by varying $\rho$.

Next we observe that bounds on the line coherences translates into bounds for the corresponding block coherences.

\begin{lemma} \label{Lem:lineToBlock}
Let $U=[(B_1,\rho),(B_2,\tau)]$. Suppose $\mu(\pi_NU)=\Theta(f(N))$ for some decreasing function $f:\bbN \to \bbR_{>0}$. Then $\mu(R_NU)=\Theta(f(N))$. Likewise for $\mu(U \pi_N)$ and $\mu(UR_N)$.
\end{lemma}
\begin{IEEEproof}
The lower bound is immediate since $\mu(R_NU)\ge\mu(\pi_NU)$ by definition. The upper bound follows by observing that
\be{ \label{Eq:PassToBlock}
\mu(R_NU) =\max_{M\ge N}\mu(\pi_MU) \le C_2 \max_{M \ge N} f_2(M)=C_2f(N).
}
\end{IEEEproof}

Throughout this paper we would like to define an ordering according to a particular property of a basis but this property may not be enough to specify a unique ordering. To deal with this issue we introduce the notion of consistency:

\begin{definition}[Consistent ordering]\label{consistent_ordering}
Let $F: S \to \mathbb{R}$ where $S$ is a set. We say that an ordering $\rho: \mathbb{N} \to S$ is `consistent with respect to F' if
\[ F(f)  <  F(g)  \quad \Rightarrow  \quad \rho^{-1}(f)  <  \rho^{-1}(g), \qquad \forall f,g \in S. \]
\end{definition}

\section{Bases \& Ordering} \label{Sec:Bases}

\subsection{Fourier Basis} \label{SubSec:FourierBasis}
We recall the definition of the Fourier basis $B_\rf(\epsilon)$ from (\ref{Eq:FourierDefineMain}).

\begin{definition}[Frequency ordering]\label{Def:frequency_ordering}
We define $F_\rf:B_\rf \to \mathbb{N} \cup \{0\}$  by $F_\rf(\chi_k)=|k|$ and say that an ordering $\rho: \mathbb{N} \to B_\rf$ is a `frequency ordering' if it is consistent with $F_\rf$. 
\end{definition}
For convenience in what follows we shall identify $B_\rf(\epsilon)$ with $\mathbb{Z}$ by the function $\lambda:B_\rf \to \mathbb{Z}, \ \ \lambda(\chi_k):=k$ which means that for any ordering $\rho$ of $B_\rf(\epsilon)$ we have
\[ \rho(m)( x ) \ = \ \sqrt{\epsilon} \exp{(2 \pi \ri \epsilon \cdot \lambda \circ \rho(m) x)} \ \cdot \ \mathds{1}_{[(- 2 \epsilon)^{-1},(2 \epsilon)^{-1}]} (x), \qquad \forall m \in \bbN. \]
Definition \ref{Def:frequency_ordering} says that an ordering $\rho$ of $B_\rf(\epsilon)$ is a frequency ordering if and only if the function $| \lambda \circ \rho|$ is nondecreasing. Therefore $\rho$ is a frequency ordering if and only if we have $\{ \lambda \circ \rho(2n), \lambda \circ \rho(2n+1) \} = \{+n, -n \}$ for $n \in \bbN$ and $\lambda \circ \rho(1)=0$ and consequently  $|\lambda \circ \rho(m)|=\lceil (m-1)/2 \rceil$. 

\subsection{Legendre Bases} \label{SubSec:LegendreBasis}
If $(R_N)_{n \in \mathbb{N}}$ denotes the standard Legendre polynomials on $[-1,1]$ (so $R_N(1)=1$) then the $L^2$-normalised Legendre polynomials are defined by $\tilde{p}_n=\sqrt{n-1/2} \cdot R_N$ and we write $B_\rp := (\tilde{p}_n )_{n=1}^\infty$. The basis $B_\rp$ is already ordered; call this the \emph{natural ordering} .

\subsection{Standard Wavelets}  \label{StandardWavelets}

Take a Daubechies wavelet $\psi$ and corresponding scaling function $\phi$ in $L^2(\mathbb{R})$ with \[\text{Supp} (\phi) = \text{Supp} (\psi) = [-p+1,p]. \]  
We write
\[
\begin{aligned}
 \phi_{j,k}(x) =2^{j/2} \phi(2^j x-k) , \qquad \psi_{j,k} (x)  = 2^{j/2} \psi(2^j x - k), 
 \\ V_j := \overline{\text{Span} \{ \phi_{j,k}: k \in \bbZ \}}, \quad W_j := \overline{\text{Span} \{ \psi_{j,k}: k \in \bbZ \}}.
 \end{aligned}
 \]
With the above notation, $(V_j)_{j \in \mathbb{Z}}$ is the multiresolution analysis for $\phi$, with the conventions 
\[V_j \subset V_{j+1} , \qquad V_{j+1} = V_j \oplus W_j. \]
where $W_j$ here is the orthogonal complement of $V_j$ in $V_{j+1}$. For a fixed $J \in \bbN$ we define the set\footnote{`$\rw$' here stands for `wavelet'.}
\begin{align} \label{waveletbasisdefine}
B_\rw := \left\{ \begin{array}{cc}   &  \mathrm{Supp}(\phi_{J,k}) \cap (-1,1) \neq \emptyset , \\  \phi_{J,k} , \ \psi_{j,k} :  &  \mathrm{Supp}(\psi_{j,k}) \cap (-1,1)  \neq \emptyset, \\ &  j \in \mathbb{N}, j \ge J , \ k \in \mathbb{Z}  \end{array} \right \} , 
\end{align}
Let $\rho$ be an ordering of $B_\rw$. Notice that since $L^2(\mathbb{R})= \overline{ V_J \oplus \bigoplus^{\infty}_{j=J} W_j}$ for all 
$f \in L^2(\mathbb{R})$ with $\mathrm{supp}(f) \subseteq [-1,1]$ we have
\[ f  =  \sum_{n=1}^\infty c_{n} \rho(n) \quad \text{for some} \quad (c_n)_{n \in \bbN} \in \ell^2(\bbN) .\] 

\begin{definition} [Leveled ordering (standard wavelets)]\label{leveled}
Define $F_\rw:B_\rw \to \mathbb{R}$ by 
\[ F_\rw( f) \ = \ \begin{cases} 
\ j, \  & \mbox{if }  f \in W_j\\
\ -1,  \ & \mbox{if }f \in V_J 
\end{cases} ,  \]
and say that any ordering $\tau: \mathbb{N} \to B_\rw$ is a `leveled ordering' if it is consistent with $F_\rw$.
\end{definition}
 Notice that $F_\rw(\psi_{j,k})=j$. We use the name ``leveled'' here since requiring an ordering to be leveled means that you can order however you like within the individual wavelet levels themselves, as long as you correctly order the sequence of wavelet levels according to scale.

\subsection{Boundary Wavelets} \label{BoundaryWavelets}

We now look at an alternative way of decomposing a function $f \in L^2([-1,1])$ in terms of a wavelet basis, namely using boundary wavelets  \cite[Section 7.5.3]{dDwav}. The basis functions all have support contained within $[-1,1]$, while still spanning $L^2[-1,1]$. Furthermore, the boundary wavelet basis retains the ability to reconstruct polynomials of order up to $p-1$ from the corresponding standard wavelet basis. We shall not go into great detail here but we will outline the construction;  we take, along with a Daubechies wavelet $\psi$ and corresponding scaling function $ \phi$ with $ \mathrm{Supp} (\psi) = \mathrm{Supp} (\phi)=[-p+1,p]$, boundary scaling functions and wavelets (using the same notation as in \cite{dDwav} \footnote{We use $[-1,1]$ instead of $[0,1]$ as our reconstruction interval here, but everything else is the same.})

\[ \phi^{\text{left}}_n, \ \phi^{\text{right}}_n, \ \psi^{\text{left}}_n , \ \psi^{\text{right}}_n , \qquad  n =0,\cdots,p-1 .\]
Like in the standard wavelet case we shift and scale these functions,
\[ \phi^{\text{left}}_{j,n}(x) = 2^{j/2} \phi^{\text{left}}_{n}(2^j (x+1)), 
\qquad \phi^{\text{right}}_{j,n}(x)= 2^{j/2} \phi^{\text{right}}_{n}(2^j (x-1)). \]
We are then able to construct nested spaces ,
$ (V^{\text{int}}_j)_{j \ge J}$, for $J \ge \lceil \log_2 (p) \rceil $, such that \mbox{$L^2([-1,1])=\overline{ \bigoplus^{\infty}_{j=0} V^{\text{int}}_j}$} and $V^{\text{int}}_{j+1}=V^{\text{int}}_j \oplus W^{\text{int}}_j$ by defining
\begin{equation*}
 V^{\text{int}}_j = \overline{ \text{Span} 
\left \{ \begin{aligned}
 \phi^{\text{left}}_{j,n} & , \phi^{\text{right}}_{j,n}  \\ & \phi_{j,k} 
 \end{aligned} 
 :  
\begin{aligned}
 & n =0 , \cdots , p-1 \ \\  & k \in \mathbb{Z} \ s.t. \ \mathrm{Supp}(  \phi_{j,k} ) \subset [-1,1]
 \end{aligned} 
 \right \} } ,
 \end{equation*}
 
 \begin{equation*} 
 W^{\text{int}}_j = \overline{ \text{Span} 
\left \{ \begin{aligned}
 \psi^{\text{left}}_{j,n} & , \psi^{\text{right}}_{j,n}  \\ & \psi_{j,k} 
 \end{aligned} 
 : 
\begin{aligned}
 & n =0 , \cdots , p-1 \ \\  & k \in \mathbb{Z} \ s.t. \ \mathrm{Supp}(  \psi_{j,k} ) \subset [-1,1]
 \end{aligned} 
 \right \} } .
 \end{equation*}
 
We then take the spanning elements of $V^{ \text{int}}_J$ and the spanning elements of $W^{\text{int}}_j$ for every $j \ge J$ to form the basis $B_{\rb \rw}$ ($\rb \rw$ for 'boundary wavelets').
\begin{definition}[Leveled ordering (boundary wavelets)]
Define $F_w: B_{\rb \rw} \to \mathbb{R}$ by the formula
\[ F_{\rb \rw}( f) \ = \ \begin{cases} 
\ j, \  & \mbox{if }  f \in W^{\text{int}}_j\\
\ -1,  \ & \mbox{if }f \in V^{\text{int}}_J 
\end{cases} . \]
Then we say that an ordering $\tau: \mathbb{N} \to B_{\rb \rw}$ of this basis is a `leveled ordering' if it is consistent with $F_{\rb \rw}$.
\end{definition}

\begin{remark} \label{Rem:CorrectEpsilon}
Let $U=[(B_\rf(\epsilon),\rho),(B_\rw,\tau)]$. If we require $U$ to be an isometry we must impose the constraint $(2\epsilon)^{-1} \ge 1+2^{-J}(p-1)$ otherwise the elements in $B_\rw$ do not lie in the span of $B_\rf(\epsilon)$. For convenience we rewrite this as $\epsilon \in I_{J,p}$ where 
\[I_{J,p}:=(0,(2+2^{-J+1}(p-1))^{-1}]. \]
If $B_\rw$ is replaced by $B_{\rb \rw}$, we only require $\epsilon \le 1/2$, since every function in $B_{\rb \rw}$ has support contained in $[-1,1]$. For the rest of this section, we shall assume these constraints on $\epsilon$ hold.
\end{remark}

\section{1D Fourier-Wavelet Case} \label{Sec:1DFW}

Let $U=[(B_\rf(\epsilon),\rho),(B_2,\tau)]$ with either $B_2=B_\rw$ or $B_{\rb \rw}$. The key observation for handling the entries of $U$ are 
\begin{equation} \label{Udef}
\begin{aligned}
U_{m,n} = \langle \tau(n) , \rho(m) \rangle & = \int_{\mathbb{R}} \sqrt{\epsilon} \exp(- 2 \pi i \epsilon x \cdot \lambda \circ \rho(m)) \cdot  \tau(n)(x) \ dx
\\ & = \sqrt{\epsilon} \mathcal{F} \tau(n)(\epsilon \cdot \lambda \circ \rho(m)),
\end{aligned}
\end{equation}
where $\mathcal{F}$ denotes the 1D Fourier Transform. We also observe that
\begin{equation} \label{Uuse}
\begin{aligned}
\mathcal{F}\phi_{j,k}( \omega ) & = e^{-2\pi \ri  2^{-j} k \omega} 2^{-j/2} \mathcal{F} \phi(2^{-j} \omega), \qquad
\mathcal{F}\psi_{j,k}( \omega )  = e^{-2\pi \ri 2^{-j} k \omega} 2^{-j/2} \mathcal{F} \psi(2^{-j} \omega), \\
\mathcal{F}\psi^{\text{left}}_{j,n}( \omega )  & =  2^{-j/2} e^{2 \pi \ri} \mathcal{F} \psi^{\text{left}}_n(2^{-j} \omega), \qquad
\mathcal{F}\psi^{\text{right}}_{j,n}( \omega)  = 2^{-j/2} e^{-2 \pi \ri}  \mathcal{F} \psi^{\text{right}}_n(2^{-j} \omega). 
\end{aligned} 
\end{equation}
\textbf{Outline of Argument:} Suppose $\rho$ is a frequency ordering and $\tau$ a leveled ordering. When we bound $\mu(\pi_NU)$ and $\mu(U\pi_N)$ we rely heavily on the formula
\be{ \label{Eq:FourierWaveletRough}
|U_{m,n}|^2 =  \epsilon  2^{-j(n)} |\mathcal{F} \psi(2^{-j(n)} \epsilon \cdot \lambda \circ \rho(m))|^2,
}
where $j(n)$ denotes what wavelet level we are on and we also ignore scaling function terms. 

If we want to bound $\mu(U \pi_N)$, i.e. looking at a single wavelet function and maximising over all frequencies, then the $|\mathcal{F} \psi(\cdot)|^2$ term varies little as $j(N)$ gets large and we just have $\mu( \pi_N U) = \epsilon  \Theta(2^{-j(N)})$. There are roughly $2^{j(N)}$ functions in the wavelet basis for each value of $j(N)$, implying that $2^{-j(N)} =\Theta(N^{-1})$ and therefore $\mu(U \pi_N ) = \epsilon  \Theta(N^{-1})$.

When upper bounding $\mu(\pi_N U)$, one uses the decay property $|\psi(\omega)| \le K \cdot |\omega|^{-1/2}$ to remove the dependence on $j(n)$ in (\ref{Eq:FourierWaveletRough}), leaving us with $\mu( \pi_N U)= \mathcal{O}(|\lambda \circ \rho(m)|^{-1})$. Observing $|\lambda \circ \rho(m)|=\Theta(m)$ gives us $\mu(\pi_NU)=\mathcal{O}(N^{-1})$. For the lower bound we look near the diagonal of the matrix $n=m$. Recalling $2^{j(m)},|\lambda \circ \rho(m)| =\Theta(m)$ and noticing that $ 2^{-j(m)} \epsilon |\lambda \circ \rho(m)|$ is roughly constant, this suggests that $|\mathcal{F} \psi(2^{-j(m)} \epsilon |\lambda \circ \rho(m)|)|^2$ is nearly constant and we're left with $ \mu(\pi_NU) \gtrsim 2^{-j(N)}=\Theta(N^{-1})$.

We now come to our first concrete example of bounding line coherences

\begin{theorem} \label{Thm:1dWaveletFourier}
Let $U=[(B_\rf(\epsilon),\rho),(B_\rw,\tau)]$ where $\tau$ is a leveled ordering of a standard wavelet basis. Then there are constants $C_1, C_2>0$, dependent on the choice of wavelet, such that for all $\epsilon \in I_{J,p} $ and $N \in \mathbb{N}$, we have
\begin{equation} \label{1sttheoremtarget}  
\frac{ \epsilon \cdot C_1}{N}  \le  \mu(U \pi_N  )  \le  \frac{ \epsilon \cdot C_2}{N} . 
\end{equation}
Furthermore, suppose instead $U=[(B_\rf(\epsilon),\rho),(B_{\rb \rw},\tau)]$ where $\tau$ is a leveled ordering of a boundary wavelet basis. Then there are constants $C_1, C_2>0$, dependent on the choice of wavelet, such that for all $\epsilon \in (0,1/2] $ and $N \in \mathbb{N}$,  (\ref{1sttheoremtarget}) holds.
\end{theorem}

\begin{IEEEproof} 
By equation (\ref{Udef}) we know that (since $\lambda \circ \rho:\bbN \to \bbZ$ is bijective)
\[ \mu( \pi_N U )  =  \sup_{  m \in \mathbb{N}}{ \epsilon | \mathcal{F} \tau(N)(\epsilon \cdot \lambda \circ \rho(m))} |^2 = \sup_{  m \in \mathbb{Z}}{ \epsilon | \mathcal{F} \tau(N)(\epsilon m)} |^2. \]
 
\textbf{Case 1 (Standard wavelets):} 
In this case we define $j(N):=F_\rw(\tau(N))$ and let $a:=2p-1 \in \mathbb{N}$ denote the length of the support of the scaling function $\phi$ corresponding to $B_\rw$. Notice that for a leveled ordering of $B_\rw$, the functions belonging to $V_J$ come first, and there are of $2^{J+1}+a-1$ of these functions. Therefore, for $N \le 2^{J+1}+a-1$ we have that, by (\ref{Uuse}),
\begin{equation} \label{phiterms}  
\mu( U \pi_N ) =  \epsilon \sup_{m \in \mathbb{Z}}{ 2^{-J} |\mathcal{F} \phi( 2^{-J} \epsilon m)|^2} .
\end{equation}
Furthermore, for the wavelet terms in $B_\rw$, which correspond to $N \ge 2^{J+1}+a$, we have that, by (\ref{Uuse}),
\begin{equation} \label{psiterms}  
\mu( U \pi_N) =  \epsilon \sup_{m \in \mathbb{Z}}{ 2^{-j(N)} |\mathcal{F} \psi(2^{-j(N)} \epsilon m)|^2}. 
\end{equation}
Since the wavelet is compactly supported and in $L^2(\mathbb{R})$ it is in $L^1(\bbR)$ and so its Fourier transform is continuous. Notice that by continuity and the Riemann-Lebesgue Lemma, we see that $\sup_{\omega \in \mathbb{R}}| \mathcal{F} \psi(\omega)| = | \mathcal{F} \psi( \hat{\omega} )| $ for some $\hat{\omega} \in \mathbb{R}$. Therefore, since $j(N) \to \infty$ as $N \to \infty$ because the ordering $\tau$ is leveled, we find that
\begin{equation} \label{FTtwiddle}
 \frac{\sup_{m \in \mathbb{Z}} |\mathcal{F} \psi( \epsilon 2^{-j(N)} m)|^2}{ \sup_{\omega \in \mathbb{R}} |\mathcal{F}{\psi}(\omega)|^2} \longrightarrow 1 \quad \text{as} \quad N \to \infty. 
\end{equation}
Furthermore, this convergence is uniform in $\epsilon \in I_{J,p}$ as $N \to \infty$.
We are therefore left with handling the $2^{-j(N)}$ term, which means estimating $j(N)$ as $N \to \infty$. 

Notice that for each value of $j(N)\ge J$ there are $2^{j(N)+1}+a-1$ functions in our wavelet basis with this value of $j(N)$. For simplicity we shall use the simple bounds $ 2^{j(N)+1} \le 2^{j(N)+1}+a-1 \le  2^{j(N)+a}$. Now for every $N \in \bbN$ with $j(N)>J$, we must have had all the terms of the form $f \in B_\rw, F_\rw(f)=j(N)-1$ come before $N$ in the leveled ordering and there are at least $2^{j(N)}$ of these terms. If $j(N)=J$ we instead have $N > 2^{J+1}+a-1 > 2^{J}$. Likewise for every $N \in \bbN$ with $j(N)\ge J$ there can be no more than $\sum_{i=J}^{j(N)} 2^{(i+a)}+2^J+a-1 \le 2^{j(N)+a+2}$ terms that came before $N$. Therefore, for $j(N) \ge J$, we have the inequality
\begin{equation} \label{j2n}
2^{j(N)} \le  N  \le  2^{j(N)+a+2} . 
\end{equation}
From here we will tackle the upper and lower bounds of (\ref{1sttheoremtarget}) separately: 

\textbf{Upper Bound:} We will show that $ \mu(U \pi_N )  \le  \frac{ \epsilon \cdot C_2}{N}$. Notice from (\ref{j2n}) we have the upper bound $ 2^{-j(N)} \le \ 2^{a+2}N^{-1} $ for $j(N) \ge J$ and therefore for these terms we can bound  (\ref{psiterms}) by 
\[  \epsilon \sup_{m \in \mathbb{Z}}{ 2^{-j(N)} |\mathcal{F} \psi(2^{-j(N)} \epsilon m)|^2} \leq \epsilon \frac{2^{a+2}}{N} \cdot \sup_{\omega \in \mathbb{R}} | \mathcal{F} \psi (\omega) |^2, \]
For the $j(N)=-1$ terms (i.e. $N \le 2^{J+1}+a-1$) we also have the simple bound
\[ \epsilon \cdot  2^{-J} \sup_{\omega \in \mathbb{R}} | \mathcal{F} \phi (\omega) |^2 \le \epsilon 2^{-J} \frac{2^{J+1}+a-1}{N} \cdot  \sup_{\omega \in \mathbb{R}} | \mathcal{F} \phi (\omega) |^2, \]
and so the upper bound is complete.

\textbf{Lower Bound:} First notice that from (\ref{j2n}) that we have the lower bound $2^{-j(N)} \ge  N^{-1}$. Next notice that from (\ref{FTtwiddle}) there is an $N' \in \mathbb{N}$ independent of $\epsilon \in (0,1/2p]$ such that for all $N \ge N'$ we have
\[ \sup_{m \in \mathbb{Z}} |\mathcal{F} \psi( \epsilon 2^{-j(N)} m)|^2 \ge \frac{1}{2}\sup_{\omega \in \mathbb{R}} |\mathcal{F}{\psi}(\omega)|^2.  \]
Consequently for $N \ge N'$ we have the lower bound
\[ \mu(\pi_N U) \ge \epsilon 2^{-j(N)} \cdot \frac{\sup_{\omega \in \mathbb{R}} | \mathcal{F}\psi(\omega)|^2}{2} \ge \frac{\epsilon }{2N} \cdot \sup_{\omega \in \mathbb{R}} | \mathcal{F}\psi(\omega)|^2 .\]
Therefore, in order to prove the lower bound, we need only show there exists a constant $C>0$ such that every $N <N'$ we have $ \mu(\pi_N U) \ \ge \ \epsilon \cdot C $ uniformly in $\epsilon \in I_{J,p}$. This will be satisfied if we can show that for every $j \ge J$ fixed there exists a constant $C>0$ such that for all $\epsilon \in I_{J,p}$
\[ \sup_{m \in \mathbb{Z}} | \mathcal{F} \phi ( 2^{-J} \epsilon m)|^2 , \,\sup_{m \in \mathbb{Z}} | \mathcal{F} \psi ( 2^{-j} \epsilon m)|^2 \ge C. \]
We will deal with latter term since the scaling function term is handled similarly. We know that for every $\epsilon \in I_{J,p}$ fixed, 
$\sup_{m \in \mathbb{Z}} | \mathcal{F} \psi ( 2^{-j} \epsilon m)|^2 >0$
since if it were not the case we would find that $ \langle \chi_m, \psi_{j,0} \rangle = 0$  for every $m$, contradicting the $\chi_m$ forming a basis of $L^2([(-2 \epsilon)^{-1},(2 \epsilon)^{-1}])$. Next notice that by the Riemann-Lebesgue Lemma and continuity of the Fourier transform of $\psi$, this supremum is a continuous function of $\epsilon$ and that 
\[ \sup_{m \in \mathbb{Z}} | \mathcal{F} \psi ( 2^{-j} \epsilon m)|^2 \to \sup_{\omega \in \mathbb{R}} | \mathcal{F} \psi(\omega)|^2 >0 \quad \text{as} \quad \epsilon \to 0. \]

Consequently we deduce the supremum attains its lower bound as a function of $\epsilon$ on $I_{J,p}$ and we are done.

\textbf{Case 2 (Boundary wavelets):} The method of proof is the same except that we have additional \\ $\psi^{\text{left}}, \phi^{\text{left}}, \psi^{\text{right}} , \phi^{\text{right}}$ terms to deal with. We also have slightly different behaviour of $2^{j(N)}$, i.e. for $N > 2^{J+1}$,
\begin{equation} \label{jest} 
2^{j(N)}   \le N  \le 2^{j(N)+2}.
\end{equation}
This follows from observing that for each value of $j(N)$ there are $2^{j(N)+1}$ functions in the wavelet basis, and that we are using a leveled ordering. The details are omitted for the sake of brevity.
\end{IEEEproof}
\vspace{10pt}
Next we need the following condition on our scaling function / wavelet; there exists a constant $K>0$ s.t. $\forall \omega \in \mathbb{R} \setminus \{ 0 \}$, 
\begin{equation} \label{FTdecay}
 |\mathcal{F} \phi( \omega )|  \le   \frac{K}{| \omega |^{1/2}}.  
 \end{equation}
 This condition holds for all Daubechies wavelets (see the proof of Proposition 4.7 in \cite{DaubWavelets}), in fact it even holds if we change the power of $\omega$ from $1/2$ to $1$. 
\begin{lemma} \label{decayextend}
Let $\phi$ be a Daubechies scaling function, with corresponding mother wavelet $\psi$. Then, along with (\ref{FTdecay}), we also have
\begin{equation} \label{FTdecay2}
 |\mathcal{F} \psi( \omega )|  \le  \frac{K}{| \omega |^{1/2}}.  
 \end{equation}
Furthermore in the case of boundary wavelets we also have for some constant $K >0$ and $\omega \in \mathbb{R} \setminus \{ 0 \}$
\be{ \label{FTdecayboundary}
|\mathcal{F} \phi_{n}^{\text{left}}( \omega)|, \ |\mathcal{F} \phi_{n}^{\text{right}}(\omega)| , \ |\mathcal{F} \psi_{n}^{\text{left}}(\omega)| , \ |\mathcal{F} \psi_{n}^{\text{right}}(\omega)| \le   \frac{K}{|\omega|^{1/2}}, 
}
along with (\ref{FTdecay}) and (\ref{FTdecay2}). In fact (\ref{FTdecay2}) and (\ref{FTdecayboundary}) hold with the powers of $1/2$ replaced by $1$.

\end{lemma}

\begin{IEEEproof}
 We notice that if (\ref{FTdecay}) holds then we can use the equation (from (2.14) in \cite{wav})
\begin{equation} \label{fourierscalingwavelet}
\mathcal{F}\psi(2 \omega)=\exp( 2 \ri \pi \omega) \cdot \nu(2 \omega) \cdot m_0(\omega + 1/2) \cdot \mathcal{F}\phi(\omega),
\end{equation}
 where $m_0$ is the Fourier transform of the low pass filter of the scaling function $\phi$ and $\nu$ is function whose modulus is always $1$\footnote{The equation here is not identical to that of the reference because of our choice of definition of the Fourier transform.}. Taking the modulus of this equation gives
$ | \mathcal{F}\psi(2 \omega)| \ = \ |m_0(\omega + 1/2)| \cdot | \mathcal{F}\phi(\omega) | .
$
Therefore using this along with $|m_0(\omega)| \le 1, \ \forall \omega \in \bbR$ (from (2.5) in \cite{wav}) we can show that (\ref{FTdecay}) also holds with $\phi$ replaced by $\psi.$

We now turn to the boundary wavelet estimates. We may assume $p \ge 2$ since in the Haar case boundary wavelets are redundant. First we note that the property of having a decay estimate of the form ~\eqref{FTdecay} is closed under finite linear combinations. Next observe that if we prove an estimate of the form ~\eqref{FTdecay} for the functions (see page 71 of \cite{boundary})
\[  \tilde{\phi}^k(x) = \sum_{n=k}^{2p-2} \binom{n}{k} \phi(x+n - p+1) \cdot \mathds{1}_{[0,\infty)}, \qquad k=0,\cdots,p-1,\]
then we also have the same decay (with a different constant) for the functions $ \phi^{\text{left}}_k  $ and $\psi^{\text{left}}_k$ since they are finite linear combinations of these functions. A similar argument will work for the right boundary wavelets.
Let us consider an arbitrary term from the sum 
$$
T_n(x):=\phi(x+n - p+1) \cdot \mathds{1}_{[0,\infty)}=\phi(x+n - p+1) \cdot \mathds{1}_{[0,2p-1]}.
$$ 
Now since we have expressed $T_n$ as a product of two $L^2$ functions we can apply the convolution rule on its Fourier Transform to deduce $\mathcal{F}T_n(\omega) = \big( \mathcal{F} \phi_{0,-n+p-1} * \mathcal{F}\mathds{1}_{[0,2p-1]} \big)(\omega)$. Now we make two observations:
\\
\\
1. \hspace{10pt} $| \mathcal{F}\mathds{1}_{[0,2p-1]}(\omega)|=|(\exp(-2 \pi \ri (2p-1) \omega)-1) \cdot (2 \pi \ri \omega)^{-1}| \le C_1 \cdot (|\omega|+1)^{-1} $ for some constant $C_1>0$.
\\
\\
2. \hspace{10pt}Excluding the Haar wavelet, for every Daubechies wavelet there exists constants $\alpha, C_2 >0$ such that $|\mathcal{F} \phi( \omega)| \le C_2 \cdot ( |\omega| + 1 )^{-1 - \alpha}$ (see the proof of Proposition 4.7 in \cite{DaubWavelets}).
\\
\\
We now that claim that if two functions $f, g$ satisfy 
\[ |f( \omega)| \le C_1 \cdot ( |\omega| +1)^{-1}, \quad  |g(\omega)| \le C_2 \cdot ( |\omega| + 1 )^{-1 - \alpha}, \qquad \forall \omega \in \bbR. \]  
for some constants $\alpha, C_1, C_2 >0$ then $|f * g(\omega)| \le C_3 \cdot |\omega|^{-1}$ which will prove the lemma. To see this notice that (without loss of generality $\omega>0$)
\be{ \label{integralsplit}
\begin{aligned}
 |f * g(\omega)| \cdot |\omega| & \le  C_1 C_2 \int_{ \bbR} \frac{|\omega|}{(|u|+1)(|\omega-u|+1)^{1+\alpha}} \, du  
 \\ & \le C_1 C_2 \Bigg( \int_{-\infty}^{ \omega/2} \frac{|\omega|}{(|u|+1)(|\omega-u|+1)^{1+\alpha}} \, du 
 \\
 & \qquad \qquad \qquad +  \int_{\omega/2}^{+ \infty}  \frac{|\omega|}{(|u|+1)(|\omega-u|+1)^{1+\alpha}} \, du \Bigg),
\end{aligned}
}
and notice that we would have shown the claim if we can bound the RHS uniformly in $\omega$. By noting $|\omega-u|+1 \ge |u|+1, \ |\omega-u| \ge | \omega/2|$ for $u \in (-\infty,\omega/2]$ we see that the first integral is bounded above by
\begin{align*} 
\int_{-\infty}^{\omega/2} \frac{|\omega|}{(|u|+1)^{1+\alpha/2} (|u-\omega|+1)^{1+\alpha/2}} \, du & \le \int_{-\infty}^{\omega/2} \frac{|\omega|}{(|u|+1)^{1+\alpha/2} (|\omega/2|+1)^{1+\alpha/2}} \, du 
\\& \le \int_{\bbR} \frac{2^{1+ \alpha/2}}{(|u|+1)^{1+\alpha/2}} \, du = \text{constant} < \infty.
\end{align*}
To bound the last integral in (\ref{integralsplit}) we simply use $|\omega|(|u|+1)^{-1} \le 2$ for $u \in [\omega/2,\infty)$ to give us a similar uniform upper bound, completing the proof of the claim.
\end{IEEEproof}
For our second incoherence result we will need a technical lemma.

\begin{lemma} \label{wavelower}
For any compactly supported wavelet $\psi$ with scaling function $\phi \in L^1(\mathbb{R})$ there exists an $N \in \mathbb{N}$ such that for all $q \ge N, \ (q \in \mathbb{N})$ we have 
\[ L_q := \inf_{\omega \in [2^{-(q+1)},2^{-q}] } | \mathcal{F}\psi(\omega)| \ > \ 0. \]
\end{lemma}

\begin{IEEEproof}
We recall from equation (\ref{fourierscalingwavelet}) that
\be{  \label{fourierlower}
| \mathcal{F}\psi(2 \omega)| \ = \ |m_0( \omega + 1/2)| \cdot | \mathcal{F}\phi(\omega) | . 
}
Furthermore, we also know that $|\mathcal{F}\phi(0)| = 1$  and $m_0(1/2)=0$ \cite{wav}\footnote{See Section 2 Theorem 1.7 and Equation (3.1) in the reference.}. However, since $\phi$ is compactly supported, $m_0$ is a non-zero trigonometric polynomial and so it follows that this zero at $1/2$ is isolated. Therefore, since $\mathcal{F}\phi$ is continuous, we deduce that (\ref{fourierlower}) is nonzero when $\omega>0$ is sufficiently small.
\end{IEEEproof}

Now we cover the second half of our line-coherence bounds

\begin{theorem} \label{Thm:1dFourierWavelet}
Let $U=[(B_\rf(\epsilon),\rho),(B_\rw,\tau)]$ where $\rho$ is a frequency ordering of the Fourier basis.
Then there is a constant $C_1>0$ such that for all $\epsilon \in I_{J,p} $  and $N \in \mathbb{N}$, we have the upper bound
\[  \mu(\pi_N U) \le  \frac{C_1}{N} .\]
Furthermore, there is a constant $C_2>0$ such that for all $\epsilon \in I_{J,p}$  and $N \ge 1+ 2^{J+1} \epsilon^{-1}$ we have the lower bound
\[  \mu(\pi_N U) \ge  \frac{C_2}{N} .\]
Finally, if we replace $B_\rw$ by $B_{\rb \rw}$ in the above setup, the same conclusions also hold with the constraint $\epsilon \in I_{J,p}$ replaced by $\epsilon \in (0,1/2]$ .
\end{theorem}

\begin{IEEEproof}
\textbf{Upper Bound:} Since $\rho$ is a frequency ordering if $m=1$ then $\lambda \circ \rho(m)=0$ and since $|\mathcal{F}\phi(0)|=1$, $\mathcal{F}\psi(0)=0$ (see (\ref{fourierlower}) and the line below it), in the case of standard wavelets we have $\mu(\pi_1 U)=\epsilon 2^{-J}$. In the case of boundary wavelets we have the estimate 
\[ 
\mu(\pi_1 U)  \le \epsilon \cdot 2^{-J} \cdot \max( 1,\ |\mathcal\psi^{\text{left}}(0)|, \ |\mathcal\psi^{\text{right}}(0)|, \ |\mathcal\phi^{\text{left}}(0)|, \ |\mathcal\phi^{\text{right}}(0)|)^2 ,
\]

Next let $m \ge 1$. For standard wavelets we observe that the estimate (\ref{FTdecay}) is strong enough to bound the finitely many $\phi_{J,k}$ terms as required since
\[ | \langle \phi_k , \rho(m) \rangle |^2 =  \epsilon 2^{-J} |\mathcal{F} \phi(\epsilon 2^{-J} \cdot \lambda \circ \rho(m))|^2 \le \frac{\epsilon 2^{-J} \cdot K^2 }{| \epsilon 2^{-J} \cdot \lambda \circ \rho(m)|  } \le \frac{   2K^2 }{m-1}, \]
where we used that $\rho$ is a frequency ordering in the last step (for boundary wavelets the same holds for the finitely many $V^{\text{int}}_J$ terms). Therefore we are left with the terms involving the shifts and dilations of $\psi$ (and for boundary wavelets the $ \psi^{\text{left}}_k, \psi^{\text{right}}_k$ terms as well). This is also a straightforward consequence of (\ref{FTdecay}) since we have
\begin{align*}
| \langle  \psi_{j,k} , \rho(m) \rangle |^2  \ & = \ \epsilon 2^{-j} | \mathcal{F} \psi (\epsilon 2^{-j} \cdot \lambda \circ \rho(m)) |^2
\\ & \le \  \epsilon 2^{-j} \cdot \frac{2^j K^2}{ \epsilon \cdot | \lambda \circ \rho(m) |}  \le \  \frac{K^2}{ | \lambda \circ \rho(m) |} \ \le \ \frac{ 2 K^2}{m-1},
\end{align*}
and for boundary wavelets we can tackle the $\psi^{\text{left}}_k, \psi^{\text{right}}_k$ terms in the same way. This gives the global bound for $m \ge 2$ (uniform in $n$ and $\epsilon$)
\[ | \langle \tau(n) , \rho(m) \rangle |^2  \le  \frac{ 2 K^2}{m-1} \le \frac{ 4 K^2}{m} . \] 
Combining this with our bound on $\mu(\pi_1 U)$ (we just bound $\epsilon$ by 1) we obtain the required upper bound.

\textbf{Lower Bound:} For standard wavelets, given $m \in \mathbb{N}, \ m \neq 1$, find $n \in \mathbb{N}$ such that $\tau(n)=\psi_{j,0}$ with \text{$j= \lceil \log_2 ( \epsilon | \lambda \circ \rho(m)|) \rceil + q$}, where $q \in \bbN$ is arbitrary but sufficiently large so that $j \ge J$. Notice that this means that $\epsilon 2^{-j} |\lambda \circ \rho(m)| \in (2^{-q-1},2^{-q}]$. Therefore, recalling the definition of $L_q$ in Lemma \ref{wavelower}, we see that we have 
\begin{align*} 
| \langle \tau(n), \rho(m) \rangle |^2 = \epsilon & 2^{-j} | \mathcal{F} \psi(2^{-j} \epsilon \lambda \circ \rho(m))|^2 \\ & \ge \epsilon 2^{-\lceil \log_2 ( \epsilon | \lambda \circ \rho(m)|) \rceil - q} | \mathcal{F} \psi( \epsilon \cdot 2^{- \lceil \log_2 (\epsilon |\lambda \circ \rho(m)|) \rceil - q} \cdot \lambda \circ \rho(m))|^2 \\ & \ge \frac{L^2_q \cdot 2^{-q}}{  2 |\lambda \circ \rho(m)|} \ge \frac{L^2_q \cdot 2^{-q}}{m} .
\end{align*}
We used $m \neq 1$ in the last step and the fact that the ordering $\rho$ is standard. Recall that by Lemma \ref{wavelower} there exists a $q \in \mathbb{N}$ such that $L_q>0$. We choose the same such $q$ for all $\epsilon \in I_{J,p}$. To ensure that $j= \lceil \log_2 ( \epsilon | \lambda \circ \rho(m)|) \rceil + q$ satisfies $j \ge J$ we must therefore impose the constraint that $m$ is sufficiently large. $j \ge J$ is satisfied if
\[ J \le \log_2 ( \epsilon | \lambda \circ \rho(m)|) \quad \Leftrightarrow \quad m \ge 1+ 2^{J+1} \epsilon^{-1}. \]

When using boundary wavelets the argument for the lower bound is identical.

\end{IEEEproof}
\begin{remark}
The condition $N \ge 1+ 2^{J+1} \epsilon^{-1}$ cannot be replaced by $N \in \bbN$ for the lower bound since, in the case of standard wavelets, for every fixed $N \in \bbN$ we have 
\[
\mu(\pi_N U) \le \epsilon \cdot \max \Big( \sup_{\omega \in \bbR} | \mathcal{F} \psi (\omega)|^2, \sup_{\omega \in \bbR} | \mathcal{F} \phi (\omega)|^2 \Big) = \mathcal{O} (\epsilon).
\]
\end{remark}

Summing up Theorems \ref{Thm:1dFourierWavelet} and \ref{Thm:1dWaveletFourier} without the $\epsilon$ dependence we have
\begin{corollary} \label{Cor:FourierWavelet}
Let $U=[(B_\rf(\epsilon),\rho),(B_\rw,\tau)]$ where $\rho$ is a frequency ordering, $\tau$ is a leveled ordering and $\epsilon \in I_{J,p}$. Then
\be{ \label{Eq:SumUpFourierWaveletDecay}
\mu(\pi_NU), \mu(U \pi_N)=\Theta(N^{-1}).
}
If $B_\rw$ is replaced by $B_{\rb \rw}$ and $\epsilon \in  I_{J,p}$ by $\epsilon \in (0,1/2]$ then (\ref{Eq:SumUpFourierWaveletDecay}) also holds.
\end{corollary}

\section{1D Fourier-Polynomial Case} \label{Sec:1DFP}
Let $U= [(B_\rf(\epsilon),\rho),(B_\rp,\tau)]$ where $\rho$ is frequency ordering and $\tau$ the natural ordering on the Legendre polynomials. Before we start formally covering the line coherence estimates for these two bases we shall first need to prove a preliminary result.

\textbf{Outline of Argument:} The key fact we use in this section is that the entries of $U$ are directly related to Bessel functions (see (\ref{polyest}):
\be{ \label{Eq:BesselRelations}
 \begin{aligned} |U_{m,n}|  & =2 \sqrt{ \epsilon (n-1/2)} \cdot |j_{n-1}(2 \pi \epsilon \lambda \circ \rho(m))| 
 \\  & =   \frac{\sqrt{n-1/2}}{ \sqrt{\lambda \circ \rho(m)} } \cdot |J_{n-1/2}(2 \pi \epsilon \lambda \circ \rho(m) )|,  \qquad (m \neq 1),
 \end{aligned}
}
where $J_n$ is a Bessel function of the first kind and $j_n$ is a spherical Bessel function of the first kind. From here we rely heavily upon various asymptotic results regarding $J_n,j_n$ to produce the appropriate bounds. For $\mu(U\pi_N)$ we maximise over $m$ in (\ref{Eq:CoherenceDefinitions}) suggesting that $\mu(U \pi_N)=\Theta( \epsilon (N-1/2) \cdot \sup_{\bbR} |j_{N-1}|^2)$. Since $\sup_{\bbR} |j_{N-1}|^2 =\Theta(N^{-5/6})$ this case is then complete. 
The case of bounding $\mu(\pi_NU)$ is more involved and requires breaking down suprema into the cases $m>n$ and $m\le n$ for the upper bound and then looking near the diagonal for the lower bound.
\begin{lemma} \label{sphericalbesselmax}
Let $J_n$ denote the $n$th Bessel function of the first kind and let $j'_{n,k}$ denote the $k$th non-negative root of $J'_n$. Furthermore, Let $j_n$ denote the $n$th spherical Bessel function of the first kind and let $a'_{n,k}$ denote the $k$th non-negative root of $j'_n$. Then if $n \ge 1$ we have
\[ 
\sup_{x \in \bbR} |J_n(x)|=|J_n(j'_{n,1})|, \quad \sup_{x \in \bbR} |j_n(x)|=|j_n(a'_{n,1})|.
\]
\end{lemma}
\begin{IEEEproof}
The result for $J_n$ follows from the arguments given in \cite[Section 15]{wat}. Instead of repeating them here again, we instead adapt the same approach to deduce the Lemma for $j_n$. We will be using two facts about $j_n$. First, we have the power series expansion \cite[Eqn. (10.1.2)]{stegun}
\be{ \label{sphericalbesseltaylor}
j_n(x)= \sum_{m=0}^\infty \frac{(-1)^m 2^{n+1} (n+m+1)! x^{n+2m} }{ m! (2(n+m+1))!} .
}
Second, we shall use the fact that $j_n$ is a solution to the following differential equation \cite[Eqn. (10.1.1)]{stegun}
\be{ \label{sphericalbesseldiffequation}
x^2 j_n(x)''+2x j'_n(x)+(x^2-n(n+1))j_n(x)=0.
}
We first observe that by (\ref{sphericalbesseltaylor}), $|j_n(-x)| =|j_n(x)|, \ \forall x \in \bbR$ and so we need only consider $\sup_{x \in [0, +\infty)} |j_n(x)|$.  (\ref{sphericalbesseldiffequation}) can be rephrased as
\[ \big( x^2 j_n'(x) \big)'= (n(n+1)-x^2)j_n(x). \]
Therefore, noting that by (\ref{sphericalbesseltaylor}), $j_n(x)>0$ for $x>0$ sufficiently small, we deduce that $ x^2 j_n'(x)$ is positive for $x \in (0,n(n+1)]$ and hence so is $j'_n(x)$. This tells us that $a'_{n,k} > n(n+1)$ for all $k \in \bbN$.

Now consider the function
\[ 
\Lambda_n(x):= j_n^2(x) + \frac{x^2 j_n'^2(x)}{x^2-n(n+1)}, \qquad x \in (n(n+1),+\infty).
\]
Observe that $\Lambda_n(a'_{n,k})=j_n^2(a'_{n,k})$ for all $n,k \in \bbN$. Moreover the derivative is always negative:
\be{ 
\begin{aligned}
\Lambda'_n(x) & = 2j'_n(x)j_n(x) + \frac{2 x j_n'^2(x) + 2 x^2 j_n'(x) j_n''(x)}{x^2-n(n+1)} - \frac{2 x^3 j_n'^2(x)}{(x^2-n(n+1))^2}
\\ & = \frac{ 2j'_n(x) \big( j_n(x)(x^2-n(n+1)) +  x j_n'(x) +  x^2 j_n''(x) \big) }{x^2-n(n+1)} - \frac{2 x^3 j_n'^2(x)}{(x^2-n(n+1))^2}
\\ & = - \frac{ 2 x j'^2_n(x)}{x^2-n(n+1)} - \frac{2 x^3 j_n'^2(x)}{(x^2-n(n+1))^2} <0. \qquad \text{(using (\ref{sphericalbesseldiffequation}))}
\end{aligned} 
}
This tells that $|j_n(a'_{n,1})|>|j_n(a'_{n,2})| > |j_n(a'_{n,3})|...$. To finish the proof we notice that by (\ref{sphericalbesseltaylor}), $j_n(0)=0$ for $n \ge 1$ and furthermore, by \cite[Eqn. (10.1.14)]{stegun},
\[
 j_{n}(x) = \frac{(-\ri)^n}{2} \int_{-1}^1 e^{ \ri x t} R_N(t) \,dt , 
\]
and therefore $j_n(x) \to 0 $ as $x \to +\infty$ by the Riemann-Lebesgue Lemma. We therefore know that the maxima of $|j_n(x)|$ on $[0,+\infty)$ must be attained at its first stationary point.  
\end{IEEEproof}

\begin{theorem} \label{Thm:1dPolyFourierAsymp}
Let $U=[(B_\rf(\epsilon),\rho),(B_\rp,\tau)]$ where $\tau$ is the natural ordering of the polynomial basis. Then there are constants $C_1, C_2 >0$ such that for all $\epsilon \in (0,0.45]$ and $N \in \mathbb{N}$,
\[ \frac{\epsilon \cdot C_1 }{ N^{2/3}} \le \mu(U \pi_N )  \le  \frac{\epsilon \cdot C_2}{N^{2/3}}. \]
\end{theorem}
\begin{IEEEproof}
\textbf{Upper Bound:} First notice that
\begin{equation} \label{polyest}
 \begin{aligned} U_{m,n}  & =  \langle \rho(m) , \tilde{p}_n  \rangle_{L^2([-1,1])} \\  & =  \sqrt{\epsilon} \cdot \sqrt{n-1/2} \int_{-1}^1 e^{2  \pi \ri \lambda \circ \rho(m) \epsilon t} R_N(t) \,dt \ 
\\ & =  \ri^{n-1} 2 \sqrt{ \epsilon (n-1/2)} \cdot j_{n-1}(2 \pi \epsilon \lambda \circ \rho(m)) 
 \\  & =   \ri^{n-1} \frac{\sqrt{n-1/2}}{ \sqrt{\lambda \circ \rho(m)} } \cdot J_{n-1/2}(2 \pi \epsilon \lambda \circ \rho(m) ),  \qquad (m \neq 1)
 \end{aligned}
 \end{equation}
 where on the third line we have used \cite[Eqn. (10.1.14)]{stegun} and on the fourth line we have used the following formula connecting the spherical Bessel function to the standard Bessel function:
\begin{equation} \label{sphericalbessel}
 j_n(z)  =  \sqrt{\frac{\pi}{2z}} J_{n+1/2}(z). 
 \end{equation}
 Therefore, we find 
\begin{equation} \label{polytospherical}
 \mu(U \pi_N )  \le  4 \epsilon (N-1/2) \sup_{t \in \bbR} j^2_{N-1}(t). 
\end{equation}
We therefore need to estimate $\sup_{t \in \bbR} |j_{n}(t)|$. By Lemma \ref{sphericalbesselmax}, we know that $\sup_{t \in \bbR} |j_{n}(t)|=|j_n(a'_{n,1})|$ for $n \ge 1$, where $a'_{n,1}$ denotes the first positive root of $j'_n$.

Thus, we only need to have estimates for $|j_n(a'_{n,1})|$. But we also know \cite[Eqn. (10.1.61)]{stegun},  that the following asymptotic expansion holds
\begin{equation} \label{sphericalmax}
\begin{aligned}
j_n(a'_{n,1}) \sim \gamma (n+1/2)^{-5/6} + \mathcal{O} ((n+1/2)^{-3/2}),
\end{aligned}
\end{equation}
for some positive constant $1/2<\gamma<1$.
Therefore we know there exists $N' \in \mathbb{N}$ such that for all $N>N'$ we have
\[ \sup_{x \in \mathbb{R}}|j_N(x)| \le (N+1/2)^{-5/6} . \]
Applying this bound to (\ref{polytospherical}) we get the upper bound
\[
\begin{aligned}
\mu(U \pi_N ) \ & \le \  4 \epsilon (N-1/2) \cdot (N-1/2)^{-5/3}
\\ & \le \frac{4 \epsilon}{(N-1/2)^{2/3}} \le  \frac{8 \epsilon}{N^{2/3}}.
\end{aligned}
\]
Therefore the upper bound is complete for the case $N>N'$ (and notice that $N'$ is independent of $\epsilon$). However since $\sup_{x \in \mathbb{R}}|j_N (x)|< \infty$ for every $N$ we can use (\ref{polytospherical}) to cover the case $N \le N'$, completing the upper bound.

\textbf{Lower Bound:} We focus on the following equation taken from (\ref{polyest})
\be{ \label{standardbessel}
|U_{m,n}|= \frac{\sqrt{n-1/2}}{ \sqrt{|\lambda \circ \rho(m)|} } \cdot |J_{n-1/2}(2 \pi \epsilon \lambda \circ \rho(m) )|,  \qquad (m \neq 1).
}
Let $j'_{\nu,1}$ denote the first positive zero of $J'_\nu$. From \cite[Eqns. (9.5.16), (9.5.20)]{stegun}, we have the asymptotic estimates 
\begin{align}
 \label{standardbesselmaxima}
j'_{\nu,1} \sim \nu + \zeta \nu^{1/3} + \mathcal{O}(\nu^{-1/3}),
\\
\label{standardbesselmax}
J(j'_{\nu,1}) \sim \kappa \cdot \nu^{-1/3}+ \mathcal{O}(\nu^{-1}),
\end{align}
where $\kappa, \zeta>0$ are some constants. Next let $k_m$ denote the nearest integer multiple of $2 \pi \epsilon$ to $j'_{n-1/2,1}$, which means that $|k_n-j'_{n-1/2,1}| \le \pi \epsilon$. We shall first prove a lower bound for $|J_{n-1/2}(k_n)|$. Before we do so, we need the following two results:
\begin{enumerate}
\item $2J'_{\nu}(x)=J_{\nu-1}(x)-J_{\nu+1}(x), \qquad \text{\cite[p. 45]{wat}}$,  
\item $\sup_{x \in \bbR} |J_\nu(x)|=|J_\nu(j'_{\nu,1})|, \qquad$ using Lemma \ref{sphericalbesselmax} .
\end{enumerate}
These two results can be combined to give us $\sup_{x \in \bbR}|J''_{\nu}(x)| \le |J_\nu(j'_{\nu,1})|$, which we will use in (\ref{doubleint}) below. 

By the triangle inequality $|J_{n-1/2}(k_n)| \ge |J_{n-1/2}(j'_{n-1/2,1})|-|J_{n-1/2}(k_n)-J_{n-1/2}(j'_{n-1/2,1})|$ and we bound the latter term by using integrals\footnote{The use of the second integral is valid since $J_{n-1/2,1}'(t)= J_{n-1/2,1}'(t) - J_{n-1/2,1}'(j'_{n-1/2,1})$ by the definition of $j'_{n-1/2,1}$.}:
\be{ \label{doubleint}
\begin{aligned}
|J_{n-1/2}(k_n) & -J_{n-1/2}(j'_{n-1/2,1})|  = \Bigg| \int_{j'_{n-1/2,1}}^{k_n} J_{n-1/2,1}'(t) \,dt \Bigg|
\\ & = \Bigg|  \int_{j'_{n-1/2,1}}^{k_n} \int_{j'_{n-1/2,1}}^{t} J''_{n-1/2,1}(u) \,du \,dt \Bigg|
\\ & \le \Bigg| \int_{j'_{n-1/2,1}}^{k_n} \int_{j'_{n-1/2,1}}^{t} |J_{n-1/2}(j'_{n-1/2,1})| \,du \,dt \Bigg|
\\ & \le \frac{|j'_{n-1/2,1}-k_n|^2}{2} \cdot |J_{n-1/2}(j'_{n-1/2,1})|
\\ & \le \frac{ (\pi \epsilon)^2}{2} \cdot |J_{n-1/2}(j'_{n-1/2,1})|.
\end{aligned}
}
Notice that there is a constant $1>d>0$ such that for all $\epsilon \in (0,0.45]$ we have $(\pi \epsilon)^2/2 \le d$ and therefore (\ref{doubleint}) becomes
\[ |J_{n-1/2}(k_n)  -J_{n-1/2}(j'_{n-1/2,1})| \le d \cdot |J_{n-1/2}(j'_{n-1/2,1})|, \]
and therefore we deduce 
\[
\begin{aligned}
|J_{n-1/2}(k_n)| & \ge |J_{n-1/2}(j'_{n-1/2,1})|-|J_{n-1/2}(k_n)-J_{n-1/2}(j'_{n-1/2,1})|
\\ & \ge (1-d) \cdot |J_{n-1/2}(j'_{n-1/2,1})|. 
\end{aligned}
\]
Combining this inequality with (\ref{standardbesselmaxima}), (\ref{standardbesselmax}) gives us the following bound:
\[
\begin{aligned}
\frac{\sqrt{n-1/2}}{ \sqrt{k_n} } & \cdot |J_{n-1/2}(k_n )|  \ge \frac{\sqrt{j'_{n-1/2,1}}}{\sqrt{k_n}} \cdot \frac{\sqrt{n-1/2}}{ \sqrt{j'_{n-1/2,1}} } \cdot (1-d) |J_{n-1/2}(j'_{n-1/2,1})|
\\ & = \frac{\sqrt{j'_{n-1/2,1}}}{\sqrt{k_n}} \cdot \frac{\sqrt{n-1/2}}{ \sqrt{n-1/2 + \mathcal{O}(n^{1/3}) }} \cdot (1-d)(\kappa (n-1/2)^{-1/3} + \mathcal{O}(n^{-1})).
\end{aligned}
\]
The first two fractions on the last line converge to $1$ as $n \to \infty$ and therefore we deduce that there is an $M \in \bbN$ and a constant $C>0$ (independent of $\epsilon$) such that for all $n \ge M$ we have 
\be{ \label{polylocalmax}
\frac{\sqrt{n-1/2}}{ \sqrt{k_n} }  \cdot |J_{n-1/2}(k_n )| \ge C n^{-1/3}.
}
Therefore, given $n \ge M$, let $m(n) \in \bbN$  be such that $2\pi\epsilon \lambda \circ \rho(m)=k_n$. Then by (\ref{polylocalmax}) and (\ref{standardbessel}) we have
\[ |U_{m(n),n}| = \sqrt{2 \pi \epsilon} \cdot \frac{\sqrt{n-1/2}}{ \sqrt{k_n} }  \cdot |J_{n-1/2}(k_n )| \ge \sqrt{2 \pi \epsilon} \cdot C n^{-1/3}.
\]
Consequently we deduce that $\mu(U \pi_N ) \ge 2 \pi \epsilon \cdot C^2 N^{-2/3}$ for $N \ge M$.

 For $N \le M$ we observe that from (\ref{polyest})
 \[ 
 \mu(U \pi_N) =  4 \epsilon (N-1/2) \sup_{m \in \bbZ} |j_{N-1}(2 \pi \epsilon m)|^2.
 \]
 As before we observe that since $j_{N-1}(x) \to 0$ as $x \to \infty$, the supremum $\sup_{m \in \bbZ} |j_{N-1}(2 \pi \epsilon m)|$ is a continuous function of $\epsilon$ and moreover the supremum converges to $\sup_{x \in \bbR} |j_{N-1}(x)|>0$ as $\epsilon \to 0$. Therefore by compactness of $[0,0.45]$, we know there is a constant $D_N>0$ such that for all $\epsilon \in (0,0.45]$ we have
 \[ 
 \mu(U \pi_N ) =  4 \epsilon (N-1/2) \cdot D^2_N.
 \]
 This combined with the result $\mu(U \pi_N) \ge 2 \pi \epsilon \cdot C^2 N^{-2/3}$ for $N \ge M$ gives us the required lower bound.
\end{IEEEproof}

\begin{theorem} \label{Thm:1dFourierPolyAsymp}
Let $U=[(B_\rf(\epsilon),\rho),(B_\rp,\tau)]$ where $\rho$ is a frequency ordering of the Fourier basis. Then there is a constant $C_1>0$ such that for all $\epsilon \in (0,1/2]$ and $N \in \mathbb{N}$
\[ \mu(\pi_N U  )  \le  \frac{ C_1 \epsilon^{1/3}}{N^{2/3}}.  \ \]
Furthermore, there is a constant $C_2>0$ such that for all $\epsilon \in (0,1/2]$ there exists an $M(\epsilon) \in \bbN$ such that for all $ N \ge M$ we have the bound
\[ \mu(\pi_N U  )  \ge  \frac{ C_2 \epsilon^{1/3}}{N^{2/3}}. \]
\end{theorem}
\begin{IEEEproof}
\textbf{Upper Bound:} Without loss of generality we can assume $\tau$ is the natural ordering of $B_\rp$. Recall that from (\ref{polyest}) we have 
\begin{align}  
|U_{m,n}|^2 & = \frac{n-1/2}{|\lambda \circ \rho(m)| } J^2_{n-1/2}(2 \pi \epsilon \lambda \circ \rho(m) ) \label{polysummary1} 
\\ & = 4 \epsilon (n-1/2) j_{n-1}^2 (2 \pi \epsilon \lambda \circ \rho(m)). \label{polysummary2} 
\end{align}
We shall first derive two useful bounds; notice that if we apply (\ref{sphericalmax}) to (\ref{polysummary2}) then we get the bound, for some constant $\beta>0$,
\begin{equation} \label{polyfirst}
\begin{aligned}
|U_{m,n}|^2 & \le 4 \epsilon (n-1/2) \cdot (\beta (n-1/2)^{-5/6})^2 \le 4 \epsilon \beta^2 (n-1/2)^{-2/3}.
\end{aligned}
\end{equation}
Secondly we shall use the following inequality from \cite{landau}
\be{ \label{landaubound}
 |J_{\nu}(x)| \le b \nu^{-1/3} \qquad \nu>0 , \quad x \in \mathbb{R}, 
 }
where $b>0$ is some constant. Applying this to (\ref{polysummary1}) gives the bound
\begin{equation} \label{polysecond}
\begin{aligned}
|U_{m,n}|^2 & \le \frac{n-1/2}{|\lambda \circ \rho(m)| }  (b (n-1/2)^{-1/3})^2 \le \frac{b^2(n-1/2)^{1/3}}{|\lambda \circ \rho(m)| }.  
\end{aligned}
\end{equation}
Recall that our goal is to estimate $|U_{m,n}|$ uniformly in $n$ as $m \to \infty$. We first apply the case $ n-1/2 \ge \epsilon | \lambda \circ \rho(m)| $ to (\ref{polyfirst}) to give the bound
\[  |U_{m,n}|^2 \le 4 \epsilon \beta^2 (\epsilon \lambda \circ \rho(m))^{-2/3} \le \frac{ 4 \beta^2 \epsilon^{1/3}}{|\lambda \circ \rho(m)|^{2/3}} .\]
For the other case $n-1/2 \le \epsilon |\lambda \circ \rho(m)|$ we use (\ref{polysecond}) to give the bound
\[  |U_{m,n}|^2 \le \frac{b^2(\epsilon |\lambda \circ \rho(m)|)^{1/3}}{|\lambda \circ \rho(m)| } \le \frac{b^2 \epsilon^{1/3} }{ | \lambda \circ \rho(m) |^{2/3}} = \frac{b^2 \epsilon^{1/3} 2^{2/3}}{ (m-1)^{2/3}},\]
which gives a global upper bound in terms of $m \ge 2$ and $\epsilon \in (0,1/2]$. If $m=0$, i.e. $\lambda \circ \rho(m)=0$, then since $j_n(0)=0$ for $n \ge 1$ (see (\ref{sphericalbesseltaylor})) we deduce that $\mu(\pi_1 U)= \epsilon |j_0(0)|^2= \epsilon$ which is a stronger bound than required.

\textbf{Lower Bound:} By (\ref{standardbesselmaxima}) we know that
\be{ \label{maximasteplimit}
j'_{n+1/2,1}-j'_{n-1/2,1} \to 1 \quad \text{as} \quad n \to \infty.
} 
With this in mind let $n(m)\in \bbN$ denote the nearest $j'_{n-1/2,1}$ to $|2 \pi \epsilon \lambda \circ \rho(m)|$. From (\ref{maximasteplimit}) we observe
\be{ \label{roottoindex}
|j'_{n(m)-1/2,1}-|2 \pi \epsilon \lambda \circ \rho(m)||\le 1/2 + \eta(m,\epsilon),
} 
where $\eta$ is such that $\eta(m,\epsilon) \to 0 $ as $m \to \infty$ for any fixed $\epsilon$. By using the same method as in (\ref{doubleint}) we find that
\[
\begin{aligned}
|J_{n(m)-1/2,1}(j'_{n(m)-1/2,1}) & - J_{n(m)-1/2,1}(|2 \pi \lambda \circ \rho(m)|)| 
\\ & \le \frac{|j'_{n(m)-1/2,1}-|2 \pi \epsilon \lambda \circ \rho(m)||^2}{2} \cdot |J_{n(m)-1/2,1}(j'_{n(m)-1/2,1})|
\\ & \le 2^{-1} \cdot (2^{-1} + \eta(m,\epsilon))^2 \cdot |J_{n(m)-1/2,1}(j'_{n(m)-1/2,1})|
\\ & = \xi(m, \epsilon) \cdot |J_{n(m)-1/2,1}(j'_{n(m)-1/2,1})|.
\end{aligned}
\]
Where $\xi(m, \epsilon) \to 8^{-1}$ as $m \to \infty$ with $\epsilon$ fixed. This tells us that 
\[
\begin{aligned}
 |J_{n(m)-1/2,1}(2 \pi \lambda \circ \rho(m))| = |J_{n(m)-1/2,1}(2 \pi \lambda \circ \rho(m))| &
\\  \ge |J_{n(m)-1/2,1}(j'_{n(m)-1/2,1})|-|J_{n(m)-1/2,1}(j'_{n(m)-1/2,1}) & - J_{n(m)-1/2,1}(|2 \pi \lambda \circ \rho(m)|)| 
\\  \ge (1-\xi(m,\epsilon)) |J_{n(m)-1/2,1}(j'_{n(m)-1/2,1})| &.
\end{aligned}
\]
Combining this with (\ref{polysummary1}) we see that, using (\ref{standardbesselmax}),
\be{ \label{finallower}
\begin{aligned}
|U_{m,n(m)}|^2 & =  \frac{n(m)-1/2}{|\lambda \circ \rho(m)| } |J^2_{n(m)-1/2}(2 \pi \epsilon \lambda \circ \rho(m) )|
\\ & \ge \frac{n(m)-1/2}{|\lambda \circ \rho(m)| } \cdot \Big( 1-\xi(m, \epsilon) \Big)^2 \cdot |J_{n(m)-1/2,1}(j'_{n(m)-1/2,1})|^2
\\ & \ge \frac{n(m)-1/2}{|\lambda \circ \rho(m)| } \cdot  \Big( 1- \xi(m, \epsilon) \Big)^2  \big( \kappa(n(m)-1/2)^{-1/3}+ \mathcal{O}((n(m)-1/2)^{-1}) \big)^2.
\end{aligned}
}
By (\ref{standardbesselmaxima}), (\ref{roottoindex}) and the fact that $\rho$ is a standard ordering we know that (for $\epsilon$ fixed)
\[
\frac{n(m)}{| \pi \epsilon m|} \to 1, \quad \text{as} \quad m \to \infty.
\]
Therefore we know that there is an $M(\epsilon) \in \bbN$ and a constant $C>0$ such that for all $m \ge M$ and $\epsilon \in (0,1/2]$ we have
\[ 
|U_{m,n(m)}|^2 \ge C \cdot \epsilon^{1/3} \cdot m^{-2/3}.
\]
Consequently for $N \ge M(\epsilon)$ we have $\mu(\pi_N U) \ge C \cdot \epsilon^{1/3} \cdot N^{-2/3}$. 
\end{IEEEproof}

Summarising our results in this section, while throwing away $\epsilon$ dependence again, we deduce the following

\begin{corollary} \label{Cor:FourierPolynomial}
Let $U=[(B_\rf(\epsilon),\rho),(B_\rp,\tau)]$ where $\rho$ is a frequency ordering, $\tau$ is the natural ordering and $\epsilon \in (0,0.45]$. Then
\be{ \label{Eq:SumUpFourierPolynomialtDecay}
\mu(\pi_NU), \mu(U \pi_N)=\Theta(N^{-2/3}).
}
\end{corollary}

\section{Asymptotic Incoherence and Subsampling Strategies} \label{Sec:NumericalSubsampling}

We have shown that there is faster asymptotic incoherence for the Fourier-wavelet case than for the Fourier-polynomial case. We shall demonstrate how this difference is vital for choosing an effective sampling strategy. 

Consider the problem of reconstructing the function $f \in L^2[-1,1]$ from its samples $\{ \langle f, g \rangle : g \in B_\rf (1/2) \} $, where f is defined as
\be{ \label{functiondefine}
f(x)= (1-\cos(8 \pi x)) \cdot \mathds{1}_{[0,1]} (x), \qquad x \in [-1,1].
}

The function $f$ is reconstructed as follows: Let $U:=[(B_\rf(2^{-1}),\rho), (B_2, \tau)]$ for some orderings $\rho, \tau$ and a reconstruction basis $B_2$. The number $2^{-1}$ is present here to ensure the span of $B_\rf$ contains $L^1[-1,1]$. It is assumed that $\rho$ is a frequency ordering. Next let $\Omega \subset \bbN$ denote the set of subsamples from $B_\rf(2^{-1})$ (indexed by $\rho$), $P_\Omega$ the projection operator onto $\Omega$ and $\hat{f}:=( \langle f , \rho(m) \rangle )_{m \in \bbN}$. We then attempt to approximate $f$ by $\sum_{n=1}^\infty \tilde{x}_n \tau(n)$ where $\tilde{x} \in \ell^1(\bbN)$ solves the optimisation problem
\be{ \label{basicl1full}
\min_{x \in \ell^1(\bbN)} \| x \|_1  \quad \text{subject to} \quad P_\Omega Ux= P_\Omega \hat{f} .
}
Since the optimisation problem is infinite dimensional we cannot solve it numerically so instead we proceed as in \cite{BAACHGSCS} and truncate the problem, approximating $f$ by 
$\sum_{n=1}^R \tilde{x}_n \tau(n)$ (for $R \in \bbN$ large) where $\tilde{x} = (\tilde{x}_n)_{n=1}^R$ now solves the optimisation problem
\be{ \label{basicl1}
\min_{x \in \bbC^R} \| x \|_1  \quad \text{subject to} \quad P_\Omega U P_R x= P_\Omega \hat{f} .
}
\begin{figure}[!h]
\caption{Coefficients of $f$ when decomposed into different reconstruction bases.} 
\begin{center}
\begin{subfigure}[t]{0.4\textwidth}
\begin{center}
 \hspace{0.1em} \includegraphics[width=\textwidth]{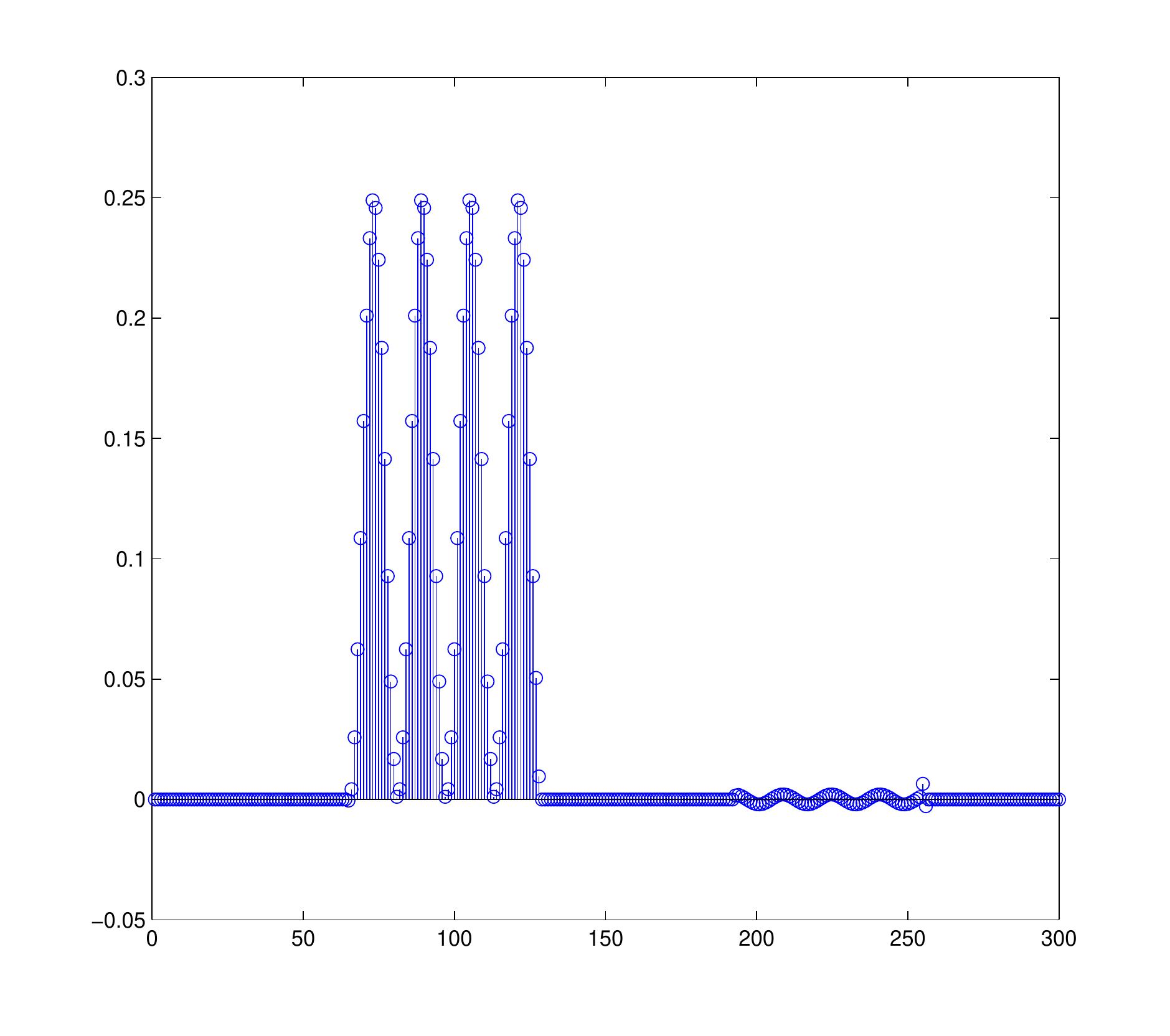}
 \caption{\footnotesize First 300 coefficients (using a leveled ordering) of $f$ in a Daubechies4 boundary wavelet expansion with J=6.}
\end{center}
\end{subfigure}
\begin{subfigure}[t]{0.4\textwidth}
\begin{center}
  \hspace{0.1em} \includegraphics[width=\textwidth]{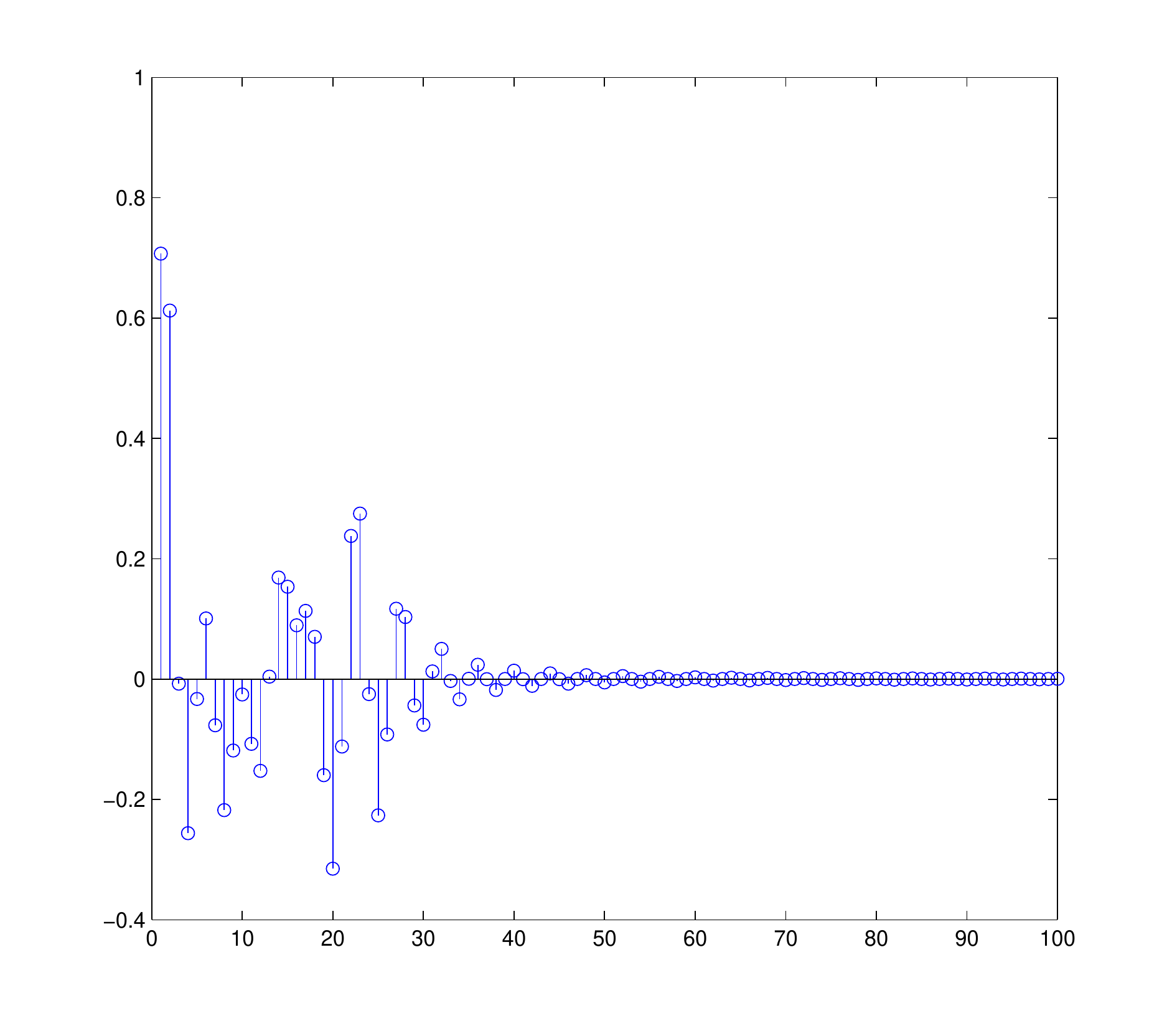} 
  \caption{\footnotesize First 100 coefficients (using a natural ordering) of $f$ in a Legendre polynomial expansion. }
  \end{center}
\end{subfigure}
\end{center}
\label{functiondecomposition}
\end{figure}
We shall be using the SPGL1 package \cite{SPGL} to solve (\ref{basicl1}) numerically. We focus on two choices of reconstruction bases:
\begin{enumerate}
 \item $B_2=B_{\rb \rw}$ with Daubechies4 boundary wavelets, $\tau$ is a leveled ordering.
 \item $B_2=B_\rp$ with Legendre polynomials, $\tau$ is a natural ordering.
\end{enumerate} 
  The coefficients of the decomposition of $f$ into these two bases is shown in Figure \ref{functiondecomposition}. The coefficients in the polynomial expansion decay quickly, but there is little sparsity in the first 40 coefficients. On the other hand in the wavelet expansion there is large number of zeros in the first block of coefficients. This, combined with asymptotic incoherence, will enable us to subsample.

\begin{figure}[!h]
\caption{Two sampling patterns and their corresponding histograms.} 

\begin{center}

\begin{subfigure}[t]{0.24\textwidth}
\includegraphics[width=\textwidth]{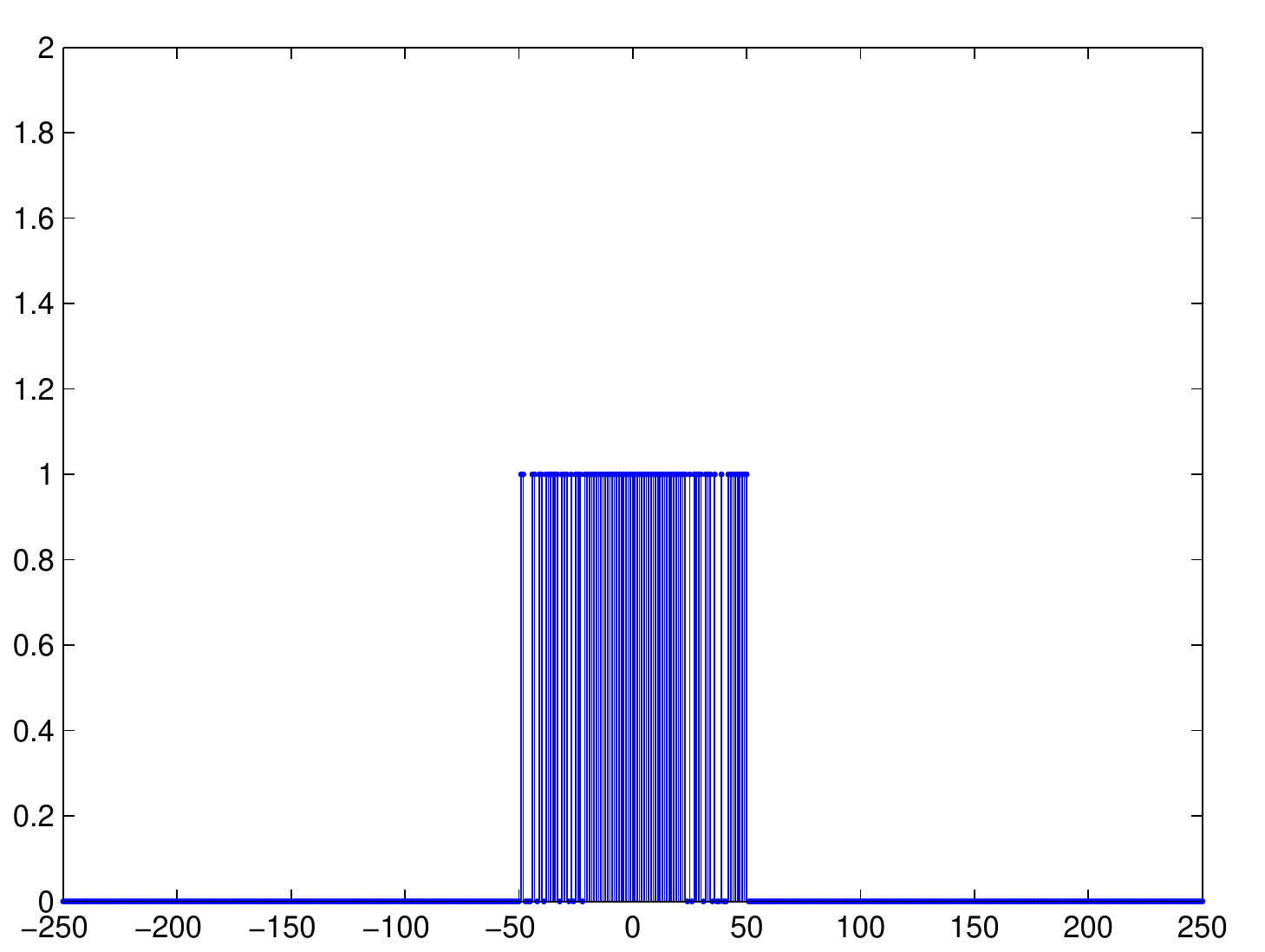}
  \caption{ \footnotesize Sampling Pattern A}
\end{subfigure}
\begin{subfigure}[t]{0.24\textwidth}
  \includegraphics[width=\textwidth]{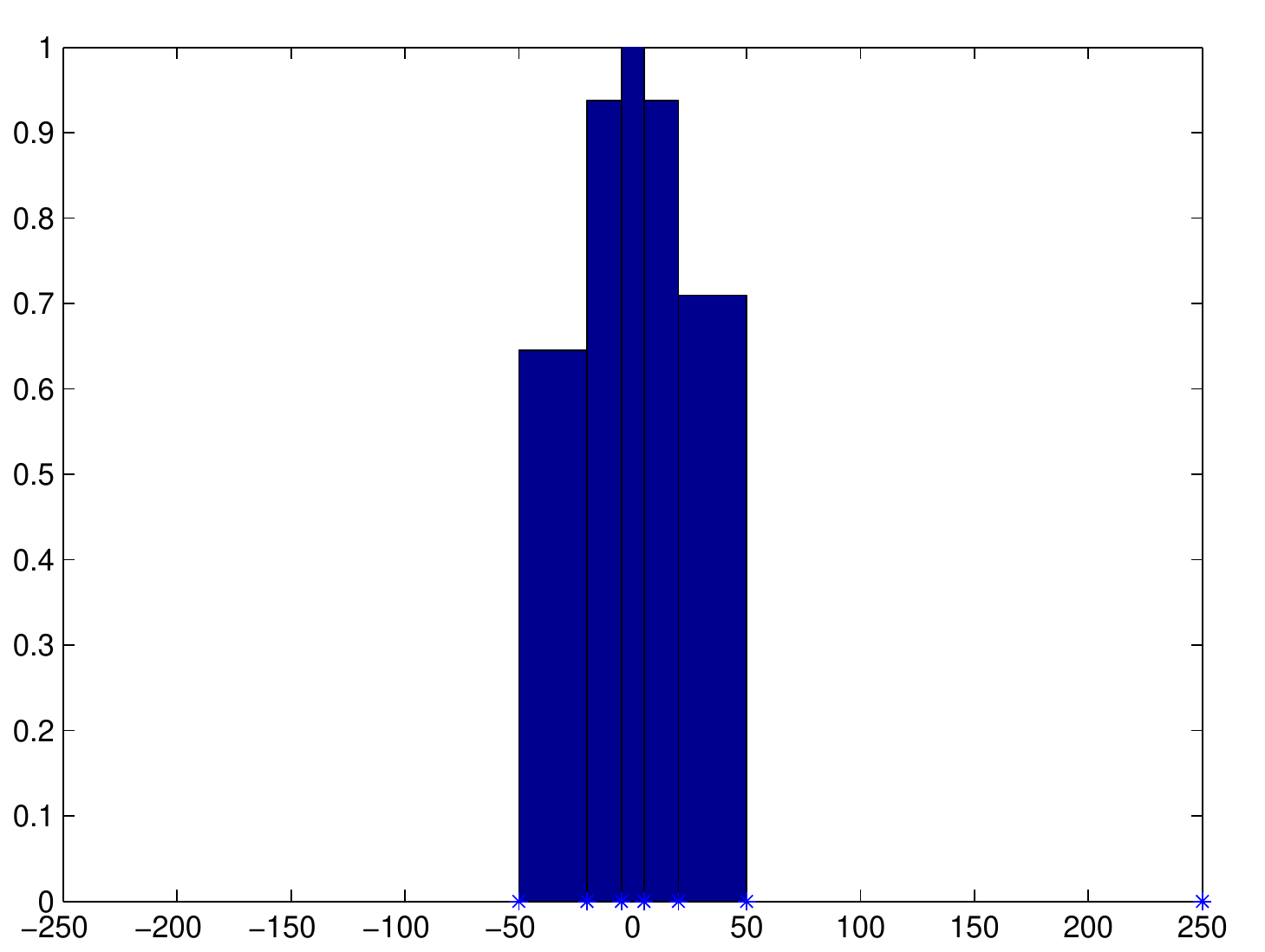} 
  \caption{ \footnotesize Histogram for Pattern A}
\end{subfigure}
\begin{subfigure}[t]{0.24\textwidth}
  \includegraphics[width=\textwidth]{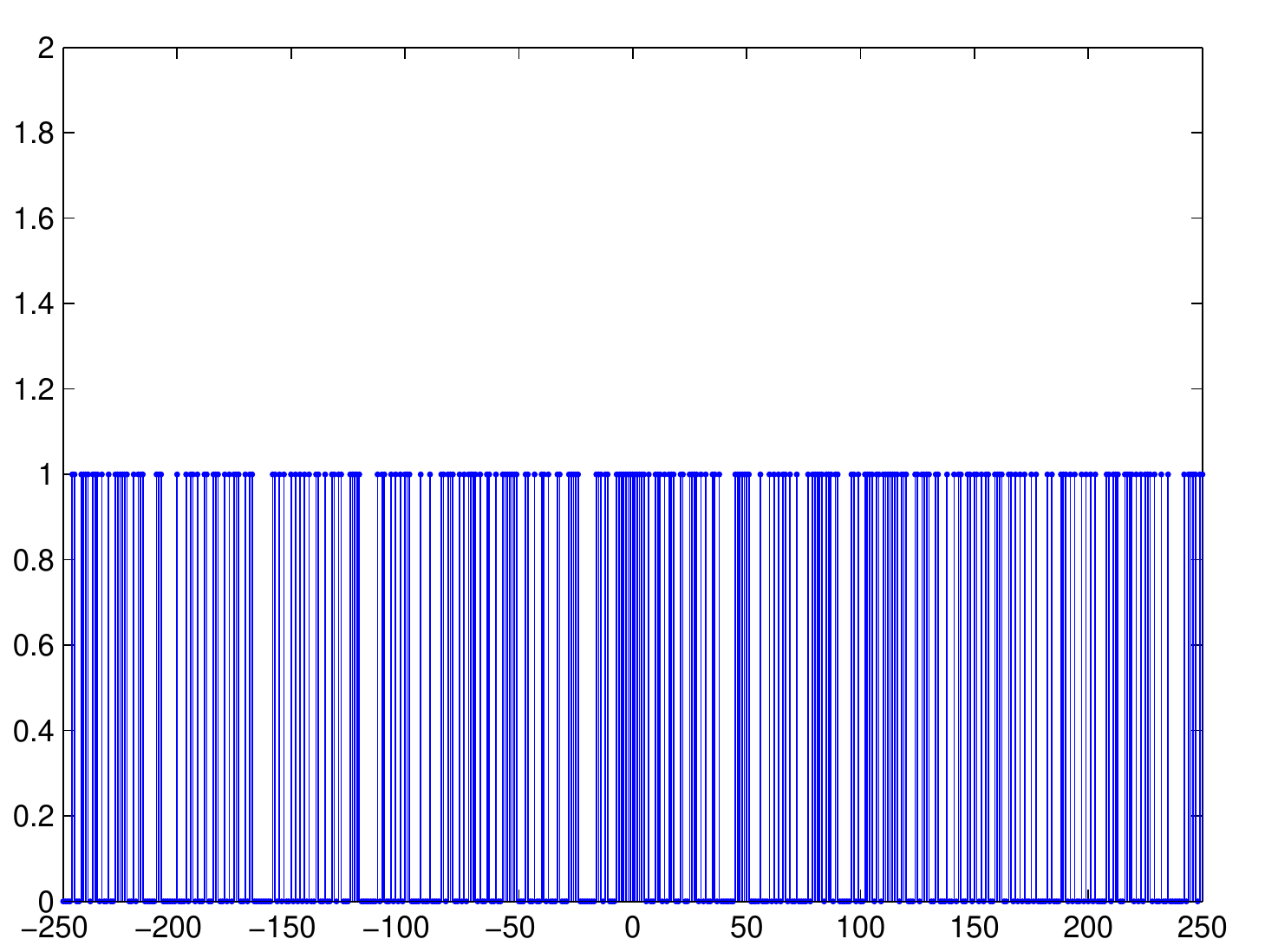}
  \caption{ \footnotesize Sampling Pattern B}
\end{subfigure}
\begin{subfigure}[t]{0.24\textwidth}
  \includegraphics[width=\textwidth]{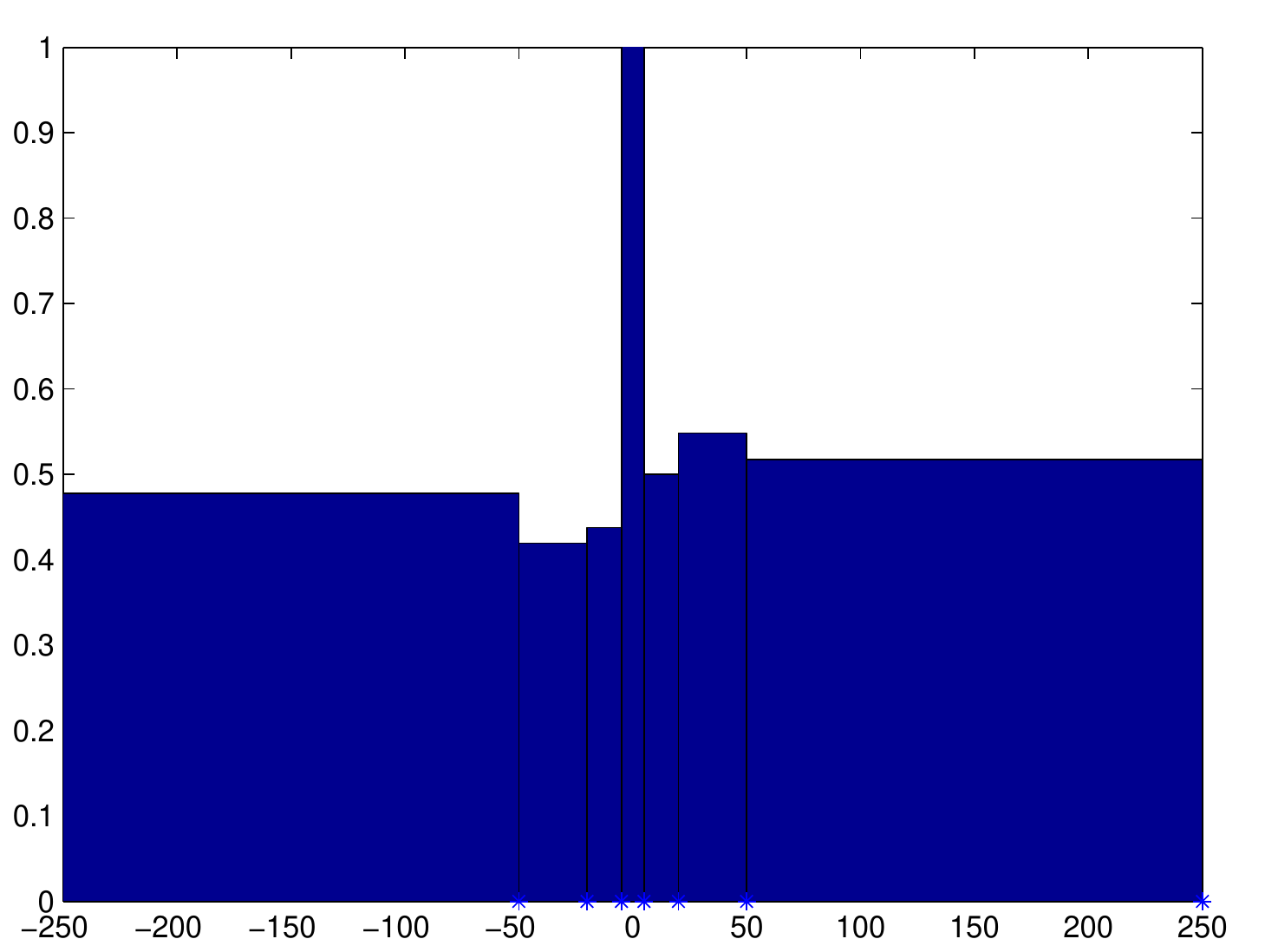}
  \caption{ \footnotesize Histogram for Pattern B}
\end{subfigure}
\end{center}
\label{samplingpatterns}
\end{figure}

We shall be looking at two simple subsamping patterns and how they perform for each reconstruction basis. We shall be subsampling from the first $501$ coefficients, and since $\rho$ is a frequency ordering this means that these coefficients correspond to
\[ 
\{\lambda \circ \rho(m): m=1, \cdots ,501 \} =  \{-250,-249, \cdots, 249,250 \}.
\]
If we were to sample all the $501$ coefficients then we would achieve a highly accurate reconstruction from both bases\footnote{For all our reconstructions we will be using $R=1024$.}. We now consider two subsampling patterns, denoted as pattern A and pattern B which are presented in Figure \ref{samplingpatterns}, and now try to use them to reconstruct in the bases $B_{\rb \rw}$, $B_\rp$.  Pattern A takes all its samples from the first $101$ coefficients and there is very little subsampling in this range. On the other hand pattern B takes around $50\%$ of the samples from across the first $501$ coefficients. Both patterns are constructed by uniformly subsampling in levels.

\begin{figure}[!h]
\caption{ Reconstructions from Pattern A (above) with errors (below).}
\begin{center}

\begin{subfigure}[t]{0.4\textwidth}
\includegraphics[width=\textwidth]{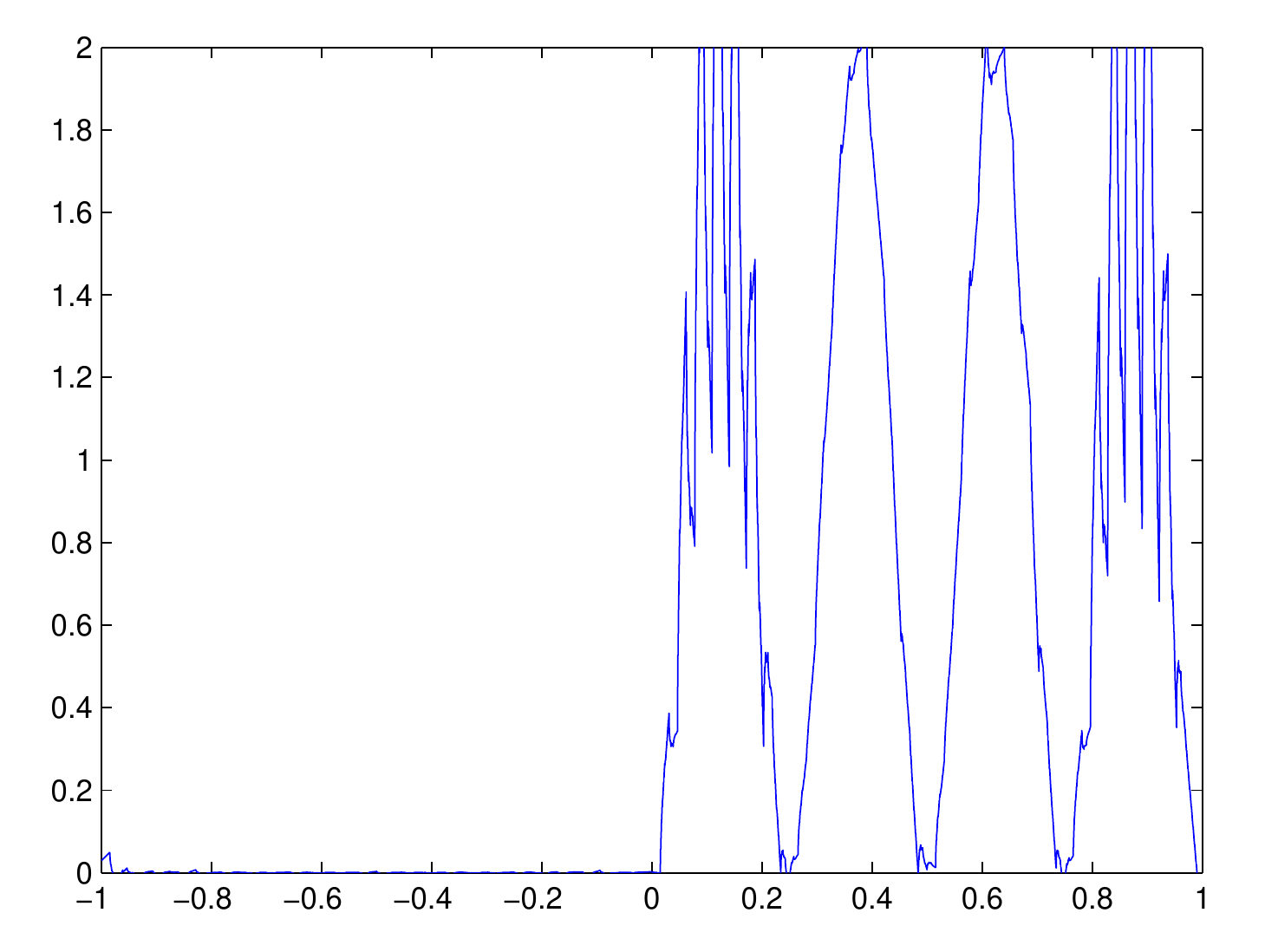}
\includegraphics[width=\textwidth]{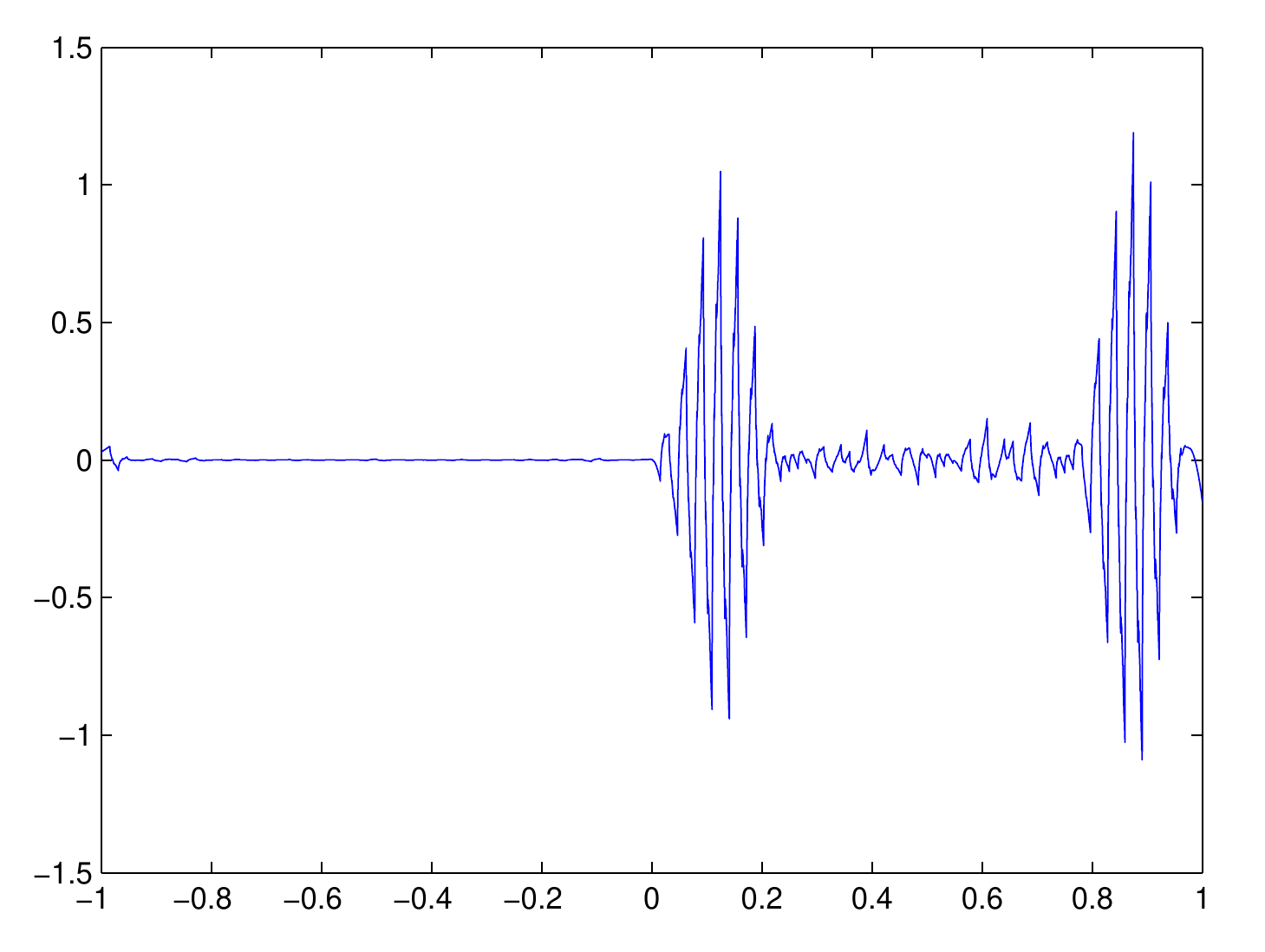} 
  \caption{ \footnotesize Wavelet Reconstruction}
\end{subfigure}
\begin{subfigure}[t]{0.4\textwidth}
  \includegraphics[width=\textwidth]{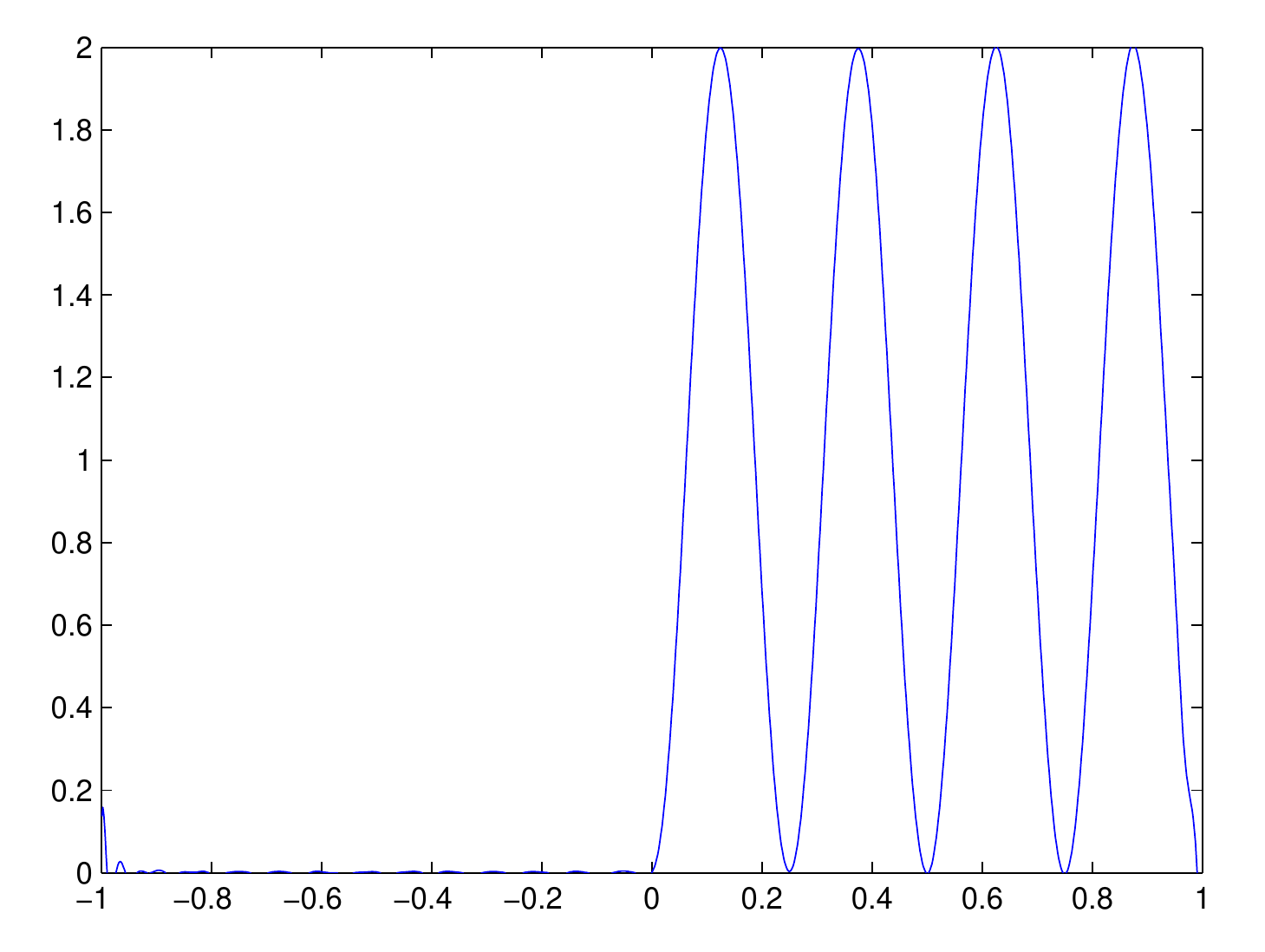}
  \includegraphics[width=\textwidth]{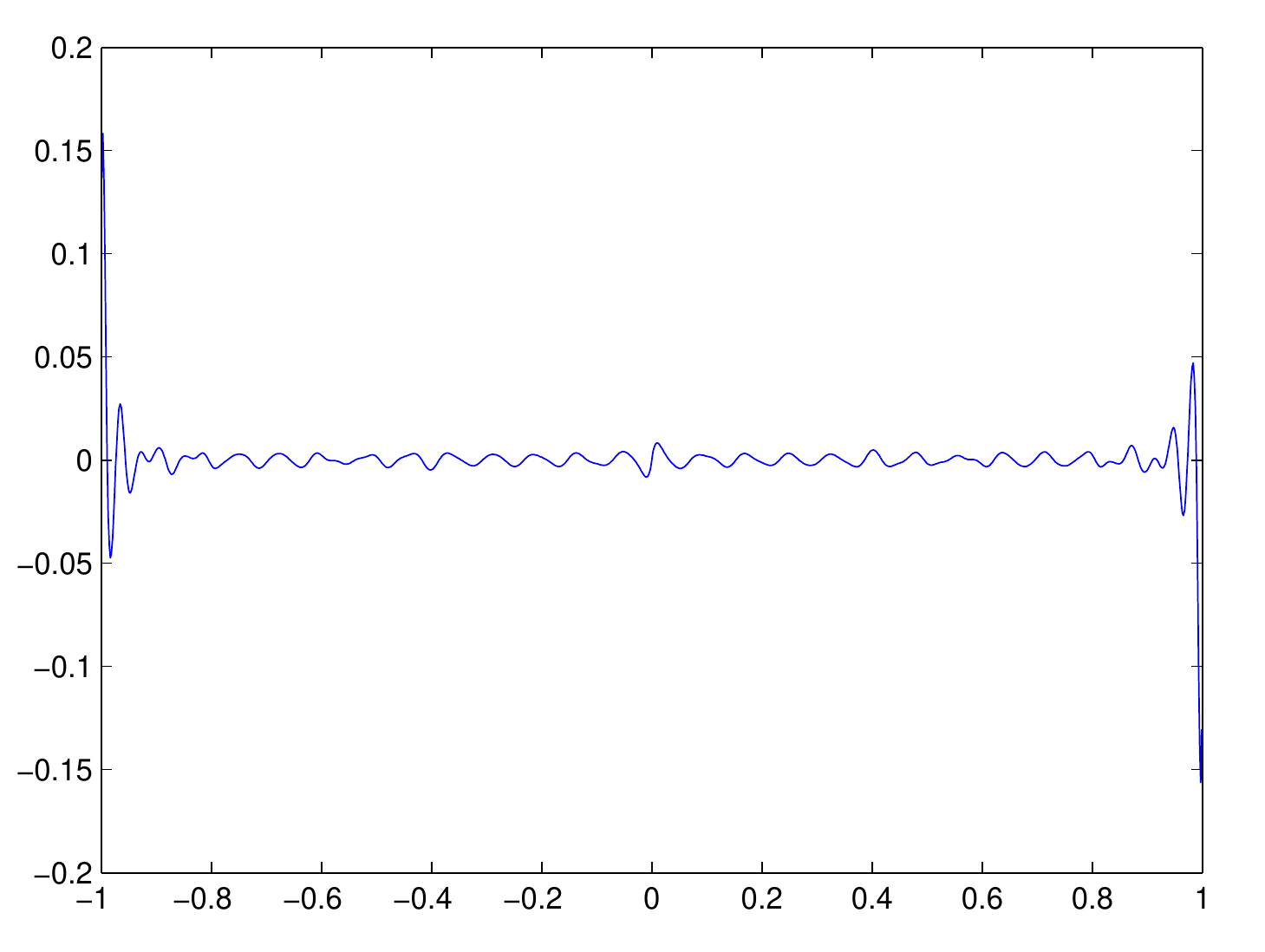}
  \caption{ \footnotesize Polynomial Reconstruction}
\end{subfigure}
\end{center}

\label{reconstructionA}
\end{figure}

Let us first consider what happens when we use subsampling pattern A, which is shown in Figure \ref{reconstructionA}. We first look at the wavelet reconstruction, which has an $L^1$ error of $1.52 \times 10^{-1}$. The reconstruction fails to reconstruct the smoothness of $f$, with the first and fourth peaks being particularly jagged. Next consider the polynomial reconstruction, which has an $L^1$ error of $8.68 \times 10^{-3}$. Since polynomials provide a relatively good linear approximation to $f$, it is unsurprising that using a near full-sampling subsampling pattern for the first $101$ Fourier coefficients would give a reasonable reconstruction.  

\begin{figure}[!h]
\caption{ Reconstructions from Pattern B with errors.}
\begin{center}

\begin{subfigure}[t]{0.4\textwidth}
\includegraphics[width=\textwidth]{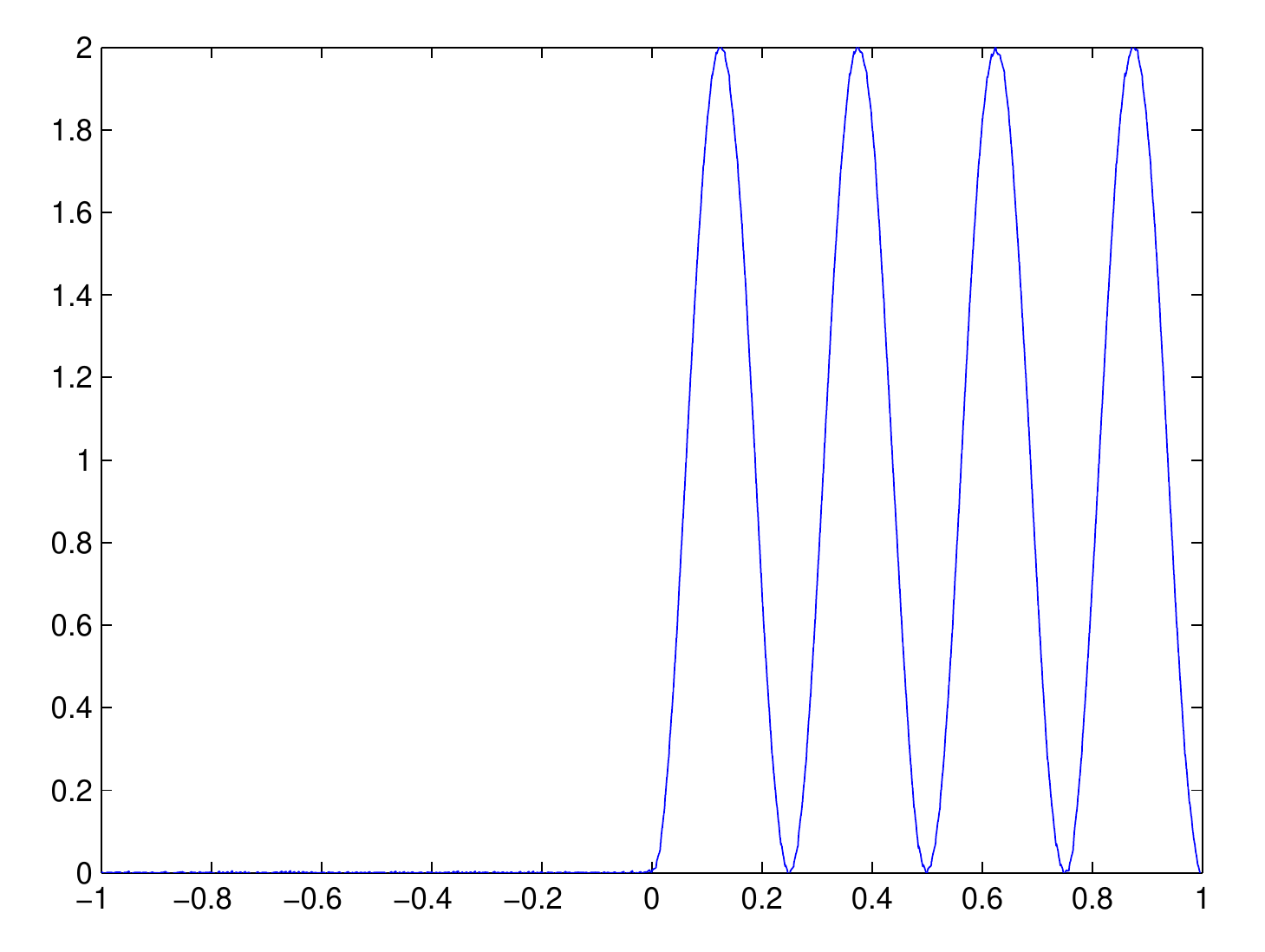}
\includegraphics[width=\textwidth]{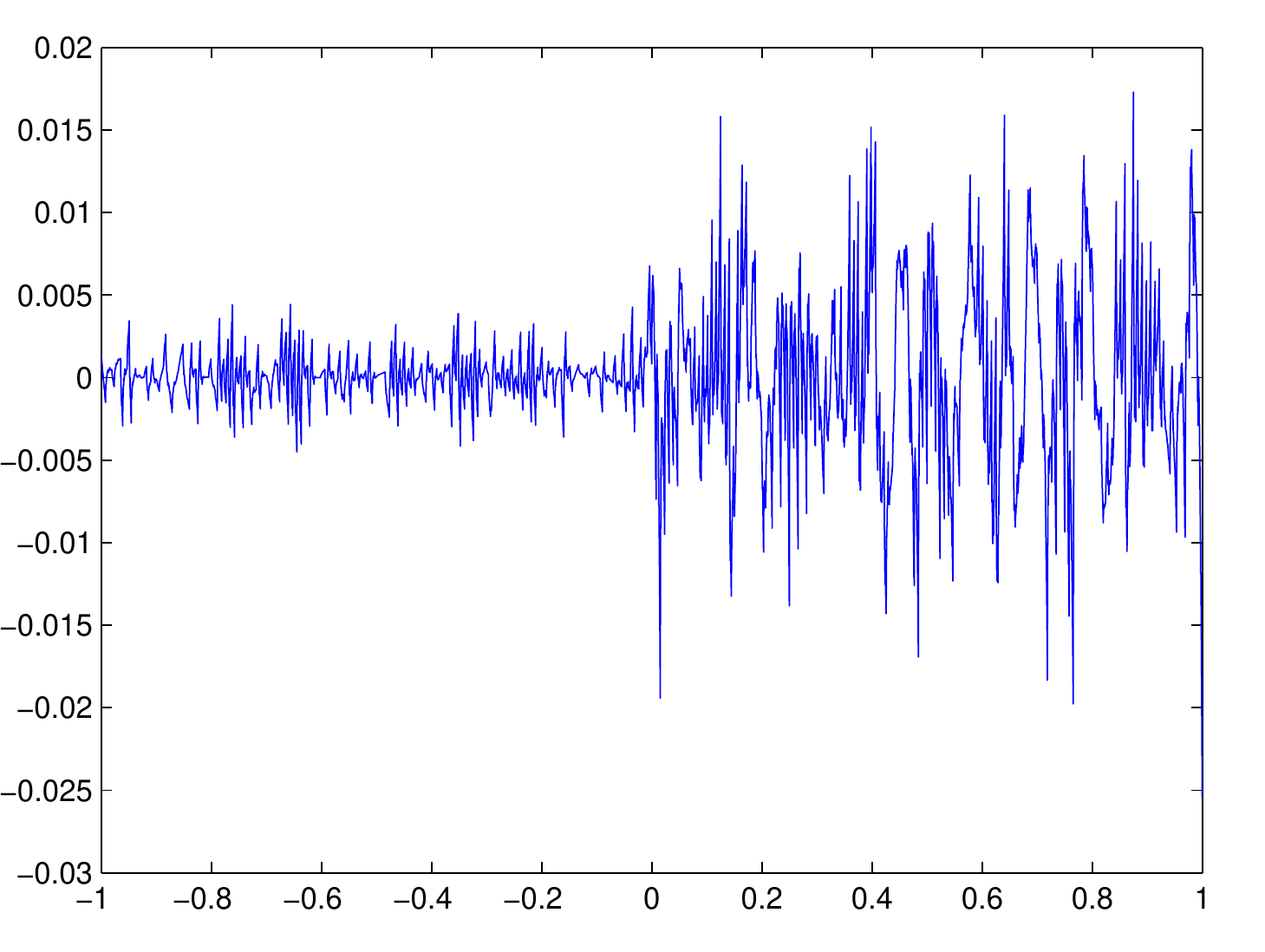} 
  \caption{ \footnotesize Wavelet Reconstruction}
\end{subfigure}
\begin{subfigure}[t]{0.4\textwidth}
  \includegraphics[width=\textwidth]{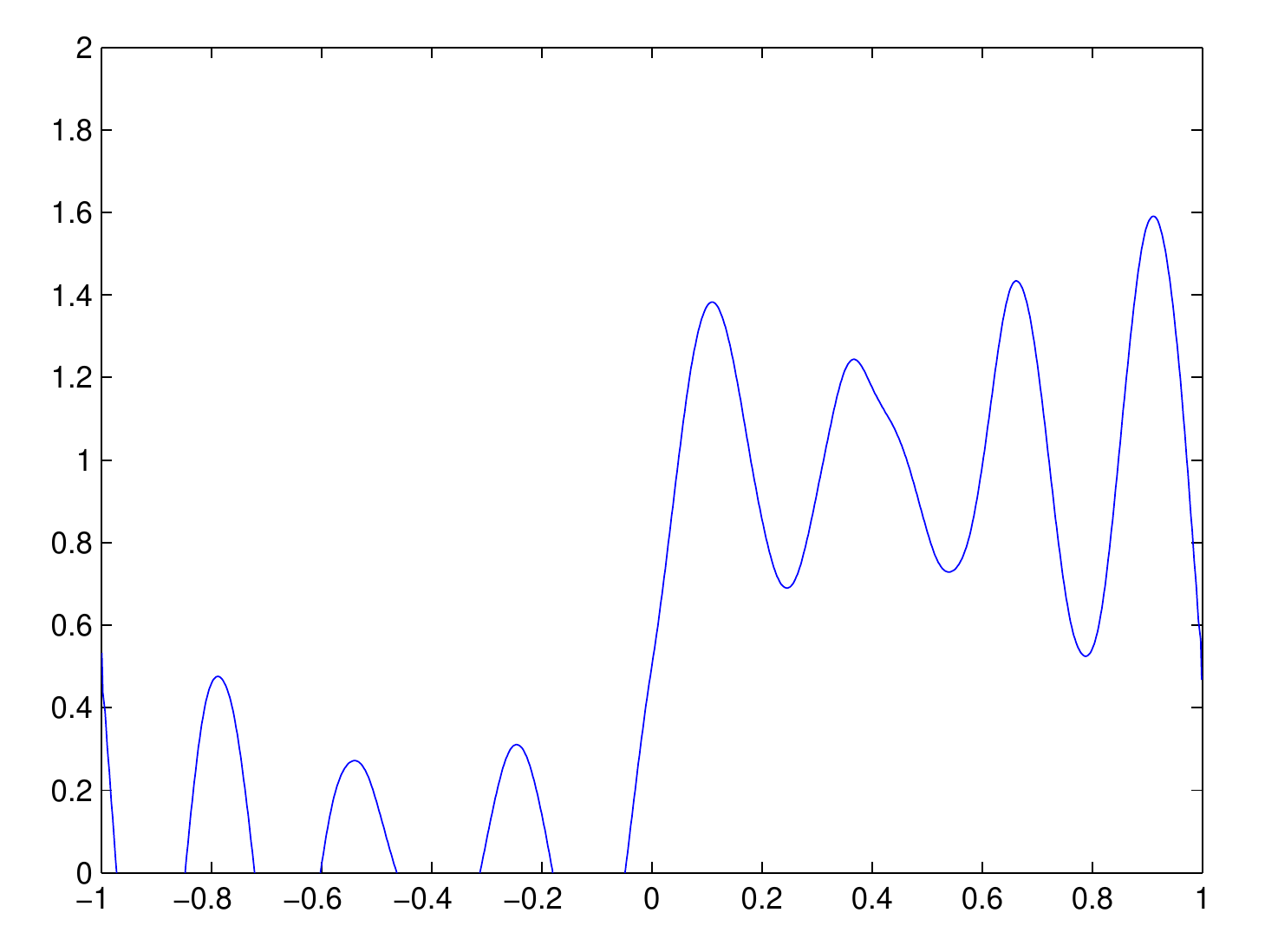}
  \includegraphics[width=\textwidth]{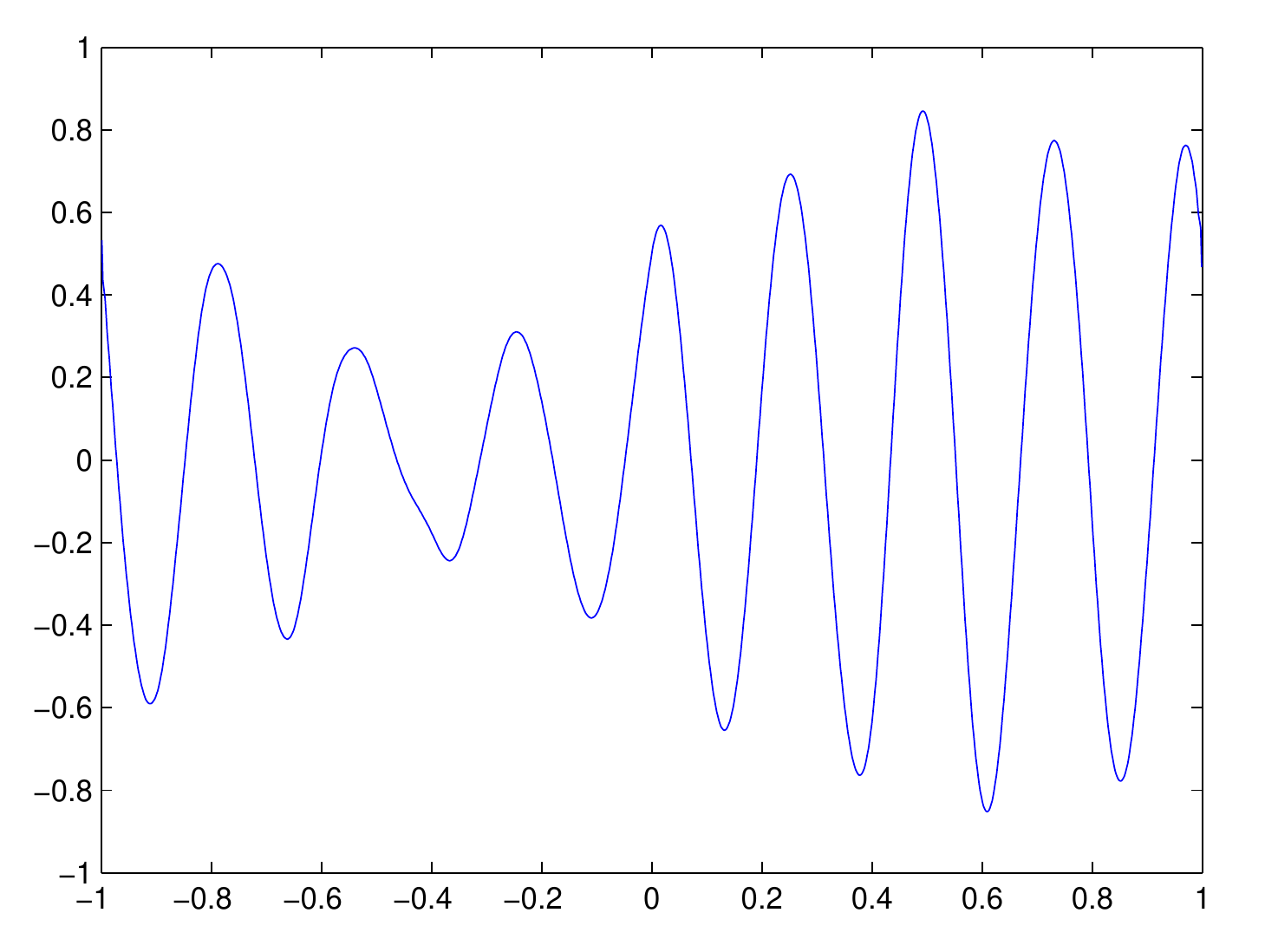}
  \caption{ \footnotesize Polynomial Reconstruction}
\end{subfigure}
\end{center}

\label{reconstructionB}
\end{figure}

Next we turn to reconstructing $f$ using sampling pattern B. Reconstructions are given in Figure \ref{reconstructionB}. First we look at the wavelet reconstruction which has an $L^1$ error of $7.14 \times 10^{-3}$. Since the wavelet basis expansion of $f$ is sparse and we have asymptotic incoherence, we see that we can obtain a good wavelet reconstruction by subsampling roughly $ 50 \%$ of the $501$ Fourier samples. Finally we consider the polynomial reconstruction, with an $L^1$ error of $7.29 \times 10^{-1}$. Due to poor sparsity and slow asymptotic incoherence, subsampling fails to be successful.

This therefore demonstrates that a subsampling pattern should not only be dependent on the function that we are trying to reconstruct, but also on the reconstruction bases that we are using. We must stress here that the ability to find two subsampling patterns, where each gives a better reconstruction in a different basis, relies crucially on the different incoherence structures of the two reconstruction problems and not simply the sparsity structure when decomposed into the two reconstruction bases; the same phenomenon can also be demonstrated if we remove $f$ completely and instead fix the sparsity structure (which means solving for a fixed $\tilde{x}$ in our optimisation setup). Asymptotic incoherence not only facilitates subsampling but also allows us to investigate the link between good subsampling patterns and reconstruction bases.

\section{Optimality} \label{Sec:Optimality}

In Sections \ref{Sec:1DFW} and \ref{Sec:1DFP} we derived bounds on the line coherences for the Fourier-wavelet case (say $U=U_\rw$) the Fourier polynomial case (say $U=U_\rp$)
\be{ \label{Eq:1DResultsSuperShort}
\begin{aligned}
& \mu(\pi_N U_\rw), \mu(U_\rw \pi_N)=\Theta(N^{-1}), \qquad \mu(\pi_N U_\rp), \mu(U_\rp \pi_N)=\Theta(N^{-2/3}),
\\ \Rightarrow \quad 
& \mu(R_N U_\rw), \mu(U_\rw R_N)=\Theta(N^{-1}), \qquad \mu(R_N U_\rp), \mu(U_\rp R_N)=\Theta(N^{-2/3}).
\end{aligned}
}
This required that we work with a frequency ordering for the Fourier basis and leveled/natural orderings for the wavelet/polynomial basis. Our next goal is to show that no other orderings can improve upon the decay rates in (\ref{Eq:1DResultsSuperShort}).

Since we want to compare decay rates with different orderings, we need a precise way of saying one ordering has a slower decay rate than another:
\begin{definition}[Relations on the set of orderings] \label{fasterdecay}
Let $U_1:=[(B_1,\rho_1) , (B_2,\tau)], \ U_2:=[(B_1,\rho_2) , (B_2,\tau)]$. If
\be{ \label{Eq:FasterDecay}
 \mu(R_NU_1) = \mathcal{O}( \mu(R_NU_2)), \quad N \to \infty, 
 }
then we write $\rho_1 \prec \rho_2$ and say that `$\rho_1$ has a faster decay rate than $\rho_2$ for the basis pair $(B_1, B_2)$'. If also $ \rho_2 \prec \rho_1 $ we write $\rho_1 \sim \rho_2$.  These relations, defined on the set of orderings of $B_1$ which we shall denote as $\cR(B_1)$, depend only on the basis pair $(B_1,B_2)$, and are therefore independent of $\tau$.
 \end{definition}
Notice that $ \prec $ is a reflexive transitive relation on $\cR(B_1)$ and $ \sim $ is an equivalence relation on $\cR(B_1)$. Furthermore, we can use the relation to define a partial order on the equivalence classes of $\cR(B_1)$ by the definition
\[ [a] \prec [b] \quad \Leftrightarrow \quad a \prec b, \]
where $[a]$ denotes the equivalence class containing $a$. Furthermore, we say an equivalence class $[a]$ is `optimal' if we have
\[ [a] \prec [b], \qquad \forall b \in \cR(B_1) .\]

\begin{definition}[Optimal ordering]
Given the setup above, then
 any element of the optimal equivalence class is called an `optimal ordering of the basis pair $(B_1,B_2)$'.
 \end{definition}
It shall be shown in Lemma \ref{bestexistence} that optimal orderings always exist . Notice that $\rho$ is an optimal ordering if and only if for every other ordering $\rho'$ we have $\rho \prec \rho'$. An optimal ordering has a corresponding optimal decay rate.

\begin{definition}[Optimal decay rate]
Suppose $\rho: \mathbb{N} \to B_1$ is an optimal ordering for the basis pair     $(B_1,B_2)$ and $U=[(B_1,\rho),(B_2,\tau)]$. Then any decreasing function $f: \mathbb{N} \to \mathbb{R}_{>0}$ which satisfies $f(N) =  \Theta(\mu(R_NU))$ is said to represent the `optimal decay rate' of the basis pair $(B_1,B_2)$. 
\end{definition}

At this point it is worth explaining why we define $\prec$ in terms of block coherences and not line coherences. This is because $\mu(\pi_NU_1) = \mathcal{O}(\mu(\pi_N U_2))$ implies $\mu(R_NU_1) = \mathcal{O}(\mu(R_N U_2))$ but not the other way around. Furthermore, $\mu(\pi_NU_1)$ is often not a decreasing function of $N$ and a statement such as  $\mu(\pi_NU_1)=\Theta(f(N))$ for a decreasing function $f$ is not possible. However, the case when this does hold is very important to us:

\begin{definition}[Optimal In lines]
Let $U=[(B_1,\rho),(B_2,\tau)]$. If $\mu(\pi_NU) =\Theta(f(N))$ for a decreasing function $f:\bbN \to \bbR_{>0}$ then $\rho$ is said to be `optimal in lines' for the basis pair $(B_1,B_2)$.
\end{definition}
From (\ref{Eq:1DResultsSuperShort}) we see that the orderings used in these cases (frequency/leveled/natural) are optimal in lines by definition. We now show that the use of the word `optimal' here is justified:

\begin{proposition} \label{Prop:lineOptImpliesOpt}
Let $U_1=[(B_1,\rho_1),(B_2,\tau)]$ and $U_2=[(B_1,\rho_2),(B_2,\tau)]$. If there exists a decreasing function $f:\bbN \to \bbR_{>0}$ such that
\be{ \label{Eq:lineUndercuts}
f(N) \le \mu(\pi_N U_1), \quad N \in \bbN,
}
then $f(N) \le \mu(R_NU_2)$ for every $N \in \bbN$.
\end{proposition}
\begin{IEEEproof}
Let $\theta(N)$ denote the smallest $m \in \mathbb{N}$ such that
 $ \rho_1(m)   \in  \{ \rho_2(k) \}_{k=N}^{\infty} $ and $m'=m'(N) \in \{ N, N+1, \hdots\}$ be such that $\rho_1(\theta(N))=\rho_2(m'(N))$.
Now notice that $\theta(N) \le N$ since $\{ \rho_2(k) \}_{k=N}^{\infty}$ can only miss at most the first $N-1$ of the $\rho_1(k)$'s. Combining this with the fact that $f$ is decreasing we see that
 \begin{align*}
   f(N) \le f(\theta(N)) & \le  \mu(\pi_{\theta(N)}U_1) = \mu(\pi_{m'(N)}U_2) \le \mu(R_N U_2).
      \end{align*}
\end{IEEEproof}

\begin{corollary} \label{Cor:Optlines2Opt}
Let $U_1=[(B_1,\rho_1),(B_2,\tau)]$ and $U_2=[(B_1,\rho_2),(B_2,\tau)]$. If $\rho_1$ is optimal in lines , i.e. $\mu(\pi_NU_1) =\Theta(f(N))$ for a decreasing function $f:\bbN \to \bbR_{>0}$, then $f(N)=\mathcal{O}(\mu(R_NU_2))$ and $\rho_1 \prec \rho_2$. Consequently, $\rho_1$ is optimal.
\end{corollary}
\begin{IEEEproof}
By definition $\mu(\pi_NU_1) =\Theta(f(N))$ implies $C \cdot f(N) \le \mu(\pi_NU_1)$ for some constant $C>0$. Applying the proposition gives us $C \cdot f(N) \le \mu(R_NU_2)$, i.e. $ f(N)=\mathcal{O}(\mu(R_NU_2))$.

Recall from Lemma \ref{Lem:lineToBlock} that $\mu(\pi_NU_1)=\Theta(f(N))$ implies $\mu(R_NU_1)=\Theta(f(N))$. Therefore $\mu(R_NU_1)=\mathcal{O}(\mu(R_NU_2))$ which by definition means $\rho_1 \prec \rho_2$.

\end{IEEEproof}

We can now use Corollary \ref{Cor:Optlines2Opt} to immediately deduce the bounds in (\ref{Eq:1DResultsSuperShort}) are optimal for their respective basis pairs:

\begin{theorem} \label{Thm:OptimalitySumUp}

\begin{enumerate}
\item \textbf{Fourier-Wavelet Case:} Let $\epsilon \in I_{J,p}$. Frequency orderings are optimal for the basis pair $(B_\rf(\epsilon),B_\rw)$. Leveled orderings are optimal for the basis pair $(B_\rw,B_\rf(\epsilon))$. In both cases the optimal decay rate is $\Theta(N^{-1})$. These statements still hold with $B_\rw$ replaced with the boundary wavelet basis $B_{\rb \rw}$ and $\epsilon \in I_{J,p}$ by $\epsilon \in (0,1/2]$.
\item \textbf{Fourier-Polynomial Case:} Let $\epsilon \in (0,1/2]$. Frequency orderings are optimal for the basis pair $(B_\rf(\epsilon),B_\rp)$. Leveled orderings are optimal for the basis pair $(B_\rp,B_\rf(\epsilon))$. In both cases the optimal decay rate is $\Theta(N^{-2/3})$.
\end{enumerate}

\end{theorem}

\section{Theoretical Limits} \label{Sec:General}

We now look at the general abstract case where $U$ is an isometry and ask; is there a universal lower bound on the block coherences?

\begin{theorem} \label{Thm:IsometryDecayLowerBound}
Let $U \in \cB(l^2(\bbN))$ be an isometry.  Then
$
\sum_{N} \mu(R_N U)
$
diverges. 
\end{theorem}
\begin{IEEEproof}
Suppose that $\sum_{N} \mu(R_N U)$ converges, Then, we can find $N' \in \bbN$ such that $\sum_{N=N'}^\infty \mu(R_N U) \le 1/4^2$. Therefore if we write $U=(u_{i,j})_{i,j \in \bbN}$ then 
\be{ \label{vectortailbound}
\sum_{N=N'}^\infty |u_{N,j}|^2 \le \sum_{N=N'}^\infty \mu(R_N U) \le 1/4^2, \qquad j \in \bbN .
}
Now define the vectors 
\[v_j:=(u_{i,j})_{i \in \bbN}, \quad v^1_j:=(u_{i,j})^{N'-1}_{i =1}, \quad v^2_j:=(u_{i,j})^\infty_{i =N'}, \qquad j \in \bbN. \]
Inequality (\ref{vectortailbound}) says that $\| v^2_j \|_2 \le 1/4$ for every $j \in \bbN$. Since $U$ is an isometry, we know its columns are normalised, i.e. $\|v_j\|_2=1,$ and so we deduce $\|v_j^1\|_2 \ge 3/4$ for every $j \in \bbN$. Let $w_j:=v^1_j/\|v^1_j \|_2, \ j \in \bbN$. Since the $w_j \in \bbC^{N'-1}$ are all finite dimensional we claim that
\be{ \label{toomanyvectors}
\sup_{\substack{j,j' \in \{1,...,M\} \\ j \neq j'}} | \langle w_j, w_{j'} \rangle | \to 1, \quad \text{as} \quad M \to \infty.
}
To see this, notice that for every $\epsilon>0$, there exists a $\delta>0$, such that for all $j \in \bbN$ the set $W_j(\epsilon):=\{ w \in \bbC^{N'-1} : | \langle w_j, w \rangle | > 1-\epsilon \} $ contains the open set $B_\delta(w_j)$ of radius $\delta$ centered at $w_j$. It must be the case that there are $j_1,j_2 \in \bbN,  \ j_1 \neq j_2$ such that $B_{\delta/2}(w_{j_1}) \cap B_{\delta/2}(w_{j_2}) \neq \varnothing ,$ else the union 
\[ \bigcup_{j \in \bbN} B_{\delta/2}(w_j) \cup \bigcup_{ \substack{w \not \in \bigcup_{j \in \bbN} B_{\delta/2}(w_j) \\ w \in \bbC^{N'-1}}} B_{\delta/4}(w) , \]
would form an open cover of the unit ball in $\bbC^{N'-1}$ with no finite subcover\footnote{Any finite subcover would miss infinitely many of the points $w_j$.}, contradicting compactness of the unit ball in $\bbC^{N'-1}$. Since $B_{\delta/2}(w_{j_1}) \cap B_{\delta/2}(w_{j_2}) \neq \varnothing , \ w_{j_1} \in  B_{\delta}(w_{j_2}) \subset W_{j_2}(\epsilon)$ and so $| \langle w_{j_1}, w_{j_2} \rangle | > 1-\epsilon$. Since $\epsilon>0$ was arbitrary we have proved (\ref{toomanyvectors}).

Therefore, by (\ref{toomanyvectors}) we know there exists $j_1, j_2 \in \bbN, \ j_1 \neq j_2 $ such that $| \langle w_{j_1}, w_{j_2} \rangle | > 1/2$ and therefore we deduce that 
\be{ \label{finitepart}
| \langle v^1_{j_1}, v^1_{j_2} \rangle | > \frac{1}{2} \|v^1_{j_1} \|_2   \|v^1_{j_2} \|_2 > \frac{3^2}{2 \cdot 4^2}.
}
Furthermore, since $\| v^2_{j_1} \|_2, \| v^2_{j_1} \|_2 \le 1/4$ we know that 
\be{ \label{infinitepart}
| \langle v^2_{j_1}, v^2_{j_2} \rangle | \le \| v^2_{j_1} \|_2  \| v^2_{j_1} \|_2
\le \frac{1}{4^2}.
}
Therefore, combining (\ref{finitepart}) with (\ref{infinitepart}) gives us
\[
\begin{aligned}
| \langle v_{j_1}, v_{j_2} \rangle | & = |\langle v^1_{j_1}, v^1_{j_2} \rangle + \langle v^2_{j_1}, v^2_{j_2} \rangle| \ge |\langle v^1_{j_1}, v^1_{j_2} \rangle| - |\langle v^2_{j_1}, v^2_{j_2} \rangle|  
\\ & \ge \frac{3^2}{2 \cdot 4^2} - \frac{1}{4^2} = \frac{7}{2 \cdot 4^2} >0.
\end{aligned}
\]
However, since $U$ is an isometry and $j_1 \neq j_2$, we know that $\langle v_{j_1}, v_{j_2} \rangle =0$ and therefore we have a contradiction.
\end{IEEEproof}
\begin{corollary}
Let $U \in \cB(l^2(\bbN))$ be an isometry.  Then there does not exist an $\epsilon > 0$ such that
$$
\mu(R_N U)  = \mathcal{O}(N^{-1-\epsilon}) , \qquad \ N \to \infty.
$$
\end{corollary}

Noting the above corollary and that $\mu(W)\ge N^{-1}$ is the best lower bound possible for any finite isometry $W \in \bbC^N \times \bbC^N$ it might be tempting to believe $\mu(R_N U) = \Theta(N^{-1})$ is the best decay rate we can achieve for an isometry $U \in \mathcal{B}(\ell^2(\bbN))$. However, it turns out that Theorem \ref{Thm:IsometryDecayLowerBound} cannot be improved without imposing additional conditions on $U$:

\begin{theorem} \label{Thm:IncoherenceCounter}
Let $f,g: \bbN \to \bbR$ be any two strictly positive decreasing functions and suppose that $\sum_N f(N)$ diverges. Then there exists $U \in \cB(l^2(\bbN))$ an isometry with 
\be{ \label{strongerthanoptimal}
\mu(R_N U) \le f(N), \quad \mu(U R_N) \le g(N), \qquad N \in \bbN .
}
\end{theorem}
\begin{IEEEproof}
The proof is constructive. We may assume without loss of generality that $f(N), \ g(N) \le 1$ for all $N \in \bbN$. We will construct a matrix $U=(u_{i,j})_{i,j \in \bbN}$ satisfying (\ref{strongerthanoptimal}) with normalised columns, $v_j:=(u_{i,j})_ {i \in \bbN}, \ j \in \bbN,$ having disjoint support. With this in mind we partition $\bbN$ as follows:
\[ \bbN = \bigcup_{i=1}^\infty \Omega_i, \quad \Omega_i:=2^{i-1} \bbN \setminus 2^i \bbN. \]
Let $j \in \bbN$ be fixed and define recursively (for\footnote{Here we use the convention that $\sum_{i=1}^{N-1}$ is an empty sum if $N=1$.} $N \in \bbN$  )
\be{ \label{Uvectdefine}
(v_j)_N=
\begin{cases}
\big( g(j) f(N) \big)^{1/2}, & \text{if} \quad \sum_{i=1}^{N-1} ((v_j)_i)^2 + g(j)f(N) \le 1, \quad N \in \Omega_j,
\\  \big(1-\sum_{i=1}^{N-1}((v_j)_i)^2 \big)^{1/2},  &  \text{if} 
 \quad \sum_{i=1}^{N-1} ((v_j)_i)^2 \le 1, \\ & \qquad \sum_{i=1}^{N-1} ((v_j)_i)^2 + g(j)f(N) \ge 1,  \quad N \in \Omega_j,
 
\\ 0, & \text{Otherwise.}
\end{cases}
}
It is immediate from the definition that $v_j$ is supported on $\Omega_j$ and $((v_j)_N)^2 \le f(N)g(j)$ for every $N, j \in \bbN$ which implies that (\ref{strongerthanoptimal}) holds. Furthermore, it easy to show by induction on $N$ that $\| v_j\|_2 \le 1$ . Since $f$ is decreasing and by the structure of the set $\Omega_j$, $\sum_{N \in \Omega_j} f(N)$ diverges for every $j$ and consequently there is an $N' \in \bbN$ such that
\[ 
 \sum_{\substack{N \in \Omega_j \\ N \le N'}} g(j)f(N) \ge 1, \qquad  \sum_{\substack{N \in \Omega_j \\ N \le N'-1}} g(j)f(N) \le 1.
\]
For $N \le N'-1, N \in \Omega_j$ we fall into the first case of (\ref{Uvectdefine}), however for $N=N'$ we fall into case 2, and therefore $\sum_{i=1}^{N'} ((v_j)_i)^2 = 1$. This means $\| v_j \|_2=1$ for every $j$ and consequently $U$ is an isometry.
\end{IEEEproof}
Although this negative result shows that we cannot define an analogue of perfect incoherence for asymptotic incoherence, if we restrict our decay function to be a power law, i.e. $f(N):= CN^{- \alpha}$ for some constants $\alpha, C >0$ then the largest possible value of $\alpha>0$ such that (\ref{strongerthanoptimal}) holds for an isometry $U$ is $\alpha=1$, which is what we achieved in the Fourier-wavelet case.

\section{Alternative Notions of Optimality} \label{Sec:AltOpt}

Before we finish, we would to discuss further why we work with Definition \ref{fasterdecay} as our notion of optimal decay and discuss a possible alternative. One argument against the definition of optimality we use is that is only unique up to constants, since it relies only on order notation. This is somewhat inconvenient if one wants to work with concrete estimates. Therefore it may be tempting to strengthen the notion of optimality in some way. One possible alternative would be the following:

\begin{definition}[Best ordering]
Let $(B_1,B_2)$ be a basis pair. Then any ordering $\rho: \mathbb{N} \to B_1$ is said to be a `best ordering' if for any other ordering $\tau$ of $B_2$ and $U=[(B_1,\rho),(B_2,\tau)]$ we have that the function $g(N):= \mu(\pi_N U)$ is decreasing.
\end{definition}

Notice that for a best ordering we have $\mu(\pi_N U)=\mu(R_N U)$. If $\rho'$ is any other ordering and $U'=[(B_1,\rho'),(B_2,\tau)]$ then since $R_N U'$ must contain one of the first $N$ lines of $U$ we must have that
\[ \mu( R_N U') \ge \min_{M=1,...N}\mu( \pi_M U) \ge \mu(\pi_N U)= \mu(R_N U), \] 
and we deduce that $\rho \prec \rho'$. This shows that any best ordering is optimal.

\begin{figure}[ht]
\begin{center}

\begin{subfigure}[t]{0.48\textwidth}
\begin{center}
 \hspace{0.1em} \includegraphics[width=\textwidth]{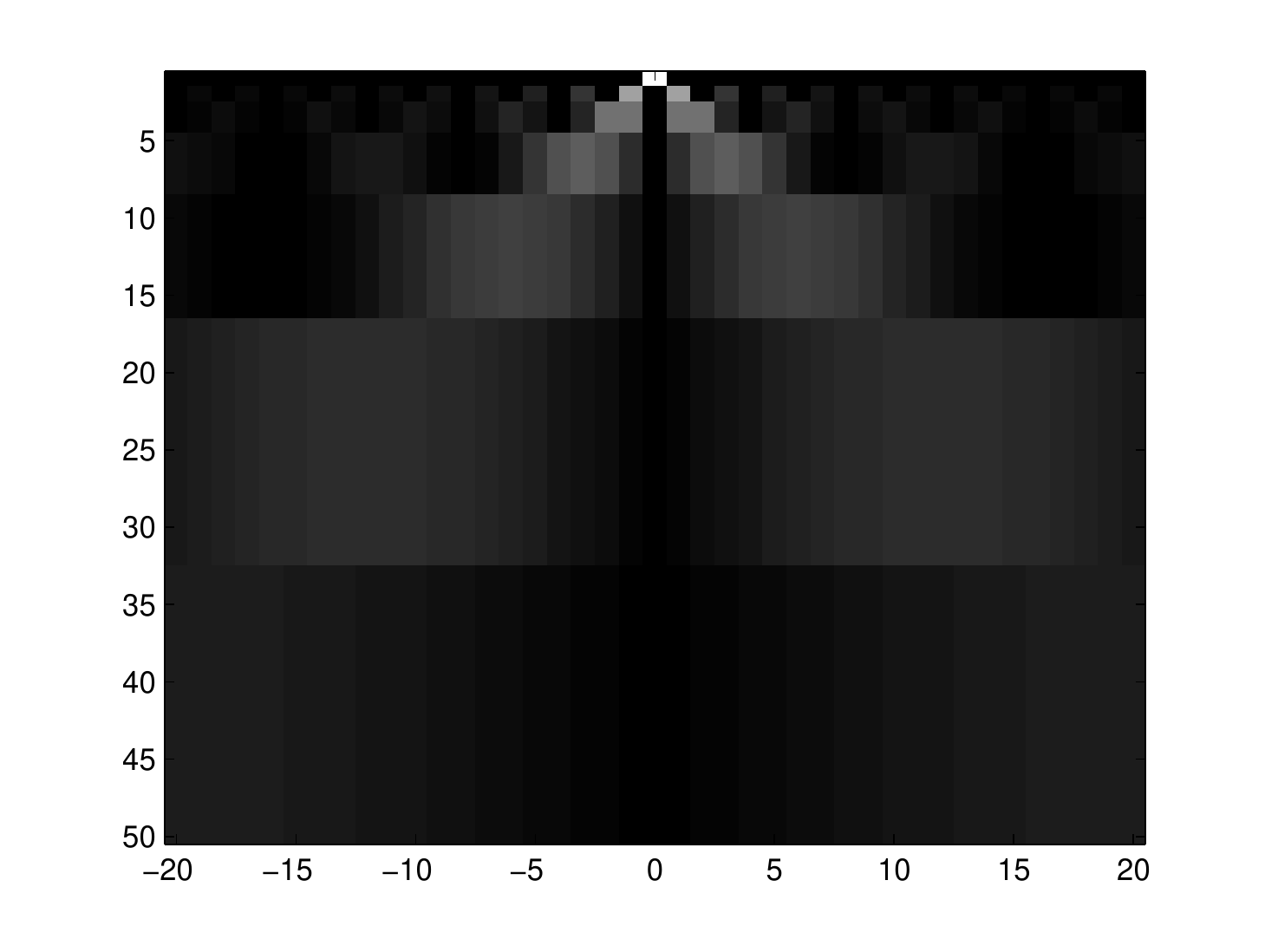}
\end{center}
  \includegraphics[width=\textwidth]{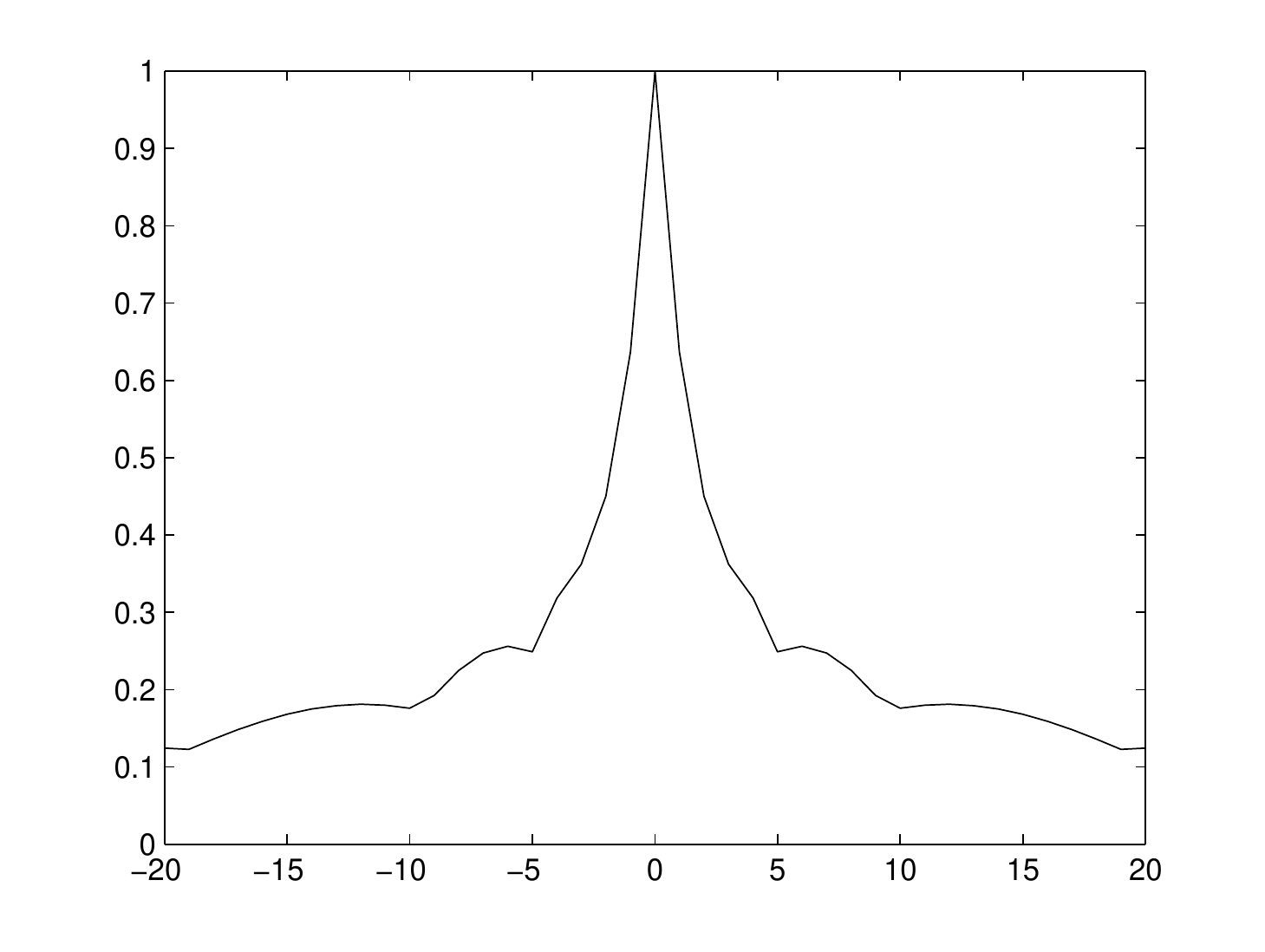}
  \caption{\footnotesize Incoherence matrix and column maxima for a Haar wavelet basis (with Fourier).}
  \label{f:glpu-256-full}
\end{subfigure}
\begin{subfigure}[t]{0.48\textwidth}
\begin{center}
  \hspace{0.1em} \includegraphics[width=\textwidth]{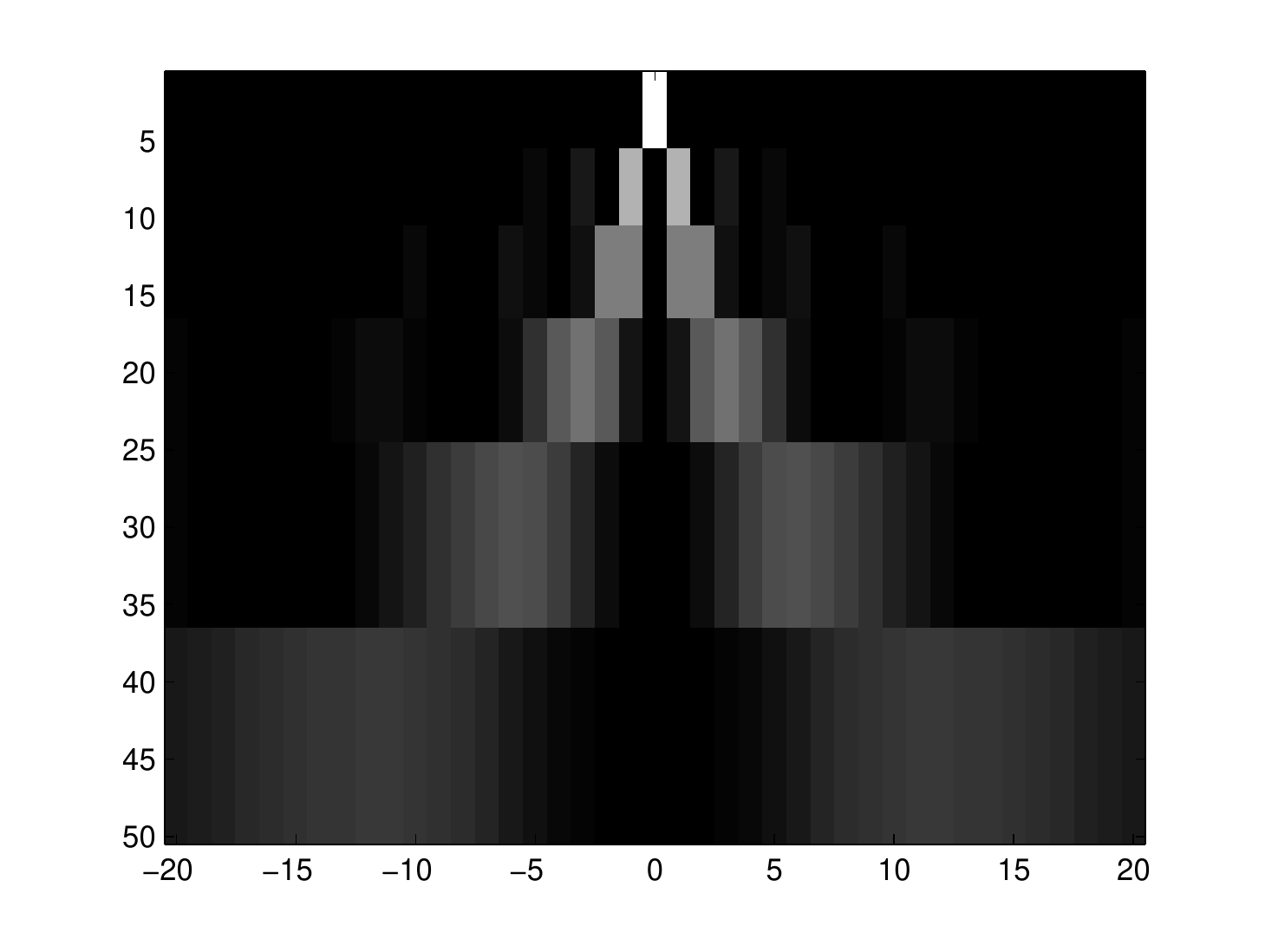} 
  \end{center}
  \includegraphics[width=\textwidth]{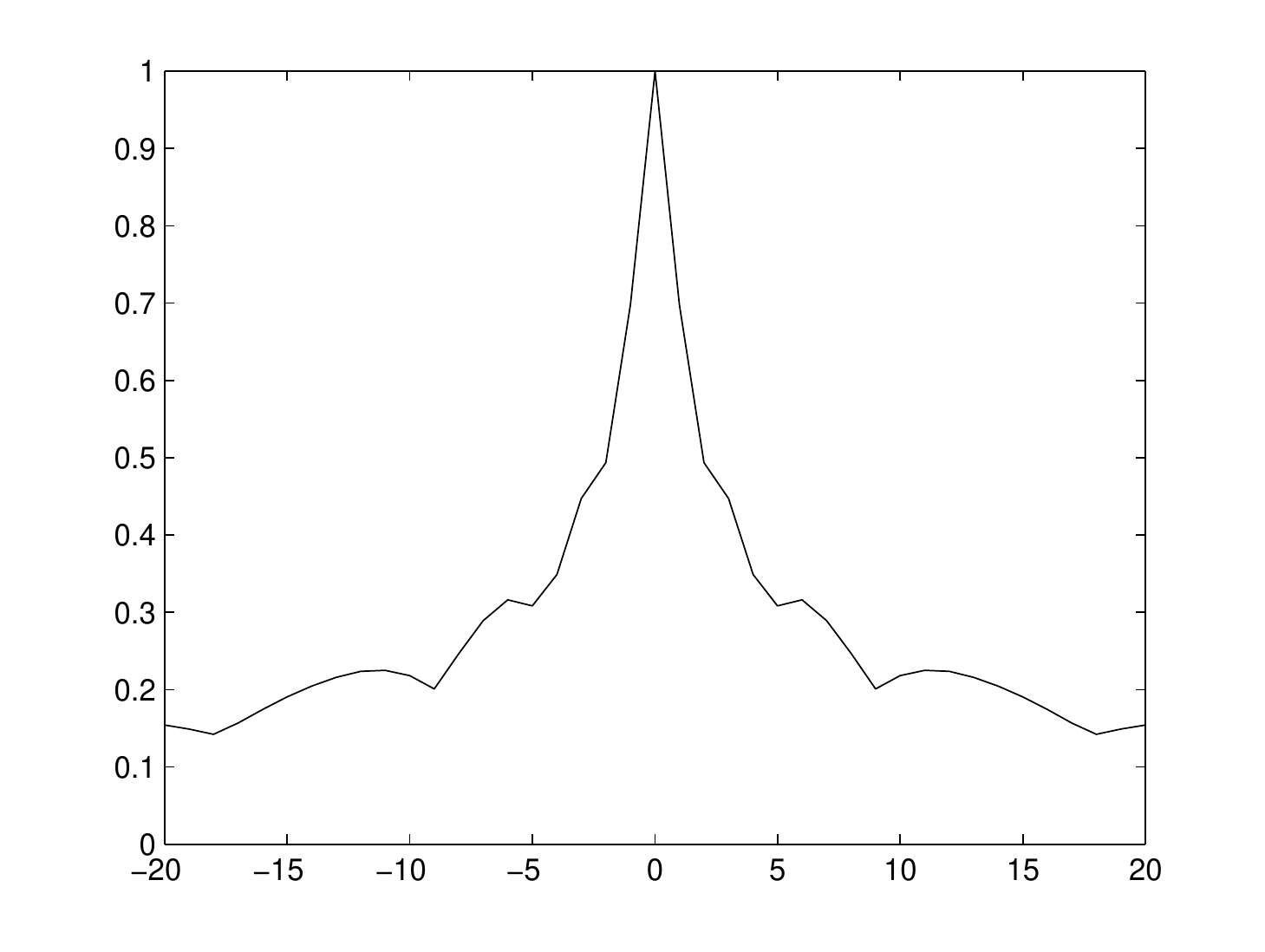}
  \caption{ \footnotesize Incoherence matrix and column maxima for Daubechies6 wavelet basis.}
  \label{DB3CoherenceGraph}
\end{subfigure}
\end{center}
\caption{Here are two ($20 \times 20$ centrally truncated) wavelet-Fourier Incoherence matrices (brighter means larger absolute value) and their corresponding column maxima. The columns denote the Fourier basis (viewed as $\mathbb{Z}$) and the rows denote the wavelet basis (ordered top to bottom). Notice that there is a slight difference in the best orderings (by looking around $-10,+10$ on the horizontal axis) even though the general decay rate is similar. The maxima are taken over a much larger matrix to ensure accuracy.} 
\label{1Dlineincoherence}
\end{figure}

\begin{lemma} \label{bestexistence}
Suppose that we have a basis pair $(B_1, B_2)$. Then one of the following two results must hold:
\begin{itemize}
\item[(1)] There is at least one best ordering.
\item[(2)] Every ordering of $B_1$ is optimal for $(B_1, B_2)$.
\end{itemize}
\end{lemma}
\begin{IEEEproof}
Let $\rho: \bbN \to B_1$, $\tau: \bbN \to B_2$ be any orderings of $B_1, B_2$ respectively and $U=[(B_1,\rho),(B_2,\tau)]$. Now first assume that for any finite subset $D \subset \bbN$  
\be{ \label{optimalsupremum}
\sup_{N \in \bbN \setminus D} \mu( \pi_N U),
}
is attained for some $N \in \bbN \setminus D$. In this case we can then construct a best ordering $\rho^*:\mathbb{N} \to B_1$ inductively by letting (for $N=1$)
\[ \rho^*(1) \in \operatorname*{argmax}_{f \in B_1} \sup_{n \in \mathbb{N}} | \langle \tau(n) , f \rangle |, \]
and for $N \ge 2$ we set
\[ \rho^*(N) \in \operatorname*{argmax}_{\substack{f \in B_1 \\ f \notin \{ \rho^*(1),...,\rho^*(N-1) \} }} \sup_{n \in \mathbb{N}} | \langle \tau(n) , f \rangle |. \]
Note that it is clear from the construction that this is an actual ordering.
Therefore if our original assumption holds we conclude that 1) must hold too.
If our assumption does not hold this means there exists a finite subset $D \subset \bbN$ such that the supremum (\ref{optimalsupremum}) is not attained for any $N \in \bbN \setminus D$. This means that if we remove finitely many elements from $\bbN \setminus D$ the supremum will remain unchanged. Therefore if $N'$ is the largest natural number in $D$ we find that 
\[ \mu(R_M U) = \sup_{N \in \bbN \setminus D} \mu( \pi_N U), \qquad \forall M > N', \]
and so $\mu(R_N U)$ is eventually constant as a function of $N$. This means that for \emph{any} ordering $\rho'$ of $B_1$ and $U'=[(B_1,\rho'),(B_2,\tau)]$, $\mu(R_N U')$ is eventually constant. If follows that any two orderings of $B_1$ are equivalent under $\sim$ and consequently 2) holds.
\end{IEEEproof}
\begin{lemma} 
Suppose that we have a basis pair $(B_1, B_2)$ with two orderings $\rho: \bbN \to B_1$, $\tau: \bbN \to B_2$ of $B_1, B_2$ respectively. If $U=[(B_1,\rho),(B_2,\tau)]$ satisfies
\[ \mu(\pi_N U) \to 0 \quad \text{as} \quad N \to \infty, \]
then a best ordering exists.
\end{lemma}
\begin{IEEEproof}
The supremum (\ref{optimalsupremum}) is always attained and therefore we fall into case 1) of the previous lemma.
\end{IEEEproof}
These two results tell us that optimal orderings always exist and best orderings exist in cases where we expect decay in the line/block coherences, i.e. every case where we want to study coherence decay. 

Therefore, why did we not work with the definition of best ordering as our notion of optimality instead? The answer to this question is that best orderings are more exotic and far less simple to describe that optimal orderings. Figure \ref{1Dlineincoherence} shows the Fourier-wavelet case where the best orderings are wavelet-dependent, even though we can describe optimal orderings in a wavelet-independent manner using frequency orderings. Although the difference between best orderings is very minor in Figure \ref{1Dlineincoherence}, this difference becomes considerable when working with the higher dimensional Fourier-wavelet cases.

\section{Outlook and Future Work} \label{Sec:Outlook}

The results presented here can be extended to higher dimensional cases, but the complexity increases significantly and is the subject of future work. Unlike the one-dimensional Fourier/wavelet case we covered here, the optimal orderings in the multidimensional Fourier/(separable) wavelet case can be wavelet dependent in more than three dimensions however the optimal decay rates are the same ($\Theta(N^{-1})$). These results rely heavily on the optimality theory developed here. An alternative direction to develop upon this work would be to cover other reconstruction bases other than the wavelet and polynomial bases presented here.

\section{Acknowledgements} \label{Sec:Ack}

A. D. Jones acknowledges EPSRC grant EP/H023348/1. B. Adcock acknowledges NSF DMS grant 1318894. A. C. Hansen acknowledges support from a Royal Society University Research Fellowship as well as EPSRC grant EP/L003457/1.

\bibliographystyle{IEEEtran}
\bibliography{IEEEabrv,bibfirst}

\end{document}